\documentclass{aa}

\usepackage[english]{babel}
\usepackage{graphicx}
\usepackage{amsmath}
\usepackage{amssymb}
\usepackage{txfonts}
\usepackage{wasysym}
\usepackage{gensymb}

\usepackage[colorlinks=true, linkcolor=blue, citecolor=blue, urlcolor=blue]{hyperref}
\usepackage[singlelinecheck=true,format=default,font=small,labelfont=bf]{caption}
\usepackage{natbib}
\usepackage{color}
\usepackage{natbib}
\usepackage{multirow}
\usepackage{array}

\bibpunct{(}{)}{;}{a}{}{,} 

\begin{document}

\title{The \textit{Herschel}\thanks{\textit{Herschel} is an ESA space observatory with science instruments provided by European-led Principal Investigator consortia and with important participation from NASA.} Exploitation of Local Galaxy Andromeda (HELGA)}
\subtitle{IV. Dust scaling relations at sub-kpc resolution}

\author{S. Viaene\inst{1} \and J. Fritz\inst{1} \and M. Baes\inst{1} \and G.J. Bendo\inst{2} \and J.A.D.L. Blommaert\inst{3,4} \and M. Boquien\inst{5} \and A. Boselli\inst{6} \and L. Ciesla\inst{7} \and L. Cortese\inst{8}  \and I. De Looze\inst{1} \and W.K. Gear\inst{9} \and G. Gentile\inst{1,4} \and T.M. Hughes\inst{1} \and T. Jarrett\inst{10,11} \and O. \L. Karczewski\inst{12} \and M.W.L. Smith\inst{9} \and L. Spinoglio\inst{13} \and A. Tamm\inst{14} \and E. Tempel\inst{14,15} \and D. Thilker\inst{16} \and J. Verstappen\inst{1}}

\institute{Sterrenkundig Observatorium, Universiteit Gent, Krijgslaan 281, B-9000 Gent, Belgium\\
	\email{sebastien.viaene@ugent.be}
	\and 
	Jodrell Bank Centre for Astrophysics, School of Physics and Astronomy, University of Manchester, Oxford Road, Manchester M13 9PL, UK
	\and 
	Instituut voor Sterrenkunde, Katholieke Universiteit Leuven, Celestijnenlaan 200D, B-3001 Leuven, Belgium
	\and
	Department of Physics and Astrophysics, Vrije Universiteit Brussel, Pleinlaan 2, 1050 Brussels, Belgium
	\and
   	Institute of Astronomy, University of Cambridge, Madingley Road, Cambridge, CB3 0HA, UK
	\and
	Laboratoire d'Astrophysique de Marseille, UMR 6110 CNRS, 38 rue F. Joliot-Curie, F-13388 Marseille, France
	\and
   	University of Crete, Department of Physics, Heraklion 71003, Greece
   	\and
	Centre for Astrophysics \& Supercomputing, Swinburne University of Technology, Mail H30 - PO Box 218, Hawthorn, VIC 3122, Australia	
	\and 
	School of Physics and Astronomy, Cardiff University, Queens Buildings, The Parade, Cardiff CF24 3AA, UK
	\and 
    Infrared Processing and Analysis Center, California Institute of Technology, Pasadena, CA 91125, USA 
	\and 
	Astronomy Department, University of Cape Town, Rondebosch 7701, South Africa
   	\and
   	Department of Physics and Astronomy, University of Sussex, Brighton, BN1 9QH, UK.
   	\and 
   	Istituto di Astrofisica e Planetologia Spaziali, INAF-IAPS, Via Fosso del Cavaliere 100, I-00133 Roma, Italy
   	\and
   	Tartu Observatory, Observatooriumi 1, 61602 T\~oravere, Estonia
   	\and
   	National Institute of Chemical Physics and Biophysics, R\"avala pst 10, Tallinn 10143, Estonia
   	\and
   	Department of Physics and Astronomy, Johns Hopkins University, 3701 San Martin Drive, Baltimore, MD 21218, US
}

\abstract{Dust and stars play a complex game of interactions in the interstellar medium and around young stars. The imprints of these processes are visible in scaling relations between stellar characteristics, star formation parameters, and dust properties.}{In the present work, we aim to examine dust scaling relations on a sub-kpc resolution in the Andromeda galaxy (M31). The goal is to investigate the properties of M31 on both a global and local scale and compare them to other galaxies of the local universe.}{New \textit{Herschel} observations are combined with available data from GALEX, SDSS, WISE, and \textit{Spitzer} to construct a dataset covering UV to submm wavelengths. All images were brought to the beam size and pixel grid of the SPIRE $500~\mu\mathrm{m}$ frame. This divides M31 in $22437$ pixels of $36$ arcseconds in size on the sky, corresponding to physical regions of $137 \times 608$ pc in the galaxy's disk. A panchromatic spectral energy distribution was modelled for each pixel and maps of the physical quantities were constructed. Several scaling relations were investigated, focussing on the interactions of dust with starlight.}{We find, on a sub-kpc scale, strong correlations between $M_\mathrm{dust}/M_\star$ and NUV--r, and between $M_\mathrm{dust}/M_\star$ and $\mu_\star$ (the stellar mass surface density). Striking similarities with corresponding relations based on integrated galaxies are found. We decompose M31 in four macro-regions based on their FIR morphology; the bulge, inner disk, star forming ring, and the outer disk region. In the scaling relations, all regions closely follow the galaxy-scale average trends and behave like galaxies of different morphological types. The specific star formation characteristics we derive for these macro-regions give strong hints of an inside-out formation of the bulge-disk geometry, as well as an internal downsizing process. Within each macro-region, however, a great diversity in individual micro-regions is found, regardless of the properties of the macro-regions. Furthermore, we confirm that dust in the bulge of M31 is heated only by the old stellar populations.}{In general, the local dust scaling relations indicate that the dust content in M31 is maintained by a subtle interplay of past and present star formation. The similarity with  galaxy-based relations strongly suggests that they are in situ correlations, with underlying processes that must be local in nature.}


\titlerunning{Dust scaling relations in M31}
\authorrunning{S. Viaene}

\maketitle

\section{Introduction}

The interstellar medium (ISM) harbours a rich variety of materials that all interact with one another through multiple chemodynamical processes. Hydrogen is by far the most abundant and occurs primarily in its neutral form, varying from warm ($\sim 8000$~K) to cold gas ($\sim 80$~K). In the very cold and dense environments, it is converted into molecular hydrogen (H$_2$). The presence of ionized hydrogen (H{\sc{ii}})  increases as new stars irradiate the neutral gas. The formation of stars and their associated nucleosynthesis are the principal processes driving the chemical enrichment of the ISM. A good fraction of the produced metals are locked up in dust grains of various sizes. The micron-sized grains are thought to be silicic and carbonaceous in nature, while there is a population of nanometre-sized polycyclic aromatic hydrocarbons (PAHs).

Although it only accounts for a relatively small fraction of the ISM mass, interstellar dust plays a vital role as a catalyst in the formation of H$_2$, which is crucial in the star formation process. At the same time,  dust tends to absorb up to $50\%$ of the optical and ultraviolet (UV) light of stars \citep{PopescuTuffs2002}, heavily affecting our view of the universe. It re-emits the absorbed energy at longer wavelengths in the mid-infrared (MIR), far-infrared (FIR), and submillimetre (submm) bands. To study the ISM in greater detail, spatially resolved FIR observations are crucial as they constrain the local dust distribution and properties. Recent space missions were able to uncover the MIR to submm window in great detail with the Spitzer Space Telescope \citep{Spitzer}, the \textit{\textit{Herschel}} Space Observatory \citep{Herschel}, and the Widefield Infrared Survey Explorer \citep[WISE,][]{WISE}.

Correlations between the main properties of dust, gas, and stars, also known as \textit{scaling relations}, define a tight link between these constituents. In the past, relations between the dust-to-gas ratio on the one hand and metallicity or stellar mass on the other hand were the only notable relations that were being investigated \citep{Issa1990,Lisenfeld1998,Popescu2002,Draine2007,Galametz2011}. Only recently were other scaling laws more systematically investigated. Namely, the ratio of dust to stellar mass, the specific dust mass, which was found to correlate with the specific star formation rate (i.e. star formation rate divided by the stellar mass, \citealt{Brinchmann2004, daCunha2010,Rowlands2012,SmithD2012}), NUV--r colour (i.e. the difference in absolute magnitude between the GALEX NUV and SDSS $r$ band), and stellar mass surface density $\mu_\star$ \citep{Cortese2012,Agius2013}.

Each of the above scaling laws was derived on a galaxy-galaxy basis, considering galaxies as independent systems in equilibrium. In order to fully understand the coupling of dust with stars and the ISM, we must zoom in on individual galaxies. This is however troublesome at FIR/submm wavelengths because of the limited angular resolution. Only in the last few years, we have been able, thanks to \textit{Herschel}, to observe nearby galaxies in the FIR and submm spectral domain whilst achieving sub-kpc resolutions for the closest galaxies ($< 5.7$~Mpc) in the $500~\mu\mathrm{m}$ band. Dust scaling relations on subgalactic scales were thus so far limited to gas-to-dust ratios for a handful of local galaxies \citep[see e.g.][]{MunozMateos2009b, Bendo2010a, Magrini2011, Sandstrom2012, Parkin2012}.

Today, exploiting PACS \citep{PACS} and SPIRE \citep{SPIRE}, plus the 3.5-metre mirror onboard \textit{Herschel}, IR astronomy has gone a leap forward. Spatial resolutions and sensitivities have been reached that allow us, for the first time, to accurately characterise the dust emission in distinct regions of nearby galaxies. 

The observed spectral energy distribution (SED) at these frequencies can be reproduced by means of theoretical models. In particular, a modified black body can be fitted to the observed FIR ($\lambda >100~\mu\mathrm{m}$) data extracted from sub-kpc regions in galaxies \citep[see e.g.][]{Smith2010,Hughes2014}. While still a step forward with respect to previous works, this approach inevitably suffers from some simplifications. For example, if dust is heated by different sources, it cannot be truthfully represented by a single temperature component. Other studies make use of the physical dust models from \citet {DraineLi2007}, which cover the $1 - 1000~\mu\mathrm{m}$ wavelength range (e.g. \citealt{Foyle2013} for M83; \citealt{Aniano2012} for NGC 626 and NGC 6946; and, more recently, \citealt{Draine2014} for M31). These studies use data at shorter wavelengths as well, being thus able to fully sample the spectral range of dust emission, including emission from warmer dust, small transiently heated grains and PAHs. They mainly focus on the distribution and properties of the dust, not including any constraints on the radiation field which heats the dust from for example, UV/optical observations.

To properly investigate scaling relations on a local scale, one would ideally need a self--consistent model to derive the desired physical quantities. The complexity of a full stellar and chemical evolution model for galaxies, however, requires some simplifications. The models should treat both stellar and dust components, taking into account their influence on each other (the so--called dust energy balance). Panchromatic emission modelling of subgalactic regions has been carried out by \citet{MentuchCooper2012} for the Whirlpool galaxy (M51). However, the stellar and dust components were treated separately.  \citet{Boquien2012,Boquien2013}, have performed panchromatic pixel-by-pixel fits of nearby star forming galaxies using CIGALE \citep{Noll2009}, which does include a dust energy balance.  They showed that most of the free parameters could accurately be constrained, given a sufficiently large range of priors.

We will perform panchromatic SED fitting, using the MAGPHYS code \citep{Dacunha2008}. The code treats both stellar light and dust emission at the same time, forcing an energy balance. It has an extended library of theoretical SEDs based on the latest version of the stellar population models from \citet{Bruzual2003} and physically motivated, multi-component dust models. Furthermore, it applies a Bayesian fitting method including a thorough error analysis. 

The proximity of ISM regions is crucial in order to obtain the desired, sub-kpc spatial resolution. The closest giant molecular cloud systems are of course in our own Milky Way, but it is not possible to probe the entire Galaxy. The Magellanic clouds are the nearest galaxies as they are close satellites of the Milky Way. These objects are, however, quite irregular and lower in metallicity and in total mass, hence they do not represent the well--evolved ISM of virialised large galaxies.

Andromeda (M31) is the closest large galaxy at a distance $D_\mathrm{M31}=785$~kpc \citep{McConnachie2005}, which means every arcsecond on the sky corresponds to $3.8$ pc along the major axis of M31. Classified as a SAb-type LINER galaxy, M31 is a slow-star forming spiral \citep[SFR~$=0.20\;M_\odot\mathrm{yr}^{-1}$,][]{Ford2013} with an inclination of $77 \degree$ and a position angle of its major axis of $38 \degree$ \citep{McConnachie2005}.
The gas and dust components of Andromeda have been extensively studied in the past \citep[e.g.][]{WalterbosSchwering1987,Montaldo2009,Tabatabaei2010} using low--resolution data at FIR wavelengths and simplified models.

Although mapped in all wavelengths from UV to the FIR in the past, high--quality submm observations are thus far not available, yet these wavelengths are crucial to constraining the properties of the cold dust. The \textit{Herschel} Exploitation of Local Galaxy Andromeda \citep[][hereafter paper I]{HELGAI} is the first programme that mapped M31 from $100~\mu\mathrm{m}$ to $500~\mu\mathrm{m}$ with \textit{Herschel}, covering a large $5.5 \degree \times 2.5 \degree$ field centred around the galaxy. Even at the sparsest \textit{Herschel} resolution ($36^{\prime\prime}$ at $500~\mu\mathrm{m}$), physical scales of only $140$~pc are resolved. Andromeda is consequently the best suited object for studying the ISM in great detail while allowing at the same time, the comparison with global properties.

In \citet{HelgaII} (hereafter paper II), we performed a pixel-by-pixel SED fit to the \textit{Herschel} data and map the main dust properties of Andromeda. \citet{Ford2013} (hereafter paper III) investigate the star formation law in M31 on both global and local scales. A catalogue of giant molecular clouds was recently constructed by \citet{Kirk2014} (hereafter paper IV).

We aim to expand on this work by carrying out an in-depth investigation of the dust scaling relations in Andromeda. We do this by fitting panchromatic spectral energy distribution models to each statistically independent $36$--arcsecond region in the galaxy.  In this way we have  produced the largest and most complete view of the stars and ISM dust in a large spiral galaxy.

The arrangement of the paper is as follows. In Sect.~\ref{sec:data} we  give an overview of the data used and in Sect.~\ref{sec:results} we briefly discuss the processing of these data and our SED fitting method. Appendix~\ref{app:dataprocessing} goes into more detail on the processing of multi-wavelength data. The results are given in Sect.~\ref{sec:results}, along with the parameter maps of Andromeda. We analyse the dust scaling relations of Andromeda in Sect.~\ref{sec:scaling_rel}. In Sect.~\ref{sec:discussion} we present our discussion and main conclusions.

\section{The dataset} \label{sec:data}

Modelling the full spectrum of a galaxy requires a fair number of free parameters and consequently sufficient data points to sample the problem in a meaningful way. The Andromeda galaxy has been observed by many space borne telescopes such as the Galaxy Evolution Explorer \citep[GALEX,][]{Martin2005}, \textit{Spitzer}, and WISE. Recently, the \textit{Herschel} Space Observatory was added to this list and has been the main drive for this investigation. Ground--based observations from the Sloan Digital Sky Survey \citep[SDSS,][]{SDSS} complete our panchromatic dataset. A detailed account on the data treatment, including uncertainty estimates, for each of the observations is given in Appendix~\ref{app:errors}.

\subsection{Infrared data}

Far-Infrared and submillimetre observations with the \textit{Herschel} Space Observatory catch the peak in emission of the diffuse interstellar dust. This component plays an essential role in the energy balance of the SED. Andromeda was observed with both PACS and SPIRE instruments in parallel mode. Because of the large extent of the galaxy, observations were split into two fields. Both fields were combined during data reduction, resulting in $\sim 5.5 \degree \times 2.5 \degree$ maps at $100$, $160$, $250$, $350$, and $500~\mu\mathrm{m}$.  In the overlapping area of the two fields, the signal-to-noise ratio is slightly higher. The full width half maximum (FWHM) of point sources in the final PACS maps are $12.5^{\prime\prime}$ and $13.3^{\prime\prime}$ at $100~\mu\mathrm{m}$ and $160~\mu\mathrm{m}$, respectively \citep{PACSpsf}. The resulting SPIRE maps are characterised by beams with a FWHM of $18.2^{\prime\prime}$, $24.5^{\prime\prime}$, and $36.0^{\prime\prime}$ at $250$, $350$, and $500~\mu\mathrm{m}$ \citep{SPIREmanual}. Galactic dusty structures tend to cause foreground emission when observing nearby galaxies. The north-east part of the M31 disk clearly suffers from this kind of cirrus emission. Following a technique devised by \citet{Davies2010}, Galactic cirrus emission was disentangled from the light of the M31 disk. A detailed description of the data reduction process, including cirrus removal, can be found in paper I.  

The Multiband Imaging Photometre of \textit{Spitzer} \citep[MIPS;][]{MIPS} observed the mid-infrared and far-infrared light of M31 in its three bands ($24$, $70$, and $160~\mu\mathrm{m}$). \citet{Gordon2006} made a complete data reduction of the observations, covering a $1\degree \times 3\degree$ area along the major axis of the galaxy. The images have standard MIPS FWHM values of $6.4^{\prime\prime}$, $18.7^{\prime\prime}$, and $38.8^{\prime\prime}$ at $24$, $70$, and $160~\mu\mathrm{m}$, respectively \citep{MIPS}. Both MIPS and PACS cover a wavelength range around $160~\mu\mathrm{m}$. While this could be used to more accurately estimate the uncertainties at this wavelength, it limits our working resolution. Both MIPS and PACS measurements come with a total uncertainty of $\sim10\%$ in their $160~\mu\mathrm{m}$ band so they can be considered equally sensitive. We therefore opted to omit the MIPS $160~\mu\mathrm{m}$ image from our sample. 

The same area of M31 was also mapped in all four bands of the \textit{Spitzer} Infrared Array Camera \citep[IRAC;][]{IRAC}. The complete data reduction, including background subtraction, was carried out by \citet{Barmby2006}. Their final, background subtracted frames have the standard FWHM values of $1.6$, $1.6$, $1.8$, and $1.9$ arcseconds in the $3.6$, $4.5$, $5.8$, and $8~\mu\mathrm{m}$ wavebands, respectively.

Complementary to the IRAC/MIPS observations, the mid-infrared part of M31 has been observed by WISE as part of an all-sky survey at $3.4$, $4.6$, $12$, and $22~\mu\mathrm{m}$. High--quality mosaics of M31 were provided by the WISE Nearby Galaxy Atlas team \citep{Jarrett2013}. Recent results from these authors have proven the possibility of enhancing the resolution of WISE using deconvolution techniques. Here, however, we use the mosaics with the standard beams because we will have to degrade the resolution to the SPIRE $500~\mu\mathrm{m}$ beam in order to remain consistent. The FWHM of the WISE beams are $6.1$, $6.4$, $6.5$, and $12.0$ arcseconds  at $3.4$, $4.6$, $12$, and $22~\mu\mathrm{m}$, respectively \citep{WISE}.

Several WISE and Spitzer bands lie close to each other in central wavelength. This overlap improves our sampling of the ambiguous MIR SED and will reduce the dependence of the SED fit on a single data point, which is important in the coarsely sampled wavelength ranges, e.g. around $24\;\mu$m. At the same time, this serves as a sanity check of the measurements of both instruments. We found no strong outliers between WISE and Spitzer fluxes.

Efforts to observe M31 in the NIR bands include the 2MASS survey \citep{2MASS, Beaton2007} and the ongoing ANDROIDS project \citep{Sick2013}, all of them covering the $J$, $H$, and $K$ bands. The main difficulty of NIR imaging is the brightness of the sky. At these wavelengths the brightness can vary significantly between pointings, making it extremely hard to produce a large--scale mosaic with a uniform background. To meet the goals of our paper, it is important to have a reliable and consistent absolute flux calibration over the entire disk of M31. No $JHK$ bands were included in our dataset for this reason. The NIR part of the SED is, however, sufficiently covered by the WISE, IRAC, and SDSS $i$ and $z$ bands.

\subsection{UV/optical data} \label{sec:SDSS}

The Sloan Digital Sky Survey mapped M31 at superb resolution (FWHM~$\sim1.2$ arcsec) in its optical \textit{u},\textit{g},\textit{r},\textit{i}, and \textit{z} filters. Background estimation for these observations proved difficult because of the great extent of the galaxy and the narrow field of view of the telescope. \citet{tempel2012} created detailed mosaics from the separate SDSS tiles, taking special care of background subtraction and flux preservation. The resulting frames span a stunning $2.5\degree \times 8\degree$ field with a pixel scale of $3.96$ arcseconds. The mosaics are contaminated by several artefacts around the brightest sources, especially in the $u$ and $z$ bands. They are most likely ghost projections as they are slightly smaller and appear on each of the four sides of brightest sources along the pixel grid. After masking (see Sect.~\ref{sec:masking}), the images proved sufficiently reliable for SED fitting at SPIRE resolutions.
	
Unattenuated ultraviolet photons are the main tracers of very recent star formation. Most of the emitted UV light, however, is heavily attenuated by interstellar and circumstellar dust and consequently important to constrain the dust distribution in our spatially resolved SED. \citet{Thilker2005} created images using separate observations from the Galaxy Evolution Explorer (GALEX) in both near-UV (NUV) and far-UV (FUV) filters. The number of frames has recently been expanded to 80, almost fully covering a $5\degree \times 5\degree$ field around the centre of M31. Their mosaics have FWHM values around $5^{\prime\prime}$. Because of the co-adding of separate tiles, background variations were visible at the edges of each tile. Additionally, the UV sky around M31 is clouded with scattered light from Galactic cirrus structures. Both features will be taken into account as background variations in the uncertainty estimation for the fluxes.

\section{Method and results} \label{sec:results}

The data was processed in several steps to create a homogeneous set. We give a brief description here. For a complete account, we refer the reader to appendix \ref{app:processing}.

First, the background was subtracted from the images. This proved necessary for the GALEX, WISE, and \textit{Herschel} frames. The average background level was already zero for the SDSS and Spitzer subsets, hence no further background subtraction was needed. Second, foreground stars and background galaxies were masked and replaced by the local background. The GALEX and SDSS frames were masked using UV colours, while MIR colours were used to mask the WISE, IRAC, and MIPS images. As a third step, all frames were convolved to the resolution of the SPIRE $500~\mu\mathrm{m}$ point spread function (PSF) using the convolution kernels from \citet{Aniano2011}. Finally, the pixel scales were resized to match the pixel grid of this frame, which was rebinned to a $36$ arcsec/pixel scale.

A detailed uncertainty analysis was performed for each pixel as well. Therefore, we did not start from the original errors relating to the observations and data reduction, but they were estimated directly from the convolved and rescaled images. Three sources of uncertainty were considered: background variations in the frame, calibration uncertainties, and Poisson noise due to incoming photons. The last was only considered in the UV and optical bands, where they are known to be significant.

The above procedures yield a panchromatic SED for thousands of pixels, each corresponding to a sub-kpc region in Andromeda. A complete UV-to-submm spectral energy distribution will be fitted to each of these regions to investigate their underlying properties.

	\subsection{MAGPHYS} \label{subsec:magphys}

We make use of the Bayesian SED fitting code MAGPHYS \citep{Dacunha2008} to perform the strenuous task of modelling panchromatic SEDs. The program determines the best fit from a library of optical and infrared SEDs, taking special care of the dust-energy balance when combining the optical and infrared part of the spectrum. This library is derived from one general multi-component galaxy-SED model, characterised by a number of parameters. 

The stellar emission is computed by assuming a \citet{Chabrier2003} initial mass function (IMF) and evolved in time using the latest version of the stellar population synthesis (SPS) model of \citet{Bruzual2003}. The obscuring effects of interstellar and circumstellar dust are computed using the \citet{charlot2000} model.

A multi-component dust model is used to calculate the infrared and submm emission from the reprocessed starlight. The model consists of five modified black bodies, three of which have fixed temperatures ($850$~K, $250$~K, and $130$~K) representing the hot dust. The other two have variable temperatures and embody the warm and cold dust components in thermal equilibrium. The PAHs are modelled using a fixed template based on observations of the starforming region M17. Although MAGPHYS keeps the emissivity index of the modified black body, $\beta$, fixed at $2$ for the coldest dust component, this is partially compensated by adding multiple dust components at multiple temperatures, broadening the FIR-submm peak.
The total amount of dust is distributed in two different geometries: (1) the diffuse dust in the ISM, which consists of all ingredients of the aforementioned dust model, and (2) the circumstellar dust, which resides in the birth clouds of new stars and consists of all ingredients except for the cold dust component.

The library of template SEDs is derived from this multi-parameter model for the FUV--submm SED. Each free parameter comes with a physically motivated probability distribution. From these distributions, a random parameter set is drawn to create a template SED. The standard MAGPHYS library consists of 25000 UV--optical templates and 50000 IR--submm templates. When modelling the observed SED of a galaxy, maximum likelihood distributions are created for each of the free parameters in the model. This is done by weighing the parameters of each template fit with its respective $\chi^2$ value. The different output parameters are summarised in Table~\ref{tab:magphysparams} and are briefly discussed below.

\begin{itemize}
\item The contribution of the dust component in the diffuse ISM to the total infrared luminosity $f_\mu = L_\mathrm{dust}^\mathrm{ISM} / L_\mathrm{dust}^\mathrm{Tot}$ is derived from both the absorption of starlight and from infrared emission.
\item The total stellar mass  $M_{\ast}$ is derived from the population synthesis models and is proportional to the flux in the UV to NIR wavebands.  
\item The standard star formation rate (SFR) expresses the number of stars formed per year, averaged over the last $100$ Myr. The process of star formation is modelled with an exponentially declining SFR law starting from the birth of the galaxy. Superimposed are bursts with a random chance of occurring throughout its lifetime.
\item The specific star formation rate (sSFR) is then simply the ratio of the SFR and the stellar mass and compares the number of stars formed during the last 100~Myr with the total number of stars formed throughout the lifetime of the galaxy.
\item Dust attenuation is expressed by the optical depth parameter, which is evaluated in the $V$ band: $\tau_V = \tau_V^\mathrm{BC}+\tau_V^\mathrm{ISM}$. Starlight of young stars in their birth clouds (BC) experiences extinction from circumstellar dust and from the diffuse interstellar dust. This is parametrised by $\tau_V^\mathrm{BC}$. Most of the stars however, only irradiate the interstellar dust, modelled by $\tau_V^\mathrm{ISM}$.
\item The bulk of the dust mass is contributed by warm and cold dust in the diffuse ISM and by warm circumstellar dust. A factor of $1.1$ takes into account the contributions of hot dust and PAHs to the total dust mass:
\begin{equation}
M_\mathrm{dust} = 1.1(M_\mathrm{W}^\mathrm{BC}+M_\mathrm{W}^\mathrm{ISM}+M_\mathrm{C}^\mathrm{ISM}).
\end{equation}
\item The equilibrium temperature for the warm circumstellar dust and cold ISM dust is left free for the FIR/submm modified black--body components. They are represented in  $T_\mathrm{W}^\mathrm{BC}$ and $T_\mathrm{C}^\mathrm{ISM}$, respectively.
\item Each dust component produces infrared emission, which is summed in the total dust luminosity $L_\mathrm{dust}$. The relative contributions of the dust components are quantified in fractions to the total BC or ISM dust luminosity. We refer the reader to Sect. 2.2.1 of \citet{Dacunha2008} for a detailed explanation of the infrared emission parameters. Two important components will be discussed in this work: $L_\mathrm{C}^\mathrm{Tot}$, the total luminosity of the cold dust in the diffuse ISM and $L_\mathrm{PAH}^\mathrm{Tot}$, the total luminosity from PAHs in the ISM and around young stars.
\end{itemize}

\begin{table}
\caption{Overview of the output parameters from a MAGPHYS SED fit.}
\label{tab:magphysparams}
\centering     
\begin{tabular}{>{\centering\arraybackslash}m{1.1cm}>{\centering\arraybackslash}m{1.1cm}>{\raggedright\arraybackslash}m{5.5cm}}
\hline 
\hline
Symbol & Unit & Description \\ [1ex]
\hline 
 $f_\mu$ 						&  						& ISM dust to total dust luminosity \\ [1ex]
 $\tau_V$ 						& 						& Total $V$ band optical depth \\ [1ex]  
 $\tau_V^\mathrm{ISM}$ 			&  						& ISM dust contribution to $\tau_V$ \\ [1ex]  
 SFR 							& $M_{\odot}$yr$^{-1}$ 	& Star formation rate  \\ [1ex]  
 sSFR 							& yr$^{-1}$  			& Specific star formation rate \\ [1ex]  
 $M_{\ast}$ 						& $M_{\odot}$ 			& Total stellar mass\\ [1ex]  
 $L_\mathrm{dust}$ 				& $L_{\odot}$ 			& Total luminosity of emitting dust\\ [1ex]  
 $M_\mathrm{dust}$ 				& $M_{\odot}$ 			& Total dust mass \\ [1ex]  
 $T_\mathrm{W}^\mathrm{BC}$ 		& $K$ 					& Dust temperature in birth clouds \\ [1ex]  
 $T_\mathrm{C}^\mathrm{ISM}$ 	& $K$ 					& Dust temperature in ISM \\ [1ex]  
 $L_\mathrm{C}^\mathrm{Tot}$ 	& $L_{\odot}$			& Total cold dust luminosity  \\ [1ex]  
 $L_\mathrm{PAH}^\mathrm{Tot}$	& $L_{\odot}$			& Total PAH luminosity \\ [1ex]
\hline
\end{tabular} 
\end{table}

The MAGPHYS SED libraries are derived from realistic, galaxy scale parameter values. The parameter space is thus optimised for objects that are orders of magnitude brighter than the sub-kpc regions to be modelled here. Pixel-by-pixel fitting makes no sense when the physical properties of a single pixel-region are out of the bounds of the MAGPHYS standard parameter space. We therefore adopted a flux scaling of $10^4$ to obtain fluxes of the order of integrated nearby galaxies and feed these higher fluxes to the code for fitting. Most of the output parameters will remain unaffected because of their relative nature. Only four parameters scale with flux and do that linearly: $M_\ast$, $L_\mathrm{dust}$, $M_\mathrm{dust}$ and the SFR. These parameters were scaled back by the same factor to obtain their true fitted value. 

\begin{figure}
	\resizebox{\hsize}{!}{\includegraphics{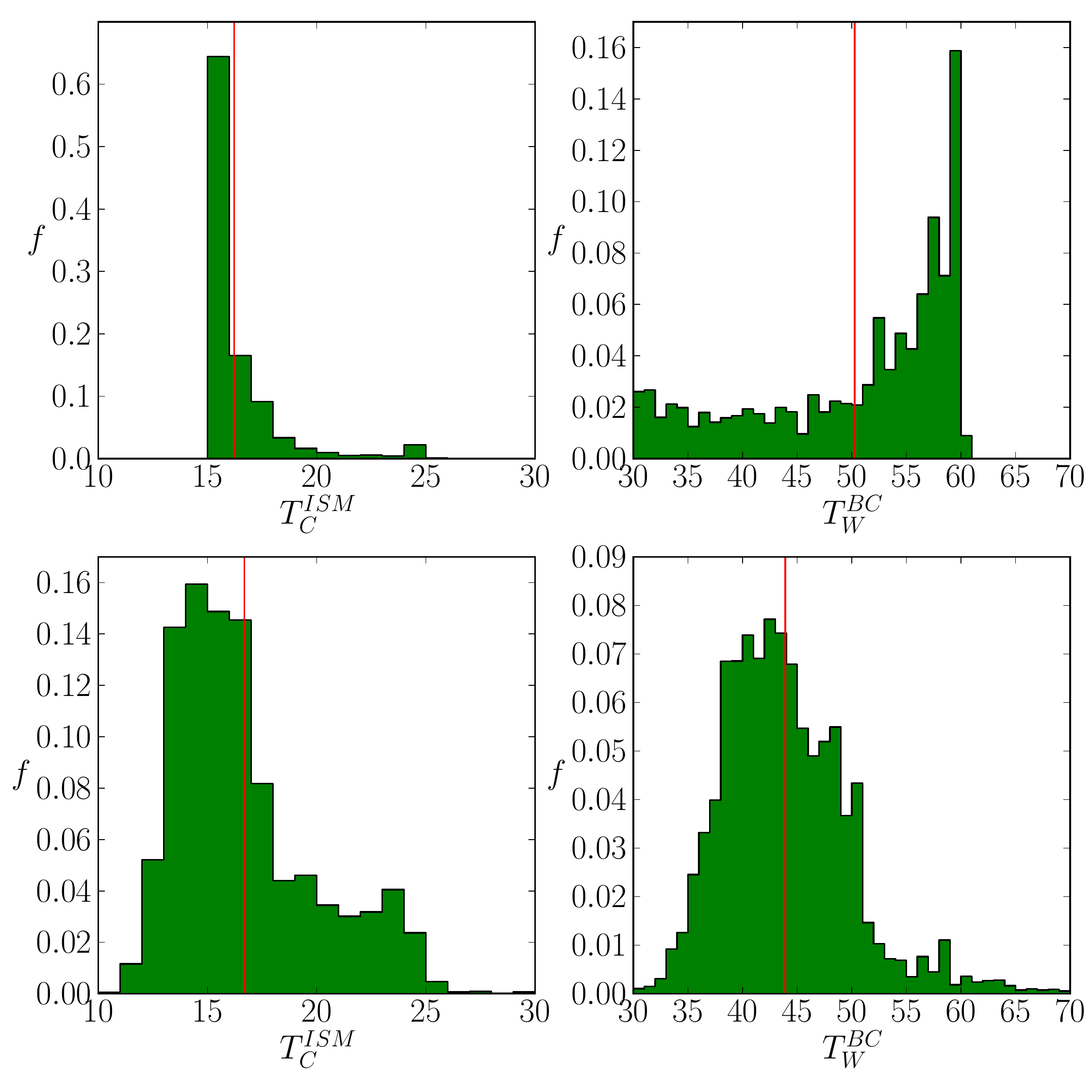}}
	\caption{Distribution of the dust temperatures for the individual pixel fits. \textbf{Top row}; are the results from the standard MAGPHYS version. \textbf{Bottom row}; are the results from the  modified version, using broader temperature ranges for the priors. The red lines indicate the average sample values.}
	\label{fig:tempdistr}
\end{figure}

Another limitation of the standard version of MAGPHYS is the range of cold dust temperatures, which is fixed between $15$~K and $25$~K. The boundaries of this interval are encountered in low (high) FIR surface brightness areas (see e.g. paper II). This causes the peak of the modified black body to be offset with respect to the observations and influences related parameters such as star formation and dust mass. We estimate that over $60 \%$ of the derived temperatures for cold dust lie outside of the standard $15-25$~K interval (see upper panels of Fig.~\ref{fig:tempdistr}). The same is true for the temperature ranges of the warm dust. Here about $15 \%$ of all regions are estimated to lie outside the $30 - 60$~K range. In order to execute reliable fits, it is thus mandatory to expand the temperature intervals and create a custom infrared library.

A custom set of infrared SEDs was constructed (da Cunha, private communication), incorporating a wider range in cold and warm dust temperatures. The new library features cold dust temperatures $T_\mathrm{C}^\mathrm{ISM}$ ranging from $10-30$~K and warm dust temperatures $T_\mathrm{W}^\mathrm{BC}$ from $30-70$~K. With this extended library, the derived cold dust temperatures are more spread out over the parameter space, also populating the coldest ($<15$~K) regions (see the lower-left panel of Fig.~\ref{fig:tempdistr}). The distribution of the warm dust temperature is also considerably changed (lower--right panel), although the parameter space was only increased by $10$~K. The peak is also not shifted towards higher temperatures as the distribution derived from the standard library would suggest. What we see here is a manifestation of the shift in cold dust temperature. As both dust components are not independent - the infrared SED is fitted entirely at the same time - the shift to lower cold dust temperatures will also cause a decrease in the warm dust temperature distribution in order to still match the flux in their overlapping area (FIR).

	\subsection{SED fits of sub-kpc regions}

We perform panchromatic fits for a total of $22437$ pixels within an ellipse with major axis of $22$~kpc and an apparent eccentricity of $0.96$, covering $2.24$ square degrees on the sky. The pixels are the same size as the FWHM of the $500\;\mu$m beam, making them statistically independent from each other. The choice of our aperture is limited to the field of view of the IRAC frames, which cover the main stellar disk of M31, but does not extend up to NGC 205 or to the faint outer dust structures as seen in the SPIRE maps. However, the field is large enough to cover over $95 \%$ of the total dust emission of the galaxy \citep{Draine2014}.

	\subsubsection{Quality of the fits} \label{subsub:filtering}
	
Instead of eliminating {\it a priori} those pixels with a non-optimal spectral coverage, we decided to exploit one of the characteristics of MAGPHYS, that is that the final results, i.e. the physical parameters we are looking for, are given as the peak values of a probability distribution function (PDF). When a pixel has a SED which is characterised by a poor spectral sampling, the parameters that are more influenced by the missing data points, whatever they are, will have a flatter PDF, showing the tendency to assume unrealistic values. For example, output parameters related to the stellar components (stellar mass, SFR, etc.) will be questionable for pixels in regions of Andromeda where the UV and optical background variations become dominant. Parameters related to interstellar or circumstellar dust turn unreliable when reaching very low flux density areas in the \textit{Herschel} bands.

To decide which pixels are to be considered reliable for a given parameter and which ones should be not considered, we evaluate the mean relative error of each fit. Each PDF comes with a median value (50th percentile), a lower limit (16th percentile), and an upper limit (84th percentile). A way to quantify the uncertainty of the median value is to look at the shape of the PDF. Broadly speaking, if the peak is narrow, the difference between the 84th and 16th percentile will be small and the corresponding parameter will be well constrained. For a broad peak or a flat distribution, the opposite is true. We define this mean relative error as follows:
\begin{equation}
\sigma_\mathrm{rel} = 0.5 \cdot (p_{84} - p_{16})/p_{50},
\end{equation} 
where the $p_x$ indicate the percentile levels of the PDF. This error only reflects the uncertainty on the modelling.

A double criterion was needed to filter out unreliable estimates for each of the parameters considered in Table~\ref{tab:magphysparams}, because the average $\sigma_\mathrm{rel}$ is quite different in each parameter. Furthermore, it proved necessary to filter out most of the parameter estimates related to the outermost pixels in our aperture. These regions have FIR emission below the \textit{Herschel} detection limits. A reliable detection at these wavelengths is crucial in constraining most of the dust-related parameters. Pixels with a non-detection at either PACS $100~\mu\mathrm{m}$ or PACS $160~\mu\mathrm{m}$ make up almost $40 \%$ of the sample. At the same time, these pixels generally have higher photometric uncertainties at shorter wavelengths as they correspond to the faint outskirts of the galaxy. Together with their poorly sampled FIR SEDs, their corresponding parameter estimates will have broad PDFs. We rank, for each parameter, all pixels according to increasing relative error for that particular parameter. Then, $40 \%$ of the parameter estimates (those with the highest mean relative errors) were removed from the sample. This corresponds to the exclusion of 8975 pixels per parameter.

Secondly, as several parameter estimates with broad PDFs were still present after the first filtering, an optimal cut was found which excludes these estimates. We chose to remove any parameter estimate with $\sigma_\mathrm{rel} > 0.32$. The combination of these filters excluded, for each parameter, all pixels with an unreliable estimate of this parameter. We note that the excluded pixels themselves might differ for each parameter set. For example, the dust mass of a particular pixel might be poorly constrained and thus removed. On the other hand, the stellar mass of that same pixel, will be kept in the sample if it meets our filter criteria. Several pixels (4384 in total) did not meet the requirements in any of the parameters and were thus completely removed from the sample (meaning they were not considered in the $\chi^2$ distribution or in any further analysis). The distribution of the best--fit $\chi^2$ values is shown in the upper--left panel of Fig.~\ref{fig:compare} and has an average value of $1.26$.

Table~\ref{tab:localglobal} lists the number of reliable pixels for each parameter. In any further analysis, only these estimates will be used. 
Generally, $T_C^\mathrm{ISM}$ and $L_\mathrm{dust}$ are the best constrained parameters in the sample, with median uncertainties of $5\%$ and $6\%$, respectively, and 13462 usable pixels. On the other hand, $\tau_V$, $T_W^\mathrm{BC}$, and sSFR are the least constrained parameters. However, in the case of sSFR, this quantity is computed from the SFR and stellar mass, the uncertainty on this parameter takes into account the uncertainty on both of its constituents, hence the relatively high median $\sigma_\mathrm{rel}$ of 0.18 for 6608 usable pixels.
Quite differently, the high mean $\sigma_\mathrm{rel}$ (0.21 for 13462 usable pixels) on estimates for $T_W^\mathrm{BC}$ stems from the poor spectral coverage in the $30-70~\mu$m regime. Other degeneracies in the fitting procedure will certainly reflect in the total number of reliable pixels per parameter, as well as in their median relative error. This is most obvious for $\tau_V$, with a median $\sigma_\mathrm{rel}$ of 0.14 and only 3753 usable pixels. This parameter is not only influenced by the amount of dust, but also by its geometry with respect to the stars. This is impossible to take into account without complex radiative transfer modelling and falls beyond the goal of this paper.

		\subsubsection{Consistency with previous parameter fits}

\begin{figure}
	\resizebox{\hsize}{!}{\includegraphics{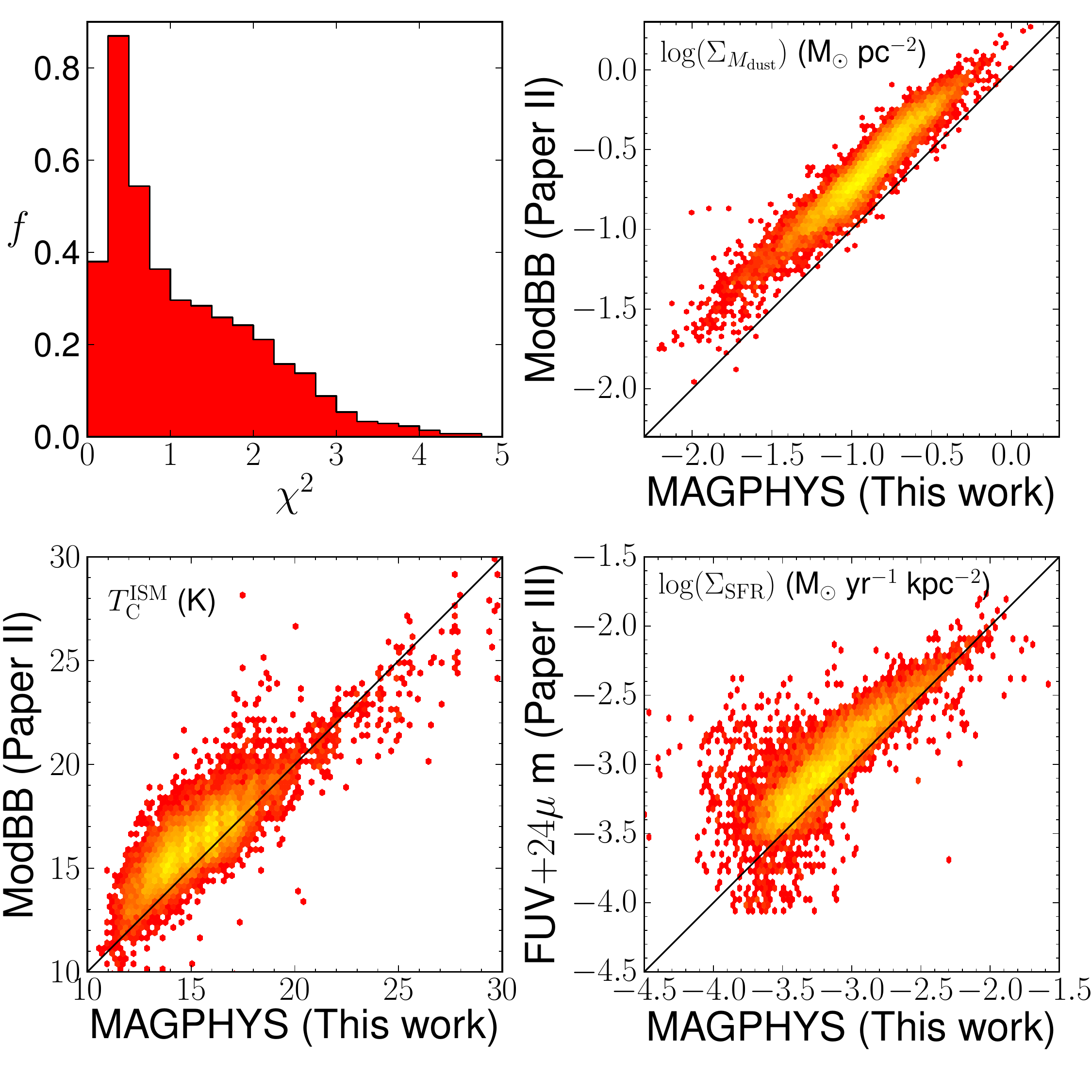}}
	\caption{Upper--left: $\chi^2$ distribution of the fits. Other panels: density plots comparing dust mass surface density, cold dust temperature, and star formation rate derived from MAGPHYS single pixel fits against modified black--body fits to \textit{Herschel} bands (paper II) and FUV$+24~\mu\mathrm{m}$ SFR tracers (paper III). Red indicates a small number of data points, yellow a large number. The black line represents the $1:1$ relation.}
	\label{fig:compare}
\end{figure}

We compare our results with previously derived values for each pixel, see Fig.~\ref{fig:compare}. The MAGPHYS dust mass and cold dust temperature are compared to the modified black--body fits from paper II. Furthermore, our star formation rate is compared to the SFR derived from FUV$+24~\mu\mathrm{m}$ fluxes in paper III. Each pixel region of the paper II and paper III maps corresponds exactly to a pixel region in our sample, hence we are comparing parameter estimates for the exact same physical region.

In general, the different approaches yield consistent results. The dust mass shows the tightest relation ($rms = 0.08$), but also the largest offset $\Delta \log(\Sigma_{M_\mathrm{dust}}) = 0.22$. It is therefore important to understand what we are comparing here. For each dust component, the flux $S_\nu$ is modelled with a modified black--body function,
\begin{equation}
S_\nu = \frac{M_\mathrm{dust}}{D^2}\kappa_\mathrm{abs}B_\nu(T_\mathrm{dust}),
\end{equation}
where $D$ is the distance to the galaxy, $B_\nu(T_\mathrm{dust})$ the Planck function, and $\kappa_\mathrm{abs}$ is the dust mass absorption coefficient, modelled as
\begin{equation}
\kappa_\mathrm{abs} = \kappa_\mathrm{abs}(\lambda_0)\times \left(\frac{\lambda_0}{\lambda}\right)^\beta
\end{equation}
with $\lambda_0$ the normalisation wavelength and $\beta$ the emissivity index.
MAGPHYS adopts the dust model from \citet{Dunne2000} (hereafter D00) who normalize the dust mass absorption coefficient at $850~\mu\mathrm{m}$: $\kappa_{850} = 0.077$ m$^2$kg$^{-1}$. At \textit{Herschel} wavelengths this becomes $\kappa_{350}= 0.454$ m$^2$kg$^{-1}$ assuming a fixed $\beta =2$. In paper II, the \citet{Draine2003} (hereafter D03) absorption coefficient, which is $\kappa_{350} = 0.192$ m$^2$kg$^{-1}$, and a variable $\beta$ were adopted. 
Lower $\kappa_\mathrm{abs}(\lambda_0)$ values are associated with more silicate-rich dust compositions \citep{Karczewski2013}. They will result in higher dust mass estimates, which is the case for paper II. On the other hand, a variable $\beta$ will generally result in higher dust temperatures. This will in turn yield lower dust masses.
As MAGPHYS computes the total dust mass from components of various temperatures, grain sizes, and compositions, this should result in more realistic dust mass estimates.

A smaller offset ($\Delta T_d = 4.47$~K), but larger scatter ($rms = 0.93$) is seen in the temperature of the cold dust. The emissivity index $\beta$ was fixed at $2$ in this paper, while left as a free parameter in paper II. As there is a known degeneracy between $\beta$ and $T_\mathrm{dust}$ \citep[see e.g.][and references therein]{Hughes2014,Tabatabaei2014,Galametz2012,HelgaII}, this could explain the scatter in the relation. Furthermore, most of the pixels in paper II had $\beta < 2$. Given this temperature-$\beta$ degeneracy, smaller $\beta$ values will yield higher dust temperatures. This probably explains the systematic offset from our sample.

The SFR shows a less clear deviation from the 1:1-relation. The rms of the scatter in the points is $0.28$ and they have an offset of $\Delta \log(\Sigma_{SFR}) = -0.42$. We tend to find systematically lower SFR values compared to the  FUV$+24~\mu\mathrm{m}$ tracer used in paper III. It must be noted that the SFR is derived in a different way in both approaches. The FUV$+24~\mu\mathrm{m}$ tracer is empirically derived from a sample of starforming galaxies \citep{Leroy2008}. Several regions in M31 exhibit only low star forming activity, far from the rates of starforming galaxies. Additionally, this formalism assumes a stationary star formation rate over timescales of $100$ Myr. In M31, we resolve sub-kpc structures, where star formation may vary on timescales of a few Myr \citep{Boselli2009}. MAGPHYS does allow variations in SFR down to star formation timescales of 1 Myr.

Most of the outliers in the SFR plot of Fig.~\ref{fig:compare} correspond to pixels surrounding the bulge of M31. It is in these areas that the MIR emission from old stars is modelled quite differently. Overall, we expect that our SED fits give a more realistic estimate of the SFR because they take information from the full spectrum and are derived from local star formation histories.

		\subsubsection{Total dust and stellar mass}

One of the most straightforward checks we can perform between our modelling technique and other techniques, is a comparison between the total number of stars and dust in M31. With respect to the first component, we find a value of $\log(M_\star/M_\odot)= 10.74$ from the fit to the integrated fluxes (see Sect.~\ref{subsec:localglobal}). This value, calculated by exploiting simple stellar population (SSP) models assuming a \citet{Chabrier2003} IMF, lies a factor of 1.5--3 below dynamical stellar mass estimates for Andromeda \citep[e.g.][]{Chemin2009,Corbelli2010}. This discrepancy can be easily accounted for if we consider that often in dynamical models the derived stellar M/L ratio are consistent with more heavyweight IMFs. \citet{Tamm2012} did exploit SSP models to calculate the stellar mass, and found values in the $\log(M_\star/M_\odot)= 11-11.48$ range, using the same IMF as we do. This discrepancy might arise from a number of possible causes: first of all, the aperture used to extract the total fluxes is slightly smaller in our case and secondly, a different masking routine was applied to remove the foreground stars. The likely most effective difference, however, could be due to the fact that MAGPHYS considers an exponentially declining star formation history to which star formation bursts are added at different ages and with different intensities. This can cause the M/L to decrease, and might explain the difference in stellar mass. 

Instead, when we compare the stellar mass in the inner $1$~kpc with the value calculated with MAGPHYS by \citet{Groves2012}, we find a remarkably good agreement ($\log(M_\star/M_\odot)= 10.01$ vs $\log(M_\star/M_\odot)= 9.91$ in our case).

The dwarf elliptical companion of Andromeda, M32, also falls in our field of view. We find a total stellar mass of $\log(M_\star/M_\odot)= 8.77$. Of course, as the light of M32 is highly contaminated by Andromeda itself, mass estimates of this galaxy are highly dependent on the aperture and on the estimation of the background flux of M31. Nevertheless, our estimate is only a factor of 2--3 lower then dynamical mass estimates \citep[e.g.][]{Richstone1972}. This discrepancy is of the same order as the difference in total mass we find for M31.

As a total dust mass, we find $M_\mathrm{dust} = (2.888^{+0.006}_{-0.005})\cdot 10^{7} M_\odot$, for the sum of all pixel-derived dust masses, using the D00 dust model. This estimate is preferred over the dust mass from a fit to the integrated fluxes as the most recent dust mass estimates for M31 are derived from the sum of pixel masses. Furthermore, it is known that dust mass estimates from integrated fluxes underestimate the total dust mass (see also Sect.~\ref{subsec:localglobal}). We note that the uncertainly on this estimate appears very small. This is because of the very narrow PDF for $M_\mathrm{dust}$ from the global fit of M31. As previously stated, this error only reflects the uncertainty on the SED fitting. The absorption coefficient $\kappa_\mathrm{abs}(\lambda_0)$ from D03 was used in the next estimates.

In paper I we estimated the dust mass from a modified black-body fit to the global flux and found $M_\mathrm{dust} = (5.05 \pm 0.45)\cdot 10^{7} M_\odot$. 
Paper II derived a total dust mass from modified black-body fits to high signal-to-noise pixels (which cover about half of the area considered) and found $M_\mathrm{dust} = 2.9 \cdot 10^{7} M_\odot$. Independent \textit{Herschel} observations of M31 \citep{Krause,Draine2014} yield $M_\mathrm{dust} = (6.0 \pm 1.1)\cdot 10^{7} M_\odot$ (corrected to the distance adopted in this paper: $D_{M31} = 0.785$ Mpc) as the sum of the dust masses of each pixel.

We find a total dust mass of the same order as these previous estimations; however, our result is somewhat lower. The reason for this discrepancy is most likely the difference in dust model, as was already clear from Fig.~\ref{fig:compare}. Determining the conversion factor $q$ between dust models is rather difficult and requires the assumption of an average emissivity index:
\begin{equation}
q = \frac{\kappa_{D00}}{\kappa_{D03}} = \frac{\kappa_{D00}^{850}}{\kappa_{D03}^{350}} \cdot \left(\frac{350}{850}\right)^{-\beta}.
\end{equation}

For M31, the mean $\beta$ was found to be in the range $1.8 - 2.1$ \citep{HelgaII,Draine2014}, yielding a conversion factor between $2.0$ and $2.6$, or dust masses in the range $5.70 - 7.44 \cdot 10^7 M_\odot$. The D00 dust model thus tends to produce dust masses that are about half of the D03 masses. Keeping this in mind, all dust mass estimates for Andromeda agree within their uncertainty ranges.		

		\subsubsection{Local vs global} \label{subsec:localglobal}

The MAGPHYS code was conceived for galaxy-scale SED fitting and it does a good job there \citep[e.g.][]{Dacunha2008,daCunha2010,Clemens2013}. At these large scales, a forced energy balance is justified because globally, most of the absorbed starlight is re-emitted by dust. When zooming in to sub-kpc regions, this assumption might not be valid any more. Light of neighbouring regions might be a significant influence on the thermal equilibrium of a star forming cloud, and if this is true, it may translate into an offset between the local parameters and the global value for that galaxy.

As a test for our extended library (see Sect.~\ref{subsec:magphys}) we compare the mean values of the physical parameters to their global counterparts (upper part of Table~\ref{tab:localglobal}). These parameters were derived from a MAGPHYS fit to the integrated fluxes of Andromeda. The fluxes are listed in Table~\ref{tab:fluxes}. In the case of additive parameters (bottom part of Table~\ref{tab:localglobal}), we compare their sum to the value derived from a global SED fit. It is important to note that we rely on our filtered set of pixels for each parameter. This means at least $40 \%$ of the pixels are excluded. Most of these badly constrained pixel values lie in the outskirts of M31. Nevertheless, their exclusion will surely affect the additive parameters. 

\begin{table*}
\caption{Comparison of the main properties for M31 as derived from the pixel-by-pixel fitting and from a fit to the global fluxes.  N$_\mathrm{pix}$ denotes the number of reliable pixels used for the analysis and $\sigma_\mathrm{rel}$ the median relative uncertainty for this sample. The mean value of the relative parameters are given in the upper part of the table, along with the standard deviation on their distribution. The sum of the additive parameters are given in the bottom part of the table, along with their uncertainty.}
\label{tab:localglobal}
\centering     
\begin{tabular}{>{\centering\arraybackslash}m{1.0cm}>{\centering\arraybackslash}m{1.0cm}>{\centering\arraybackslash}m{1.0cm}>{\centering\arraybackslash}m{1.0cm}>{\centering\arraybackslash}m{0.8cm}>{\centering\arraybackslash}m{1.6cm}>{\centering\arraybackslash}m{1.0cm}}
\hline
\hline
parameter & N$_\mathrm{pix}$ & Median $\sigma_\mathrm{rel}$ & Mean local & std. dev & Global & Unit \\ [1ex]
\hline
$f_\mu$           			&  13462 & 0.10 & $0.85$		&	$0.08$	&  $0.88^{+0.01}_{-0.02}$	& -- \\ [1ex]
$sSFR$            			 & 6608 & 0.18 & $3.45$		&	$0.02$	& $3.38 \pm 0.01	$		& $10^{-12}\mathrm{yr}^{-1}$ \\[1ex]
$T_\mathrm{W}^\mathrm{BC}$	 & 13462& 0.21 & $43$		&	$7$		&  $61^{+1}_{-9}$ 		& $K$ \\[1ex]
$T_\mathrm{C}^\mathrm{ISM}$	& 13462 & 0.05 & $15.4$		&	$2.1$	&  $16.2^{+0.3}_{-0.2}$ 	& $K$ \\[1ex]
$\tau_V$          			& 3753 & 0.14 & $0.62$		&	$0.68$	& $0.32 \pm 0.01$ 		&  -- \\[1ex]
$\tau_V^\mathrm{ISM}$    	& 13212 & 0.11 & $0.20$		&	$0.13$	& $0.16 \pm 0.01$ 		& -- \\[1ex]
$\chi^2$          			& 18053 & -- & $1.26$		&	--		& $1.35$ & -- \\[1ex]
\hline
\hline
parameter & N$_\mathrm{pix}$ & Median $\sigma_\mathrm{rel}$  & \multicolumn{2}{c}{Total local} & Global & Unit \\[1ex]
\hline
$M_\star$					 & 12853 & 0.17 & \multicolumn{2}{c}{$4.76   \pm 0.02$}		& $5.5 \pm 0.01$ 			 & $10^{10}M_\odot$ \\[1ex]
$M_\mathrm{dust}$			 & 10496 & 0.18 & \multicolumn{2}{c}{$2.89   \pm 0.06$}		& $2.7^{+ 0.4}_{-0.1}$ 		 & $10^{7}M_\odot$\\[1ex]
$L_\mathrm{dust}$			 & 13462 & 0.06 & \multicolumn{2}{c}{$4.542  \pm 0.002$}		& $4.98^{+0.6}_{-0.01}$		 & $10^{9}L_\odot$\\[1ex]
$SFR$						 & 8673 & 0.16 & \multicolumn{2}{c}{$0.1644  \pm 0.0005$}	& $0.189^{+0.002}_{-0.01}$ 	 & $M_\odot\mathrm{yr}^{-1}$\\[1ex]
$L_\mathrm{PAH}^\mathrm{tot}$ & 12563 & 0.12 & \multicolumn{2}{c}{$0.8724 \pm 0.0009$}	& $1.11^{+0.03}_{-0.16}$		 & $10^{9}L_\odot$\\[1ex]
$L_\mathrm{C}^\mathrm{tot}$	 & 10709 & 0.10 & \multicolumn{2}{c}{$2.475  \pm 0.002$}		& $2.55^{+0.15}_{-0.01}$		 & $10^{9}L_\odot$\\[1ex]
\hline 
\end{tabular} 
\end{table*}

Several parameters mimic their global counterparts quite well. This is the case for $f_\mu$, $T_\mathrm{C}^\mathrm{ISM}$, $\tau_V$, and $\tau_V^\mathrm{ISM}$, where the agreement lies within $1$ standard deviation. The distribution of the pixel-derived $\mathrm{sSFRs}$ has a broad shape. In order to compare the sSFR of the pixels to the global value, we make use of the total SFR and stellar mass as derived from the pixels. We then find the pixel-sSFR by dividing the total SFR by the total $M_\star$. This value again lies close to its global counterpart, despite the wide range of sSFRs found on an individual pixel basis.

The temperature of the warm circumstellar dust differs significantly: the average pixel value is $43$~K with a standard deviation of $7$~K, while a fit to the global fluxes reveals $T_\mathrm{W}^\mathrm{BC} = 61_{-9}^{+1}$~K. Both values still overlap at a $2 \sigma$ level, but the relative deviation is much larger than the other parameters. The peak of this dust component lies between $30~\mu\mathrm{m}$ and $70~\mu\mathrm{m}$. The MIPS $70~\mu\mathrm{m}$ band provides the only data point in this region, making it difficult to estimate this parameter accurately.

Additive parameters will also suffer because we exclude a significant number of pixels. Most of them do add up to the same order of magnitude as the global values: SFR, $M_\ast$, $L_\mathrm{dust}$, $L_\mathrm{PAH}^\mathrm{tot}$, and $L_\mathrm{C}^\mathrm{tot}$. All of them lie within $5-20\%$ below the global value. This again indicates that the excluded pixels do not contribute significantly to the light of M31. The dust mass, however, turns out to be $\sim7\%$ higher when summing all pixels. It is known that SED fitting is not a linear procedure and will depend on the employed resolution. \citet{Aniano2012} found that modelling on global fluxes yields dust masses that are up to $20\%$ lower than resolved estimates. \citet{Galliano2011} found discrepancies of up to $50\%$ depending on the applied resolution. Even taking into account another $10\%$ in flux due to the excluded pixels, our difference in dust mass estimates lies within this range.

The fact that we reproduce the global properties of M31 from our local SEDs, boosts confidence that the procedure applied here is valid, even though MAGPHYS was conceived for galaxy-scale SED fitting. 

	\subsection{Parameter maps} \label{sec:paramaps}
	
\begin{figure*}
	\centering
   	\includegraphics[scale=0.82]{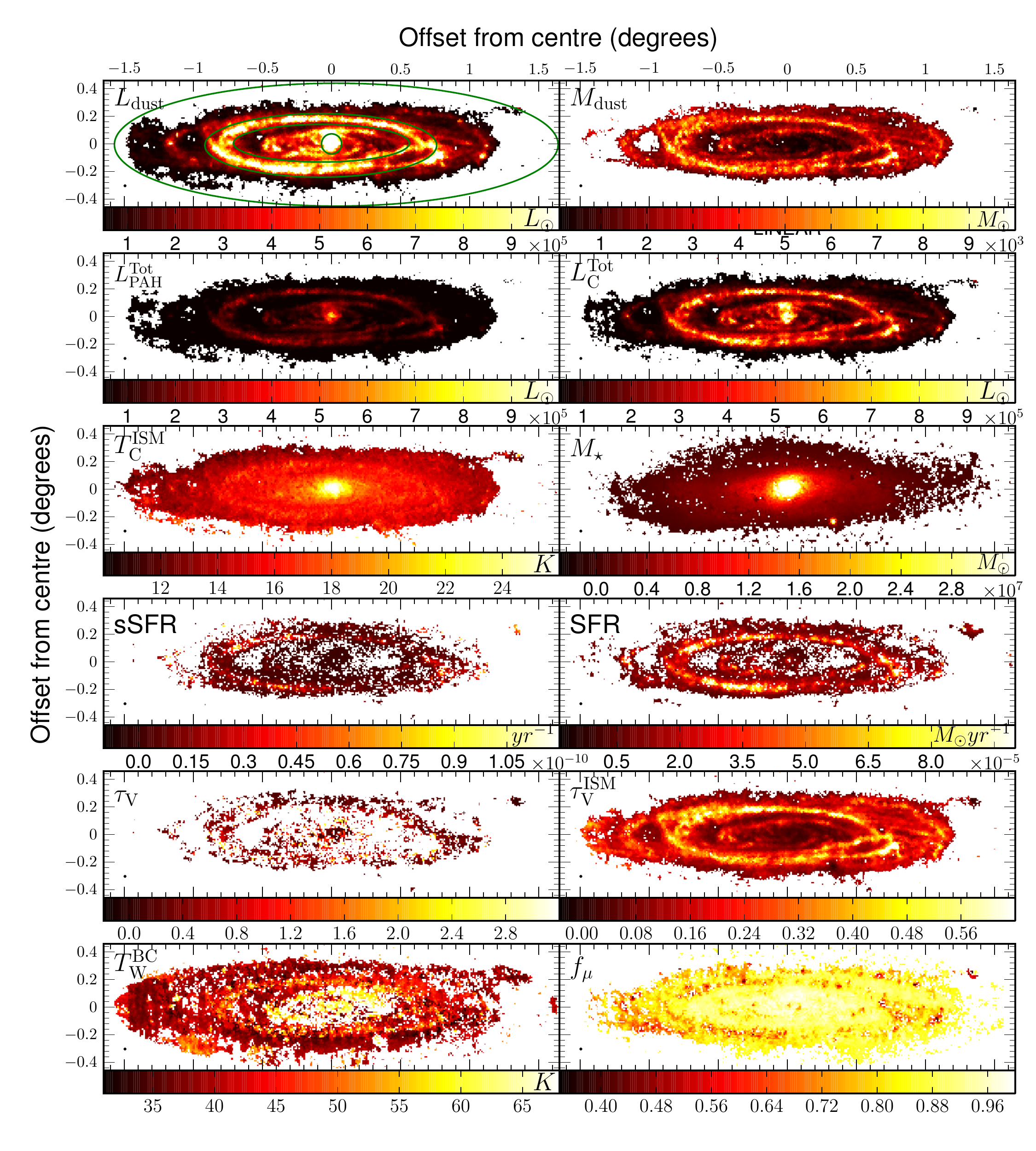}
	\caption{Parameter maps for M31, rotated from a position angle of $38\degree$. See Table~\ref{tab:magphysparams} for the meaning of the parameters. Pixels with uncertainties that were considered too large (see Sect.~\ref{subsub:filtering}) were blanked out. The green ellipses in the upper--left panel represent the apertures of the macro-regions of M31: the bulge, inner disk region, $10$~kpc ring, and the outer disk.}
	\label{fig:paramaps}
\end{figure*}

We construct detailed maps of the SED parameters from the collection of pixels with reliable parameter fits in Fig.~\ref{fig:paramaps} and briefly discuss the morphologies observed.

The dust luminosity $L_\mathrm{dust}$ closely follows the morphology seen in the PACS images (see also Fig.~\ref{fig:fitdata}). Dust emission in Andromeda is brightest in the bulge and in the $10$~kpc ring. Some fainter emission regions are seen in the outer parts of the galaxy, coinciding with a ring at $15$~kpc.

As expected, the dust mass $M_\mathrm{dust}$ map closely resembles the SPIRE images (see also Fig.~\ref{fig:fitdata}). Compared to the $L_\mathrm{dust}$ map, some intriguing distinctions can be noted. There seems to be almost no dust in the centre of M31, while the bulge is actually the brightest in dust luminosity. We hereby confirm earlier statements \citep[paper II;][]{Tempel2010, Groves2012, Draine2014} that the bulge of Andromeda holds a small amount of relatively warm ($>25$~K) dust. The south-west side is also smoother than in the $L_\mathrm{dust}$ map, pointing out that the heating of ISM dust and not the mass is crucial to the observed luminosity.

The PAH luminosity $L_\mathrm{PAH}^\mathrm{tot}$ appears relatively weak when compared to the $L_\mathrm{dust}$ map. The general morphology, however, is similar. The surface brightness at these wavelengths is the highest in the bulge of Andromeda. Furthermore, the emission is mostly concentrated in the $10$~kpc ring and in the dusty parts of the inner disk. If reprocessed UV light of recent and ongoing star formation is the only energy source for MIR emission, no bright MIR and PAH features are expected in the bulge of M31. Emission from PAHs can, however, be enhanced by increases in the diffuse ISRF \citep{Bendo2008}. This again, suggests that the radiation field of older stars in the centre of M31 is quite strong.

A similar morphology is seen when looking at the contribution of the diffuse cold dust to the total dust emission, $L_\mathrm{C}^\mathrm{tot}$. The cold dust, only found in the diffuse ISM, appears significantly more luminous than the PAH emission (the $L_\mathrm{dust}$, $L_\mathrm{C}^\mathrm{tot}$ and $L_\mathrm{PAH}^\mathrm{tot}$ maps in Fig.~\ref{fig:paramaps} have the same scale). Interestingly, the emission from this component is equally bright in the bulge and in the ring, in contrast with the PAH luminosity map.

Consequently, the temperature of the ISM dust $T_\mathrm{C}^\mathrm{ISM}$ peaks in the centre ($\sim30$~K). It follows a smooth radial decline until it reaches a plateau at $16$~K in the ring. Higher values are reached in the brightest star forming regions. This suggests the cold ISM dust is partially heated by recent and ongoing star formation. On the other hand, older stars can also contribute to the heating of the dust. In the NIR wavebands, the surface brightness is slightly enhanced in the ring, indicating a higher concentration of older stars. Outside the star forming ring, the temperature quickly drops two degrees to $14$~K.

For the warm dust temperature $T_\mathrm{W}^\mathrm{BC}$, the picture is far less clear. The map is crowded with blanked pixels due to their high uncertainties. As already mentioned in Sect.~\ref{subsec:localglobal}, the MIPS $70~\mu\mathrm{m}$ data point is the only observation in this temperature regime. Additionally, the emission of the cold dust component overlaps greatly with the SED of the warm dust, making it difficult to disentangle both components. We do find significantly higher temperatures in the bulge, where the ISRF is highest, and in the $10$~kpc ring, where most of the new stars are being formed. Outside these areas, the warm dust is relatively cold ($<45$~K).

The stellar mass $M_\star$ is one of the best constrained parameters thanks to the good coverage of the optical and infrared SED. It is highest in the bulge and the central regions around it and declines smoothly towards the outskirts of the galaxy. Interestingly, a small peak is seen where M32 resides. We do not detect this dwarf satellite in our \textit{Herschel} maps. This is caused by the overlap of emission from M32 and M31's diffuse dust emission at this location on the sky. Nevertheless, it is evident that M32 does not contain much dust.

The star formation rate map of M31 largely coincides with the dust luminosity (except in the bulge), although the distribution is more peaked in the rings. Regions where the dust emission is lower (inter-ring regions and outskirts) are mostly blanked out because it is hard to constrain very low star formation rates. Some residual star formation is seen in the bulge. However it must be noted that the high dust luminosities in the central region might cause a degeneracy between the SFR, which directly heats the dust, and $M_\star$, representing the number of older stars that have been proven to strongly contribute to dust heating.

The specific star formation rate is obtained by dividing the SFR over the last $100$ Myr by the stellar mass and gives a measure of ongoing vs past star formation. This quantity combines the uncertainties of both parameters, hence the large number of blanked pixels. The $10$~kpc and $15$~kpc rings have the highest sSFR in M31. Interestingly, the inner ring has very low values of sSFR and the bulge has close to zero. In general, we can say that stars are nowadays formed most efficiently in the rings of the galaxy.

The $V$--band optical depth is the poorest constrained parameter of the sample. Accurately estimating this value requires detailed knowledge on the dust geometry. This is not available here, so assumptions must be made based on colour criteria and an extinction law. As we are not able to probe individual star formation regions, the optical depth must be seen as an average over each resolution element. \citet{Liu2013} showed that when averaged over scales of $\sim100 - 200$~pc, the dust geometry can be approximated by a foreground dust screen. Individual stars or star forming regions are, however, likely to experience much greater optical depths than the averaged values. In Andromeda, this average varies from $0.2$ to $2$. Most of the higher optical depth regions coincide with the dusty rings of the galaxy.

The picture is more obvious for the contribution of cold ISM dust to the total optical depth $\tau_V^\mathrm{ISM}$. This parameter closely resembles the ring--like structure we also see in the $M_\mathrm{dust}$ and ranges from $0.1 - 1.1$. The ratio of those two parameters $\tau_V^\mathrm{ISM} / \tau_V = \mu$, measures the contribution of diffuse cold ISM dust to the extinction of starlight. We find a median $\mu = (54 \pm 22)\%$, but the contribution of ISM dust ranges from less than $\sim25\%$ in the centre to over $70 \%$ in the $10$~kpc ring. This suggests that diffuse dust is the main contributor to starlight extinction in the more active regions of a galaxy. This is consistent with the results of \citet{Keel2014}, who find that a greater fraction of the UV extinction is caused by the diffuse dust component. Furthermore, detailed radiative transfer models of galaxies show the importance of this diffuse component in the FIR/submm emission, and thus absorption of starlight \citep[see e.g.][]{Tuffs2004,Bianchi2008,Popescu2011}.

The contribution of the ISM dust to the dust luminosity $f_\mu$ peaks ($\sim90\%)$ in the centre, where most of the dust is in the diffuse ISM. It linearly decreases along the disk, reaching $\sim80\%$ in the star forming ring. Going beyond this ring, the ISM dust contribution declines more quickly to reach values around $50\%$ at $15$~kpc.

	\subsection{SED of the macro-regions}\label{subsec:mainregions}

As a first step towards a spatially resolved analysis, we decompose Andromeda into macro--regions, located at different galactocentric distances: the bulge, the inner disk, the star forming ring centred at a radius of $10$~kpc, and the outer part of the disk. We choose to base our definition of these regions on the morphology of the $L_\mathrm{dust}$ map of Fig~\ref{fig:paramaps}. The advantage is that, in the light of constructing dust scaling relations, each region corresponds to a separate regime in terms of SFR, radiation field, dust content, and composition. For example, the bulge is limited to the inner $1kpc$ region and is thus significantly smaller than the optical/NIR bulge, but it coincides with the zone with the hardest radiation field. The shapes of the borders between regions are apparent ellipses on the sky and do not necessarily coincide with projected circles matching the disk of M31 (see also Table~\ref{tab:physprop} for their exact definition). 

\begin{figure*}
	\centering
   	\includegraphics[scale=0.52]{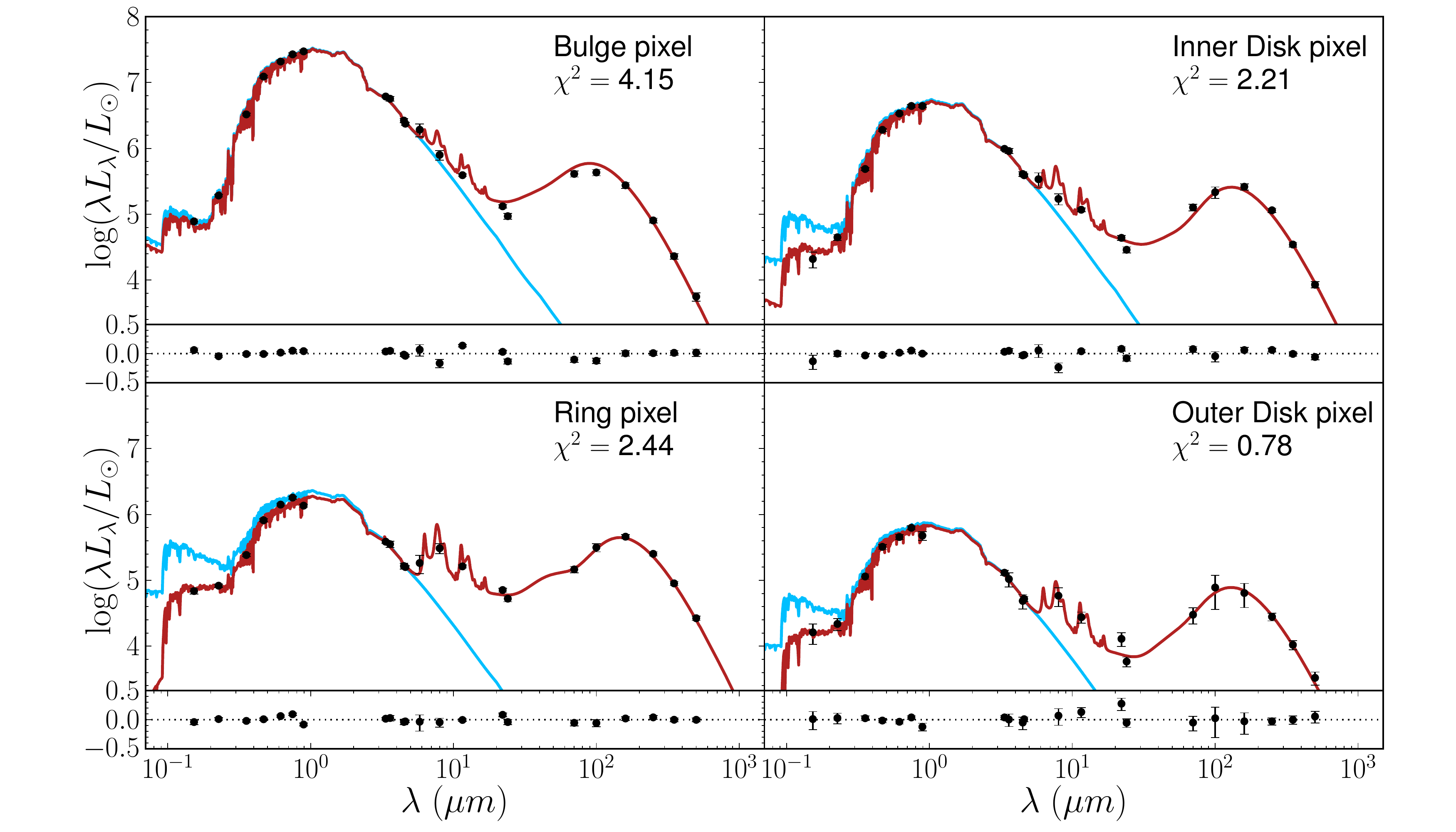}
	\caption{The panchromatic SED of four representative pixels in the bulge, inner disk, the $10$~kpc ring, and the outer disk region. The blue line represents the unattenuated SED and the red line the best fit to the observations. Residuals are plotted below each graph.The $\chi^2$ values are those for the best fitting template SED.}
	\label{fig:representativepixels}
\end{figure*}

\begin{figure*}
	\centering
   	\includegraphics[scale=0.52]{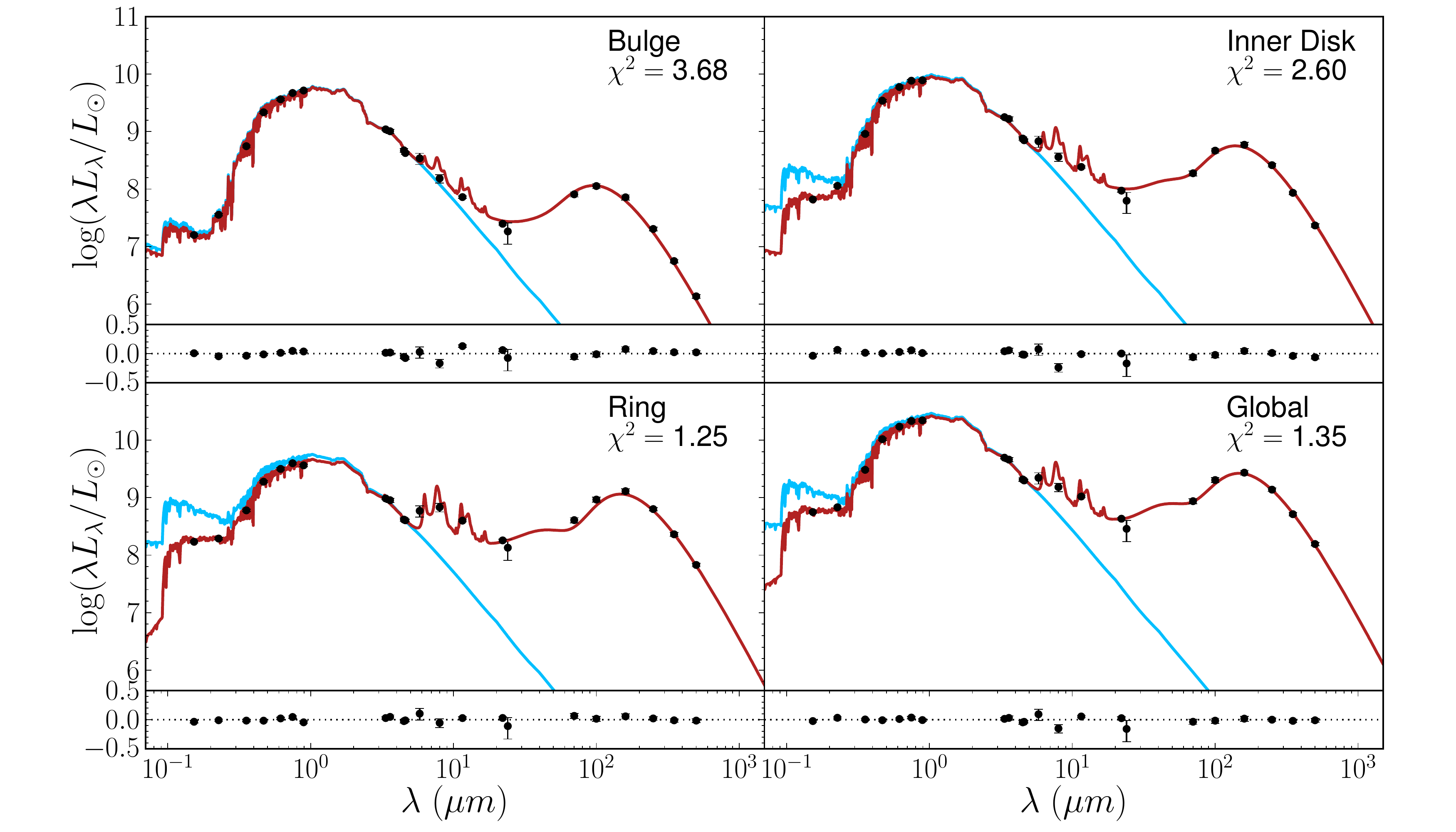}
	\caption{The panchromatic SED of four main apertures: the bulge, the inner disk region, the $10$~kpc ring, and the integrated galaxy. The blue line represents the unattenuated SED and the red line the best fit to the observations. Residuals are plotted below each graph. The $\chi^2$ values are those for the best fitting template SED.}
	\label{fig:macroRegions}
\end{figure*}

We choose a set of individual pixels, selected to represent the typical shape of an SED in these regions. They are shown in Fig.~\ref{fig:representativepixels}, together with the residual values for each wavelength band. Along with the single pixel SEDs, we also show SED fits to the integrated fluxes of these macro--regions in Fig.~\ref{fig:macroRegions}. Integrated fluxes for the separate regions can be found in Table~\ref{tab:fluxes}. The goodness of fit is here expressed by the $\chi^2$ value for the best fitting template SED. The parameter values from this SED will differ slightly from the peak values in the PDFs, which express their most likely values. Nevertheless, the template SED gives a good indication of how well the observed fluxes can be matched.

When comparing the fits in Figs.~\ref{fig:representativepixels} and \ref{fig:macroRegions}, it is clear that the macro-region fits have a systematically lower $\chi^2$ than their respective single-pixel fits. This is not surprising as one might expect greater signal-to-noise variations on smaller scales. In general, however, most of the observed fluxes are well reproduced by the best fit. Only the MIPS $24~\mu\mathrm{m}$ point seems to be systematically below the theoretical SED. The MIPS $24~\mu\mathrm{m}$ observations do come with large error bars, so that is accounted for in the determination of the parameter PDFs.

In the bulge, stars are completely dominant over the dust component in terms of mass and luminosity. This is visible in the ratio of the total dust luminosity to the $g$--band luminosity ($L_\mathrm{dust}/L_g = 0.48$), and in the small offset between the unattenuated (blue) and the attenuated (red) SED. The optical/NIR SED is much more luminous than the UV part of the spectrum, indicating a relatively low SFR and a strong interstellar radiation field (ISRF), dominated by older stars. Furthermore, the peak of the FIR SED lies at relatively short wavelengths caused by high temperatures of the cold dust. There are only weak PAH features visible in the centre of M31, although the MIR flux is relatively large compared to the other regions. Some residual star formation can be found in the bulge of M31, but the contribution to the total SFR is negligible.

The inner disk is forming stars at a slow pace ($2.30\times 10^{-2}M_\odot$yr$^{-1}$). This is confirmed by a visible offset between the unattenuated and attenuated SED in the UV regime. The FIR emission peaks at longer wavelengths than in the bulge, indicating a milder ISRF and lower dust temperatures. We consequently find a higher $L_\mathrm{dust}/L_g$ ratio of $1.50$. This less harsh environment allows PAHs to survive longer, giving rise to more prominent MIR features. The same conditions hold in the outer disk of M31, although the surface brightness is systematically lower there. The dust also gains in importance here ($L_\mathrm{dust}/L_g = 2.37$). The FIR peaks at even longer wavelengths, meaning the diffuse dust is colder in the outskirts of the galaxy.

The most active star forming region of M31 is unquestionably the ring at $\sim 10$~kpc. This region contains only $20 \%$ of the stellar mass and almost half of the total dust mass of M31 at temperatures near the galaxy's average (see also Table~\ref{tab:physprop}). The luminosity difference between the UV and the optical/NIR SED is the smallest of all regions, indicating that new stars dominate the radiation field. This also translates in strong PAH features in the MIR. The offset between the attenuated and unattenuated SED is large in the UV and even visible in the optical-NIR regime, indicating significant dust heating. Consequently, we find the highest dust--to--$g$-band ratio $L_\mathrm{dust}/L_g=4.84$ here.

\section{Dust scaling relations} \label{sec:scaling_rel}
In the following we will investigate scaling relations of stellar and dust properties in Andromeda at different sizes: we first consider the galaxy as a whole, and we will then look at its main components as separate regions. Finally, we will push the analysis down to the smallest possible scale: the hundred--pc sized regions defined by the statistically independent pixels whose SED we have modelled as explained above. 

Our results can then be compared to already known results for similar physical quantities. In this respect, the ideal sample for comparison is surely the one provided by the local dataset of the {\it Herschel} Reference Survey \citep[HRS;][]{boselli2010}. The HRS is a volume--limited, $K$--band selected survey including more than 300 galaxies selected to cover both the whole range of Hubble types and different environments. \citet{Cortese2012} have analysed in detail how the specific dust mass correlates to the stellar mass surface density ($\mu_\star$) and to NUV--r colour (see Fig.~\ref{fig:Helga_HRS}). Furthermore, \citet{daCunha2010} also found links between the dust mass and SFR and between $f_\mu$ and the specific SFR using a sample of low-redshift star forming galaxies from the SDSS survey. We will check where Andromeda is located with respect to the above relations and, more importantly, we will address the issue regarding the physical scales at which the aforementioned relations start to build up.

\subsection{Andromeda as a whole}\label{sec:wholem31}
Andromeda is classified as a SA(s)b galaxy \citep{deVaucouleurs1991} and has a prominent boxy bulge. The disk contains two conspicuous, concentric dusty rings and two spiral arms \citep[see e.g. paper IV;][]{Gordon2006}.

In Table~\ref{tab:physprop} we report a summary of the main physical properties we have derived for Andromeda from integrated fluxes. The total stellar mass was found to be $5.5\times 10^{10}M_\odot$, a typical value in local starforming galaxies \citep[see e.g. ][]{Clemens2013}.  The total dust mass was found to be $2.70\times10^{7}M_\odot$, comparable to the amount of dust in our own Galaxy \citep[e.g.][]{sodroski1997}. The distribution of dust in the Andromeda galaxy is, however, atypical for an early--type spiral. The infrared emission from early-type galaxies is usually quite compact \citep[e.g.][]{Bendo2007,MunozMateos2009}. Extended ring structures, such as the ones in M31, appear rather infrequently.

As an early--type spiral, Andromeda has a low star formation rate of $0.19$~$M_\odot$yr$^{-1}$, about ten times smaller than the value measured in the Milky Way \citep{Kennicutt2012}. In \citet{daCunha2010} a relation was derived between the SFR of normal, low redshift (z < $0.22$) SDSS galaxies and their dust content. Despite a total dust mass which is close to the sample average, the SFR in M31 is about one order of magnitude below the average relation at this dust content. Consequently, Andromeda's specific star formation rate (sSFR) of $3.38\times 10^{-12}$ yr$^{-1}$ is at the lower end compared to the trend found by \citet{daCunha2010}.

If we compare the locus of M31 in the scaling relation plots presented in \citet{Cortese2012}, we find that Andromeda follows precisely the average trends defined by HRS galaxies (see the cyan dot in the two top panels in Fig.~\ref{fig:Helga_HRS}). In general, M31 has dust and stellar masses that are typical for early--type spiral galaxies. On the other hand, the star formation activity is unusually low. This is consistent with a more active star formation history, possibly related to an encounter with M32 \citep{Block2006}.

\begin{table*}
\centering
\caption{Summary of the parameters characterising the main apertures of M31. The semi-major axis $a$ and apparent eccentricity $\epsilon$ describe the annuli on the plane of the sky. The other parameters are physical properties derived from an SED fit to the integrated fluxes inside these apertures.}
\label{tab:physprop}
\begin{tabular}{lcccccccccc}
\hline\hline
Region	& RA centre	&	DEC centre	&	$a$	&	$\epsilon$	&		$M_\star$			&	$\mu_\star$					&		$M_\mathrm{dust}$			&		SFR				&	sSFR					&	NUV--r	\\
			
	&	hh:mm:ss	&	dd:mm:ss	&	kpc	&		&	$10^{10}M_\odot$			&	$10^{8}M_\odot$kpc$^{-2}$			&		$10^{6}M_\odot$		&		$10^{-2}M_\odot$yr$^{-1}$		&  $10^{-12}$yr$^{-1}$	&	Mag		\\
			
\hline			
Global		&	10:39:52.6	&	41:14:30.1	&	22.0 &	0.96 &	$5.50$	&	$3.11$	&	$27.0$	&	$18.9$	&	$3.38$	&	$4.59$		\\
Bulge		&	10:41:10.2	&	41:16:18.4	&	1.0	 &	0.00	 &	$0.81$	&	$25.1$	&	$0.11$	&	$0.18$	&	$0.27$	&	$6.11$		\\      
Inner Disk	&	10:44:22.8	&	41:20:00.2	&	8.7	 &	0.98 &	$2.00$	&	$4.23$	&	$3.95$	&	$2.30$	&	$1.20$	&	$5.38$		\\
10~kpc Ring	&	10:45:37.6	&	41:19:39.6	&	11.5 &	0.96 &	$0.96$	&	$1.66$	&	$12.8$	&	$9.06$	&	$9.55$	&	$4.11$		\\
Outer Disk	&	10:39:52.6	&	41:14:30.1	&	22.0 &	0.96 &	$1.07$	&	$1.58$	&	$12.5$	&	$1.18$	&	$1.07$	&	$3.85$		\\
M32			&	10:40:21.1	&	40:51:52.6	& 0.636  &	0.00	 &	$0.06$  &   $4.65$	&	$0.08$	& 	$0.03$	&	$0.60$	&	$5.31$		\\
\hline			
\end{tabular}
\end{table*}

\subsection{Scaling relations in the different regions}\label{sec:components}
As outlined in Sect.~\ref{subsec:mainregions}, we have grouped our set of pixels in four macro-regions based on the morphology of the $L_\mathrm{dust}$ map. These macro-regions represent physically different components in the galaxy. It is well known, for example, that the bulges in spiral galaxies usually host the oldest stellar populations \citep[e.g.][]{Moorthy2006}. Bulges are usually devoid of ISM and have barely any star formation \citep[e.g.][]{Fisher2009}, closely resembling elliptical galaxies in many respects. However, unlike stand-alone elliptical galaxies, galactic bulges are intersected by galactic disks. At this intersection, a significant amount of gas, dust, and many young stars are present. In this respect, Andromeda is again an atypical early-type spiral as none of these components are prominently visible at the bulge-disk intersection.

\begin{figure}
	\resizebox{\hsize}{!}{\includegraphics{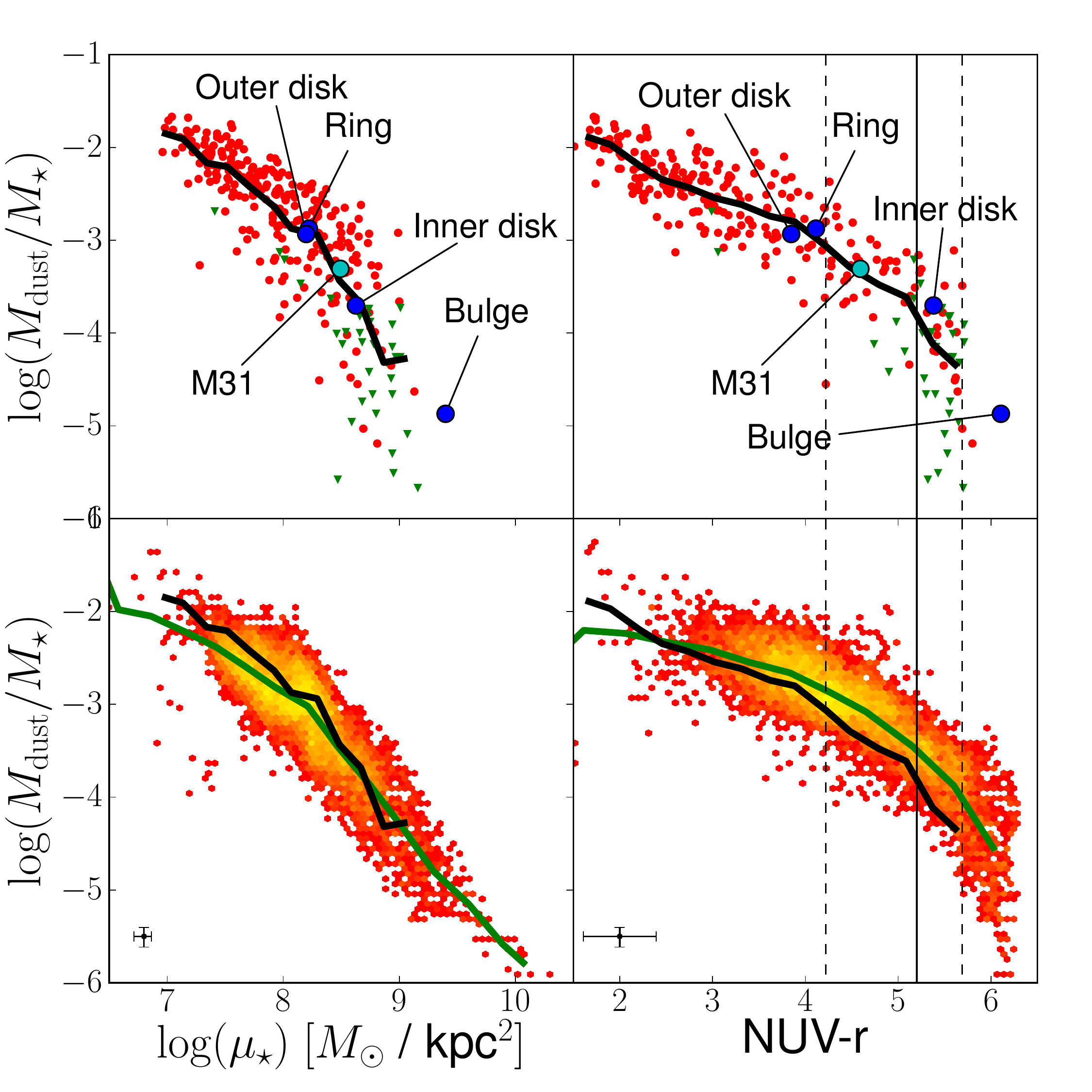}}
	\caption{\textbf{Top:} dust scaling relations from the \textit{Herschel} Reference Survey \citep{Cortese2012}; the specific dust mass $M_\mathrm{dust}/M_\star$ as a function of the stellar mass surface density $\mu_\star$ (left) and NUV-r colour (right). NUV-r serves as an inverse tracer of the sSFR. Red are \textit{Herschel} detected galaxies, green are upper limits for the undetected ones. The blue dots represent the macro-regions of M31 and the cyan dot indicates the position of the galaxy itself. The thick black line is the HRS mean trend. The vertical solid line indicates our division between late-type and early-type galaxies. The dashed vertical lines indicate the bluest early-type galaxy and the reddest late-type galaxy, respectively. \textbf{Bottom:} Same scaling relations, but now represented in a density plot using single pixel regions from M31. Yellow points indicate a higher density of points, red a lower density. The black line is the HRS mean trend, the green line is the M31 mean trend.}
	\label{fig:Helga_HRS}
\end{figure}

In the top panels of Fig.~\ref{fig:Helga_HRS}, we plot the physical quantities derived from fitting the integrated fluxes of Andromeda's four macro-regions (blue points), together with the relations for HRS galaxies (red dots) and their average trend (black line). Quite remarkably, they all fall on (or very close to) the average HRS relation, each of those components lying on a specific part of the plot which is typical for a given Hubble (morphological) type. On average, the outer parts of M31 closely resemble the physical properties of late--type HRS galaxies, while its bulge has, instead, characteristics similar to those of elliptical galaxies. 

In the remainder of the paper, we define objects as early-type when NUV--r $> 5.2$. This is the transition where most of the HRS objects are either E or S0 galaxies. Consequently, we define objects as late-type bluewards from this line. In Fig.~\ref{fig:Helga_HRS}, we also indicate the bluest S0 galaxy (NUV--r $= 4.22$) and the reddest Sa galaxy (NUV--r $= 5.69$) of the sample to indicate the spread of the transition zone.

In this respect, it is worth noting that Andromeda's bulge is significantly redder than any of the submm detected HRS galaxies because we are picking, by definition, only its very central, hence redder, regions. There is a known colour gradient in elliptical galaxies, which have bluer outskirts with respect to their inner parts, and this difference can be as high as $\sim 1$ mag \citep[see e.g.][]{Petty2013}. This can easily explain the offset in the bulge colour with respect to the HRS elliptical galaxies, whose colours are instead calculated from global apertures. For similar reasons, Andromeda's bulge is found at larger stellar mass surface densities values if compared to global elliptical galaxies, where also the outer, less dense parts are included in the measurements.

\begin{figure}
	\centering
	\includegraphics[width=0.35\textwidth]{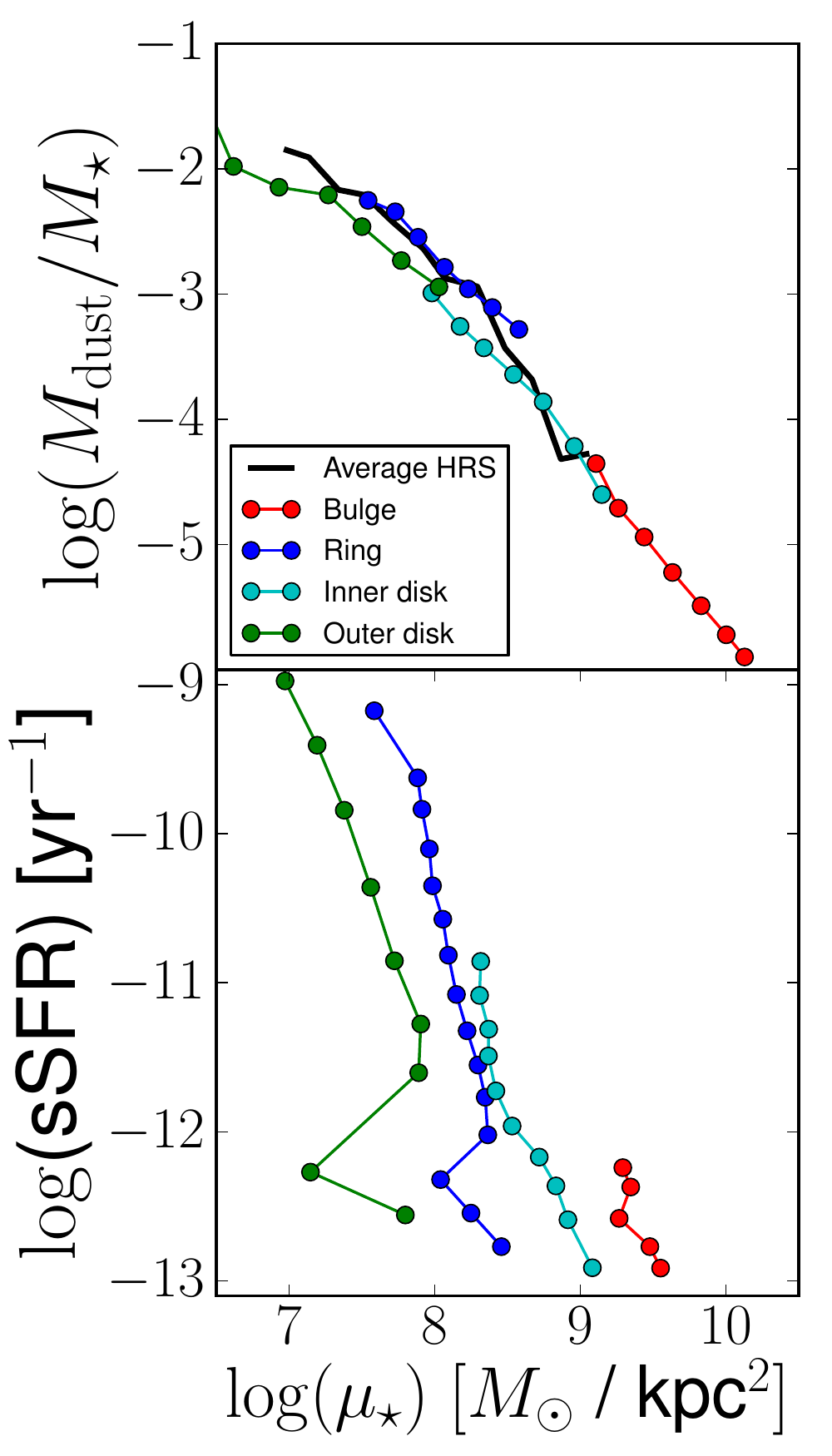}
	\caption{Average trends of the scaling relations for the stellar mass surface density $\mu_\star$, separated by the main morphological regions of M31: the bulge (red), the inner disk (cyan), the $10$~kpc ring (blue), and the outer disk (green). \textbf{Top:} $\mu_\star$ vs $M_\mathrm{dust}/M_\star$. All pixel values are binned in $\mu_\star$; each point is the average of a bin. \textbf{Bottom:} $\mu_\star$ vs sSFR, where all pixels are binned in sSFR.}
	\label{fig:regions}
\end{figure}

From a geometrical perspective, these regions follow a pattern that is determined by the average galactocentric distance: going from the centre outwards, we find the macro-regions at progressively bluer colours or, equivalently, lower stellar mass surface density. This reflects a typical inside-out formation of the bulge-disk geometry \citep{White1991,Mo1998}. This is consistent with the results from \citet{Perez2013} and \citet{Delgado2013}, where they confirm an inside-out growth pattern for a sample of local starforming galaxies. Inside the disk, the situation might be more complex. An increasing sSFR for the macro-regions, from the inner disk to the $10$~kpc ring supports this scenario. However, it does drop down significantly in the outer disk (see Table~\ref{tab:physprop}).

\citet{MunozMateos2007} have derived sSFR gradients for a sample of nearby galaxies. They found a slope that is, on average, positive and constant outwards of one scale length. When considering the very coarse sampling provided by the four macro-regions, Andromeda shows hints of a similar behaviour, with the sSFR declining in the outer disk with respect to the inner one. However, the $10$~kpc ring exhibits a clear peak in SFR and sSFR. 

This can also be seen from the top panel of Fig.~\ref{fig:regions}, where the scaling relations of the individual pixels are separated in the different macro-regions, and the pixel-derived properties are binned within each macro-region: the bulge (red), inner disk (cyan), ring (blue), and outer disk (green). In the $\log(\mu_\star)$ vs $\log(M_\mathrm{dust}/M_\star)$ plot, all regions but the ring follow a continuous relation which coincides with the HRS scaling relation. The ring is clearly offset in this relation, indicating a dustier environment compared to the average in the disk.

The specific dust mass was found to correlate with both the stellar mass surface density and NUV--r colour. It is important to verify whether both relations are not connected by a tighter correlation between $\mu_\star$ and NUV--r. In the bottom panel of Fig.~\ref{fig:regions}, the relation between $\mu_\star$ and sSFR is displayed. We prefer sSFR over NUV--r as the former allows a direct comparison of physical quantities. We find that both quantities are tightly anti-correlated \citep[see also][]{Salim2005}, so any evident link with one of them directly implies a link with the other. It must be noted, however, that sSFR values below $10^{-12}\;\mathrm{yr}^{-1}$ should be interpreted with care, these values correspond to low star formation rates for any stellar mass and are subject to model degeneracies. It is immediately evident, as seen before, that the regions can be separated in stellar mass surface density. All of them lie within a small interval in $\mu_\star$ of about $1$ order of magnitude. In sSFR, however, the spread within one region covers over three orders of magnitude, except for the bulge. This implies that, no matter the density of stars, a wide range of star formation rates are possible. The bulge, which is the region with the lowest SFR, is an obvious exception to this trend. 
Across the regions, the correlation sSFR and $\mu_\star$ is not particularly tight. In comparison, the link between both parameters and the specific dust mass is much more compact. The correlation between sSFR and $\mu_\star$ is therefore not likely to be the main driver of the other dust scaling relations.

\subsection{Scaling relations at a sub-kpc level}\label{sec:subkpc}

We can now go a step further and see if and how the aforementioned scaling relations still hold at a sub-kpc level. We are not yet able to reach the resolution of the molecular clouds, the cradles of star formation, but we are instead sampling giant molecular cloud aggregates or complexes. Hence, we cannot yet say if the scaling relations will eventually break, and at which resolution.

In the lower panels of Fig.~\ref{fig:Helga_HRS} we show the $M_\mathrm{dust}/M_\star$ ratio as a function of both the stellar mass surface density and NUV--r colour, for the statistically independent pixels. Regions with a higher number density of pixels in the plot are colour--coded in yellow. Displayed as a green line is the average relation for the pixels. The relations found between the dust--to--stellar mass ratio and NUV--r and $\mu_\star$ confirm, from a local perspective, the findings of \citet{Cortese2012}. As NUV--r traces sSFR very well, we also confirm the local nature of the results from \citet{daCunha2010}, who use the same spectral fitting tool as we do, and found a strong correlation between the sSFR and the specific dust content in galaxies. 

As already mentioned in Sect.~\ref{sec:wholem31}, M31 has a significantly lower SFR compared to the galaxies in the \citet{daCunha2010} sample. In the relation between the dust mass and the SFR for each pixel, we recover an average displacement with respect to the extrapolation of the empirical law found by \citet{daCunha2010}. Nevertheless, the slope of the average relation defined by individual pixels is remarkably similar. This suggests that the intrinsic star formation relation (i.e. more dust equals more star formation, or vice versa) still holds, but is less efficient and hence scaled down to lower SFR in M31.

\begin{figure}
	\centering
	\includegraphics[width=0.4\textwidth]{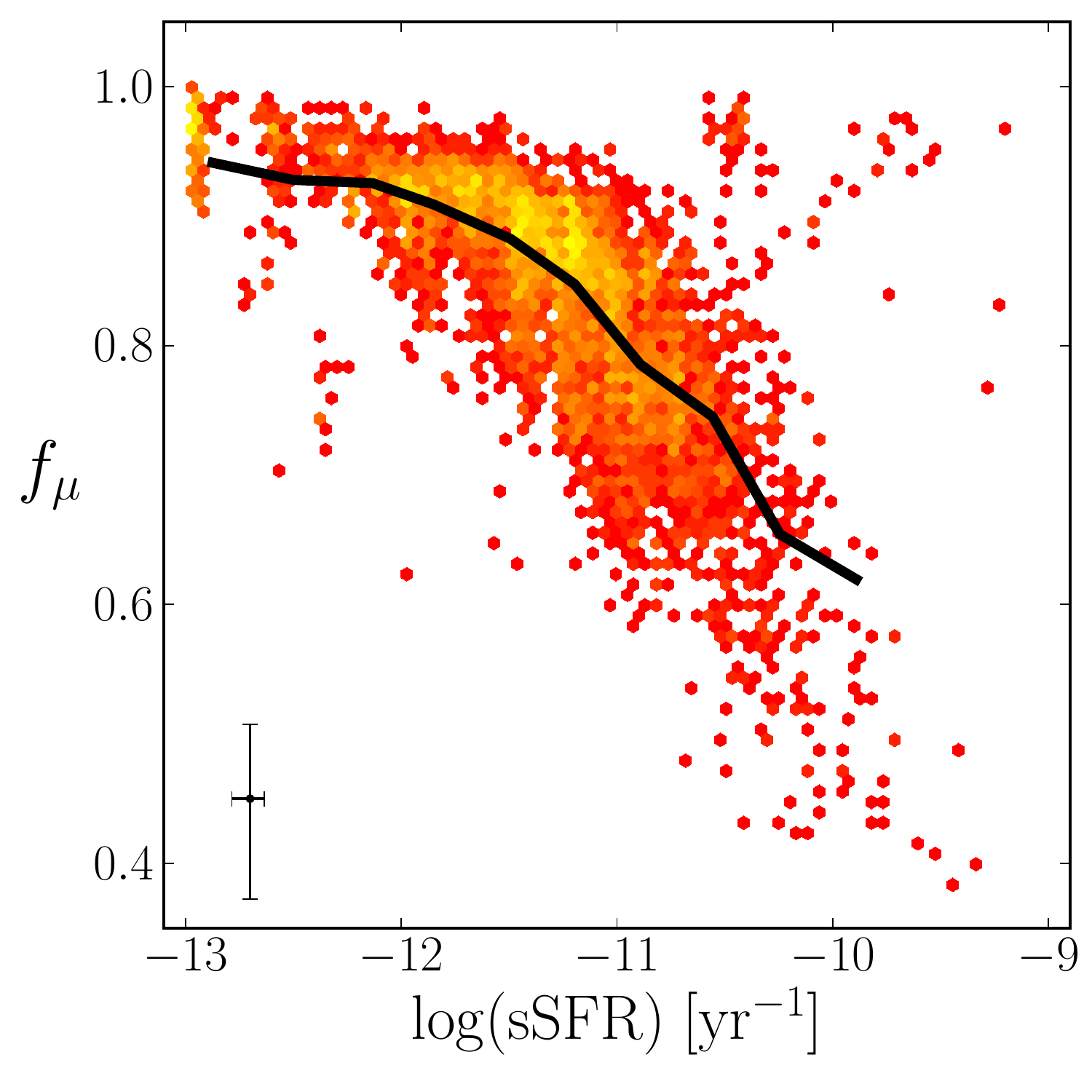}
	\caption{Density plot of the sSFR against $f_\mu$, the luminosity fraction of  ISM dust to the total dust. Yellow points indicate a higher density of points, red a lower density. The black line represents the mean trend of the points.}
	\label{fig:plotfmusSFR}
\end{figure}

Following \citet{daCunha2010} we can compare $f_\mu$, the fraction of the total IR luminosity contributed by dust in the diffuse ISM, to the specific SFR for each pixel. In Fig.~\ref{fig:plotfmusSFR}, we show how this relation, found for galaxies as a whole, is recovered on a local basis as well, with two main differences. First, the region corresponding to values of  $f_\mu$ close to 1 is more densely populated. It turns out that these regions correspond to the very red bulge of M31. Secondly, systematically lower sSFR values are found for a given $f_\mu$ at these local scales. This is likely a manifestation of the low star formation activity of Andromeda with respect to the galaxies in the sample of \citet{daCunha2010}.

More specifically, these results combined together show that the interplay between the ISM, the radiation field and the stellar content (so the SFR as well) takes place on a sub-kpc scale. That the scaling relations were local in nature was already suggested in the previous section, when we subdivided M31 in morphologically distinct regions. We can now state that, globally, the main scaling relations are built up from the properties of the local galactic environment. 

Furthermore, even on the smallest scales accessible with our data, the inside--out trend is, on average, preserved. The bulge is dominated by very red sub-kpc regions with very low sSFR values and high stellar densities. Pixels in the inner disk region are on average slightly bluer, but still have low sSFR values. Star forming pixels are mostly found in the $10$~kpc ring and also have the highest specific star formation. In the outer part of M31 the star forming activity drops again. 

It must be stressed that the above considerations are the average trends constructed from the large set of pixels within each macro-region. It is clear from the sub-kpc regions plotted in Fig.~\ref{fig:Helga_HRS} and from the trends within each region in Fig.~\ref{fig:regions}, that a wide range of physical properties are present within each of the macro-regions. For example, the star forming ring does hold a small number of early-type pixels, while it is dominated by the starforming pixels. Vice versa, the inner disk of M31 holds a significant amount of late-type pixels, while the bulk of its sub-kpc regions are red and have low sSFR values. The exception to these findings is the bulge of M31, which consists purely of early-type regions with high stellar densities and only a minor SFR.

\subsection{Radiation field and dust heating}
We can now analyse the local characteristics of the radiation field in Andromeda and study the dependence of the dust temperature as a function of the stellar characteristics. This exercise has already been performed in \citet{HelgaII}, but we repeat it here using a physically consistent model, which simultaneously takes into account information from the whole spectral domain. 

\begin{figure}
	\centering
	\includegraphics[width=0.35\textwidth]{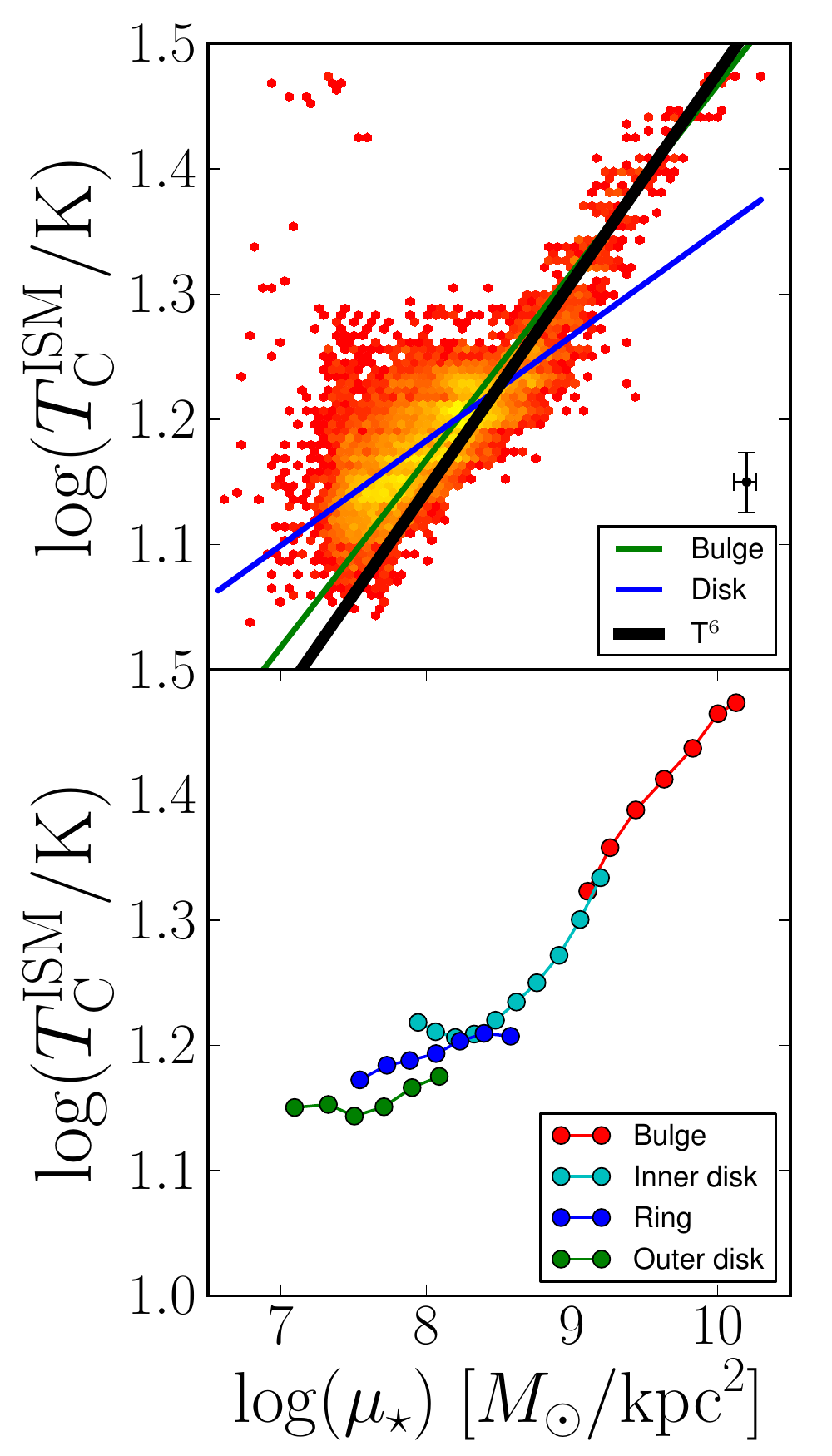}
	\caption{\textbf{Top:} Density plot of the stellar mass surface density $\mu_\star$ vs $T_\mathrm{C}^\mathrm{ISM}$ for the individual sub-kpc regions of M31. Yellow points indicate a higher density of points, red a lower density. A linear function was fit to the bulge (green line) and disk (blue line) regions. The black line is a theoretical function for pure heating by the old stellar populations. \textbf{Bottom:} Same relation, but separated in the macro-regions for M31: the bulge (red), the inner disk (cyan), the $10$~kpc ring (blue), and the outer disk (green). All pixel values are binned in $\mu_\star$; each point is the average of a bin.}
	\label{fig:plotTcMs_regs}
\end{figure}

In Fig.~\ref{fig:plotTcMs_regs} we plot the cold dust component temperature as a function of the stellar mass surface density, for each single independent pixel (upper panel) and for each of the macro-regions of M31 (lower panel). We observe a clear bimodality in this trend, which can be characterised by two slopes, with a break at $\log(T) \sim 1.25$ or T $\sim18$~K. The upper part of the relation, towards the higher temperature regime, is entirely and exclusively populated by pixels of the bulge. In the intermediate regime, the inner disk region dominates, while the ring and outer disk correspond to the lowest regime of temperatures and stellar densities. Interestingly, the transition takes place at stellar surface densities of $\log(\mu_\star)\sim8.5$. In the case of integrated galaxies, this is exactly the value that indicates the transition from disk to bulge dominated systems \citep{Schiminovich2007}.

As described in \citet{HelgaII}, we expect a slope of $6^{-1}$ in a $T_d$ vs $\mu_\star$ plot (as $\mu_\star\sim T^{4+2}$ from the Stefan--Boltzmann law, weighted with the dust emissivity index $\beta=2$) if the heating of diffuse dust is purely due to the ISRF of the old stellar populations. The slope of the linear fit to the bulge pixels is $6.68^{-1}$, which compares to the value of $4.61^{-1}$ that was calculated by \citet{HelgaII}, correlating the dust temperature from black--body fitting and the $3.6~\mu\mathrm{m}$ surface density. Figure~\ref{fig:plotTcMs_regs} visually shows how that our value is close to the expected one. The slope of the ``low--temperature'' regime is instead $11.95^{-1}$.

The bimodal relation we find is clearly indicative of two different heating regimes. This duality lies in the line of previous investigations of dust heating sources on sub-kpc scales \citep[e.g.][]{Boquien2011,Bendo2012}. In the disk of Andromeda, the dust is heated by both old and new stars, giving rise to a rather flat slope and a significant amount of scatter. The bulge of M31 is instead dominated by the light of old stars, making them the dominant dust heaters. This further supports the results of \citet{HelgaII}, \citet{Groves2012}, and \citet{Draine2014}.

\section{Discussion and conclusions} \label{sec:discussion}

In the present work, we have performed SED fitting of a panchromatic dataset, collected for our neighbour galaxy M31. New \textit{Herschel} observations were combined with GALEX, SDSS, WISE, and \textit{Spitzer} data, covering UV to submm wavelengths and allowing us to derive, by exploiting a physically self--consistent model, some physical parameters both on a global and on a local scale. To create statistically independent regions, all the data were convolved to a resolution matching that of the SPIRE $500~\mu\mathrm{m}$ waveband, the lowest in our dataset, which allowed us to probe physical scales of  $\sim 137\times 608$ pc in the plane of M31. In this paper we concentrate on the analysis of the scaling relations linking the dust and stellar properties.  

We have fitted a multi-component theoretical SED to each pixel which allowed us to estimate several physical properties of that region. Every physical parameter for every pixel was given an uncertainty estimation based on the broadness of its corresponding PDF. Physical quantities that could not be sufficiently constrained were removed from the sample. Furthermore, 2-D parameter maps are constructed for each physical quantity.

Additionally, we have decomposed Andromeda in four macro-regions: the bulge, the star forming ring (the so--called $10$~kpc ring), and the inner and outer disk regions. The same fitting routine was applied to the integrated fluxes of each of these main components, as well as to the global observed fluxes. 

From the point of view of the dust scaling relations, M31 is an average galaxy when compared to the local galaxies in the HRS sample. On the other hand, it lies above the average $M_\mathrm{dust}$ vs SFR relation of \citet{daCunha2010}; despite a dust mass close to the sample average, Andromeda is forming stars significantly less efficiently than the other galaxies.

By investigating the properties of the distinct morphological components, we find strong hints for an inside--out star formation scenario. In this evolutionary model, the bulk of stars are being formed at early epochs in the bulge, and the more recent star formation happens at larger galactocentric distances (i.e. $\gtrsim 3$~kpc). In particular, the bulk of star formation is currently taking place in the 10~kpc ring, the morphological structure that contains most of the dust in the galaxy. While the analysis presented in this paper can give no strong clues regarding the past star formation history, we do find an enhancement of dust and star formation in the ring with respect to the galactic disk. This would be consistent with a triggering due to a close encounter with the dwarf satellite M32. \cite{Block2006} used numerical N-body simulations to model the effect of the passage of M32 through Andromeda's disk. Beginning with a model with two spiral arms, they end up with a morphological structure closely matching the 10~kpc ring and the ``hole'' which is easily visible in IR images towards the south. Similarly, \citet{Gordon2006} argued that a head--on encounter with M32, might have resulted in star forming waves propagating through the 10~kpc ring. 

The bulge and inner disk region have red colours and high stellar mass surface density ($\mu_\star$). The star forming ring and outer disk region are bluer and have lower $\mu_\star$. Each of these regions lies on the average trend of the HRS scaling relations. In terms of NUV--r colour, the bulge of M31 is a remarkable exception, being redder than any of the submm detected HRS galaxies. The macro-regions thus have characteristics closely resembling those of global galaxies, where the bulge and inner disk may be seen as early-type while the ring and outer disk resemble late-type galaxies.

The results for M31 not only support an inside-out formation pattern for the bulge-disk morphology, but they also tell us something about the differences in the local environments. When looking at the dust--to--stellar mass ratio as a function of the stellar mass surface density (see Fig.~\ref{fig:regions}), we observe a smooth transition from high stellar mass/low dust regions in the bulge, to the low stellar mass/high dust content of the outermost regions. Only the starforming ring, containing a significant fraction of the dust of the whole galaxy, is slightly displaced from this relation. 

On the other hand, the four regions behave quite differently when the sSFR is plotted as a function of $\mu_\star$ (see Fig.~\ref{fig:regions}).  A tendency is found for regions of higher stellar mass surface density to host less star formation, in a way mimicking an internal downsizing process in star formation, in which the regions of highest stellar density have already stopped forming stars. 
Within each region, the variation in $\mu_\star$ spans only about 1 order of magnitude, whereas the sSFR always varies by more than 3 (with the exception of the bulge, where there is barely any star formation).

These relations seem to suggest that downsizing relations break down when considering the smallest scales. Star formation at these scales has a weak dependence on the stellar mass: small-scale environments characterised by different stellar masses can be associated with very different levels of star formation, at least as far as a quiescent galaxy like M31 is concerned. What drives the general star formation mode, which determines where the galaxy as a whole places itself on the scaling relations, must instead be the total mass of all its constituents.

When considering the modelling of the observed SED on the smallest scales, our main conclusions are as follows.
\begin{enumerate}
\item The SED of sub-kpc regions can be successfully fitted using galaxy-based models, provided that the parameter space is adequately sampled.
\item When investigating the dust heating in the bulge, we recover the theoretical $(T_\mathrm{C}^\mathrm{ISM})^6 \sim$ $\mu_\star$ relation. This indicates that old stars are the dominant heating source in this region. The dust heating is more ambiguous in the disk, where both star formation and the diffuse ISRF irradiate the dust.
\item We find strong correlations, on a pixel--by--pixel scale, between $M_\mathrm{dust}/M_\star$ and NUV--r (or, equivalently, sSFR), and between $M_\mathrm{dust}/M_\star$ and $\mu_\star$. These scaling relations, involving the dusty component of the ISM, are remarkably similar to those found for entire local galaxies. This suggests that the dust scaling relations are built {\it in situ}, with underlying physical processes that must be local in nature.
\item As already found for other galaxies, M31 seems to have undergone an inside--out evolution in its star formation process, possibly influenced by interactions with its satellites.
\item When considering the smallest scales, a wide range in dust content, sSFR, and $M_\mathrm{dust}/M_\star$ is found within Andromeda illustrating  the great diversity of sub-kpc regions. Even within the late-type ring and outer disk of M31, early-type micro-regions can be found. Vice versa, the inner part of M31 still holds a small number of late-type regions.
\end{enumerate}

The fact that we are able to reproduce the dust scaling relations on a sub-kpc scale states that these relations are not only partially a manifestation of a galaxy-wide equilibrium, but they also arise from local scales. Local evolutionary processes involving dust creation and destruction lie at the base of these relations. The balance between dust depletion and production reflects the relative presence of old and new stars, the latter being responsible for dust generation. This raises the question: at what scales does the balanced interplay between dust and stars break down? Answers to this may be found in similar studies of the Galactic ISM or by future, high--resolution FIR space missions. Zooming into the ISM of our own galaxy can unveil two very different results. If these scaling relations break down at the size of individual molecular clouds, it would indicate that non--local scattered light plays an important role in the dust energy balance. Alternatively, if the scaling relations stay intact, non--local light is negligible at each scale, which would call for a revision of the physical properties of interstellar dust.

\begin{acknowledgements}
We would like to thank Elisabete da Cunha for kindly providing the extra libraries in MAGPHYS.\\

We thank all the people involved in the construction and the launch of Herschel. SPIRE has been developed by a consortium of institutes led by Cardiff University (UK) and including Univ. Lethbridge (Canada); NAOC (China); CEA, LAM (France); IFSI, Univ. Padua (Italy); IAC (Spain); Stockholm Observatory (Sweden); Imperial College London, RAL, UCL-MSSL, UKATC, Univ. Sussex (UK); and Caltech, JPL, NHSC, Univ. Colorado (USA). This development has been supported by national funding agencies: CSA (Canada); NAOC (China); CEA, CNES, CNRS (France); ASI (Italy); MCINN (Spain); SNSB (Sweden); STFC and UKSA (UK); and NASA (USA). HIPE is a joint development (are joint developments) by the \textit{Herschel} Science Ground Segment Consortium, consisting of ESA, the NASA \textit{Herschel} Science Center, and the HIFI, PACS and SPIRE consortia.\\

GALEX is a NASA Small Explorer, launched in 2003 April. We gratefully acknowledge NASA’s support for construction, operation and science analysis for the GALEX mission, developed in cooperation with the Centre National d’Etudes Spatiales (CNES) of France and the Korean Ministry of Science and Technology.\\

Funding for the SDSS and SDSS-II has been provided by the Alfred P. Sloan Foundation, the Participating Institutions, the National Science Foundation, the US Department of Energy, the National Aeronautics and Space Administration, the Japanese Monbukagakusho, the Max Planck Society, and the Higher Education Funding Council for England. The SDSSWeb Site is http://www.sdss.org/. The SDSS is managed by the Astrophysical Research Consortium for the Participating Institutions. The Participating Institutions are the American Museum of Natural History, Astrophysical Institute Potsdam, University of Basel, University of Cambridge, Case Western Reserve University, University of Chicago, Drexel University, Fermilab, the Institute for Advanced Study, the Japan Participation Group, Johns Hopkins University, the Joint Institute for Nuclear Astrophysics, the Kavli Institute for Particle Astrophysics and Cosmology, the Korean Scientist Group, the Chinese Academy of Sciences (LAMOST), Los Alamos National Laboratory, the Max-Planck-Institute for Astronomy (MPIA), the Max-Planck-Institute for Astrophysics (MPA), New Mexico State University, Ohio State University, University of Pittsburgh, University of Portsmouth, Princeton University, the United States Naval Observatory, and the University of Washington.\\

This work is based [in part] on observations made with the \textit{Spitzer} Space Telescope, which is operated by the Jet Propulsion Laboratory, California Institute of Technology, under NASA contract 1407. We especially thank P. Barmby and K. Gordon for providing the Spitzer data.\\

This publication makes use of data products from the Wide-field Infrared Survey Explorer, which is a joint project of the University of California, Los Angeles, and the Jet Propulsion Laboratory/California Institute of Technology, funded by the National Aeronautics and Space Administration.
\end{acknowledgements}

\bibliographystyle{aa} 
\bibliography{allreferences}

\begin{thebibliography}{110}
\expandafter\ifx\csname natexlab\endcsname\relax\def\natexlab#1{#1}\fi

\bibitem[{{Agius} {et~al.}(2013){Agius}, {Sansom}, {Popescu}, {Andrae}, {Baes},
  {Baldry}, {Bourne}, {Brough}, {Clark}, {Conselice}, {Cooray}, {Dariush}, {De
  Zotti}, {Driver}, {Dunne}, {Eales}, {Foster}, {Gomez}, {H{\"a}u{\ss}ler},
  {Hopkins}, {Hopwood}, {Ivison}, {Kelvin}, {Lara-L{\'o}pez}, {Liske},
  {L{\'o}pez-S{\'a}nchez}, {Loveday}, {Maddox}, {Madore}, {Phillipps},
  {Robotham}, {Rowlands}, {Seibert}, {Smith}, {Temi}, {Tuffs}, \&
  {Valiante}}]{Agius2013}
{Agius}, N.~K., {Sansom}, A.~E., {Popescu}, C.~C., {et~al.} 2013, \mnras, 431,
  1929

\bibitem[{{Aniano} {et~al.}(2012){Aniano}, {Draine}, {Calzetti}, {Dale},
  {Engelbracht}, {Gordon}, {Hunt}, {Kennicutt}, {Krause}, {Leroy}, {Rix},
  {Roussel}, {Sandstrom}, {Sauvage}, {Walter}, {Armus}, {Bolatto}, {Crocker},
  {Donovan Meyer}, {Galametz}, {Helou}, {Hinz}, {Johnson}, {Koda}, {Montiel},
  {Murphy}, {Skibba}, {Smith}, \& {Wolfire}}]{Aniano2012}
{Aniano}, G., {Draine}, B.~T., {Calzetti}, D., {et~al.} 2012, \apj, 756, 138

\bibitem[{{Aniano} {et~al.}(2011){Aniano}, {Draine}, {Gordon}, \&
  {Sandstrom}}]{Aniano2011}
{Aniano}, G., {Draine}, B.~T., {Gordon}, K.~D., \& {Sandstrom}, K. 2011, \pasp,
  123, 1218

\bibitem[{{Azimlu} {et~al.}(2011){Azimlu}, {Marciniak}, \&
  {Barmby}}]{Azimlu2011}
{Azimlu}, M., {Marciniak}, R., \& {Barmby}, P. 2011, \aj, 142, 139

\bibitem[{{Barmby} {et~al.}(2006){Barmby}, {Ashby}, {Bianchi}, {Engelbracht},
  {Gehrz}, {Gordon}, {Hinz}, {Huchra}, {Humphreys}, {Pahre},
  {P{\'e}rez-Gonz{\'a}lez}, {Polomski}, {Rieke}, {Thilker}, {Willner}, \&
  {Woodward}}]{Barmby2006}
{Barmby}, P., {Ashby}, M.~L.~N., {Bianchi}, L., {et~al.} 2006, \apjl, 650, L45

\bibitem[{{Barmby} {et~al.}(2009){Barmby}, {Perina}, {Bellazzini}, {Cohen},
  {Hodge}, {Huchra}, {Kissler-Patig}, {Puzia}, \& {Strader}}]{Barmby2009}
{Barmby}, P., {Perina}, S., {Bellazzini}, M., {et~al.} 2009, \aj, 138, 1667

\bibitem[{{Beaton} {et~al.}(2007){Beaton}, {Majewski}, {Guhathakurta},
  {Skrutskie}, {Cutri}, {Good}, {Patterson}, {Athanassoula}, \&
  {Bureau}}]{Beaton2007}
{Beaton}, R.~L., {Majewski}, S.~R., {Guhathakurta}, P., {et~al.} 2007, \apjl,
  658, L91

\bibitem[{{Bendo} {et~al.}(2012){Bendo}, {Boselli}, {Dariush}, {Pohlen},
  {Roussel}, {Sauvage}, {Smith}, {Wilson}, {Baes}, {Cooray}, {Clements},
  {Cortese}, {Foyle}, {Galametz}, {Gomez}, {Lebouteiller}, {Lu}, {Madden},
  {Mentuch}, {O'Halloran}, {Page}, {Remy}, {Schulz}, \&
  {Spinoglio}}]{Bendo2012}
{Bendo}, G.~J., {Boselli}, A., {Dariush}, A., {et~al.} 2012, \mnras, 419, 1833

\bibitem[{{Bendo} {et~al.}(2007){Bendo}, {Calzetti}, {Engelbracht},
  {Kennicutt}, {Meyer}, {Thornley}, {Walter}, {Dale}, {Li}, \&
  {Murphy}}]{Bendo2007}
{Bendo}, G.~J., {Calzetti}, D., {Engelbracht}, C.~W., {et~al.} 2007, \mnras,
  380, 1313

\bibitem[{{Bendo} {et~al.}(2008){Bendo}, {Draine}, {Engelbracht}, {Helou},
  {Thornley}, {Bot}, {Buckalew}, {Calzetti}, {Dale}, {Hollenbach}, {Li}, \&
  {Moustakas}}]{Bendo2008}
{Bendo}, G.~J., {Draine}, B.~T., {Engelbracht}, C.~W., {et~al.} 2008, \mnras,
  389, 629

\bibitem[{{Bendo} {et~al.}(2010){Bendo}, {Wilson}, {Warren}, {Brinks},
  {Butner}, {Chanial}, {Clements}, {Courteau}, {Irwin}, {Israel}, {Knapen},
  {Leech}, {Matthews}, {M{\"u}hle}, {Petitpas}, {Serjeant}, {Tan}, {Tilanus},
  {Usero}, {Vaccari}, {van der Werf}, {Vlahakis}, {Wiegert}, \&
  {Zhu}}]{Bendo2010a}
{Bendo}, G.~J., {Wilson}, C.~D., {Warren}, B.~E., {et~al.} 2010, \mnras, 402,
  1409

\bibitem[{{Bertin} \& {Arnouts}(1996)}]{sextractor}
{Bertin}, E. \& {Arnouts}, S. 1996, \aaps, 117, 393

\bibitem[{{Bianchi}(2008)}]{Bianchi2008}
{Bianchi}, S. 2008, \aap, 490, 461

\bibitem[{{Block} {et~al.}(2006){Block}, {Bournaud}, {Combes}, {Groess},
  {Barmby}, {Ashby}, {Fazio}, {Pahre}, \& {Willner}}]{Block2006}
{Block}, D.~L., {Bournaud}, F., {Combes}, F., {et~al.} 2006, \nat, 443, 832

\bibitem[{{Boquien} {et~al.}(2013){Boquien}, {Boselli}, {Buat}, {Baes},
  {Bendo}, {Boissier}, {Ciesla}, {Cooray}, {Cortese}, {Eales}, {Koda},
  {Lebouteiller}, {de Looze}, {Smith}, {Spinoglio}, \& {Wilson}}]{Boquien2013}
{Boquien}, M., {Boselli}, A., {Buat}, V., {et~al.} 2013, \aap, 554, A14

\bibitem[{{Boquien} {et~al.}(2012){Boquien}, {Buat}, {Boselli}, {Baes},
  {Bendo}, {Ciesla}, {Cooray}, {Cortese}, {Eales}, {Gavazzi}, {Gomez},
  {Lebouteiller}, {Pappalardo}, {Pohlen}, {Smith}, \&
  {Spinoglio}}]{Boquien2012}
{Boquien}, M., {Buat}, V., {Boselli}, A., {et~al.} 2012, \aap, 539, A145

\bibitem[{{Boquien} {et~al.}(2011){Boquien}, {Calzetti}, {Combes}, {Henkel},
  {Israel}, {Kramer}, {Rela{\~n}o}, {Verley}, {van der Werf}, {Xilouris}, \&
  {HERM33ES Team}}]{Boquien2011}
{Boquien}, M., {Calzetti}, D., {Combes}, F., {et~al.} 2011, \aj, 142, 111

\bibitem[{{Boselli} {et~al.}(2009){Boselli}, {Boissier}, {Cortese}, {Buat},
  {Hughes}, \& {Gavazzi}}]{Boselli2009}
{Boselli}, A., {Boissier}, S., {Cortese}, L., {et~al.} 2009, \apj, 706, 1527

\bibitem[{{Boselli} {et~al.}(2010){Boselli}, {Eales}, {Cortese}, {Bendo},
  {Chanial}, {Buat}, {Davies}, {Auld}, {Rigby}, {Baes}, {Barlow}, {Bock},
  {Bradford}, {Castro-Rodriguez}, {Charlot}, {Clements}, {Cormier}, {Dwek},
  {Elbaz}, {Galametz}, {Galliano}, {Gear}, {Glenn}, {Gomez}, {Griffin}, {Hony},
  {Isaak}, {Levenson}, {Lu}, {Madden}, {O'Halloran}, {Okamura}, {Oliver},
  {Page}, {Panuzzo}, {Papageorgiou}, {Parkin}, {Perez-Fournon}, {Pohlen},
  {Rangwala}, {Roussel}, {Rykala}, {Sacchi}, {Sauvage}, {Schulz}, {Schirm},
  {Smith}, {Spinoglio}, {Stevens}, {Symeonidis}, {Vaccari}, {Vigroux},
  {Wilson}, {Wozniak}, {Wright}, \& {Zeilinger}}]{boselli2010}
{Boselli}, A., {Eales}, S., {Cortese}, L., {et~al.} 2010, \pasp, 122, 261

\bibitem[{{Brinchmann} {et~al.}(2004){Brinchmann}, {Charlot}, {White},
  {Tremonti}, {Kauffmann}, {Heckman}, \& {Brinkmann}}]{Brinchmann2004}
{Brinchmann}, J., {Charlot}, S., {White}, S.~D.~M., {et~al.} 2004, \mnras, 351,
  1151

\bibitem[{{Bruzual} \& {Charlot}(2003)}]{Bruzual2003}
{Bruzual}, G. \& {Charlot}, S. 2003, \mnras, 344, 1000

\bibitem[{{Chabrier}(2003)}]{Chabrier2003}
{Chabrier}, G. 2003, \pasp, 115, 763

\bibitem[{{Charlot} \& {Fall}(2000)}]{charlot2000}
{Charlot}, S. \& {Fall}, S.~M. 2000, \apj, 539, 718

\bibitem[{{Chemin} {et~al.}(2009){Chemin}, {Carignan}, \&
  {Foster}}]{Chemin2009}
{Chemin}, L., {Carignan}, C., \& {Foster}, T. 2009, \apj, 705, 1395

\bibitem[{{Clemens} {et~al.}(2013){Clemens}, {Negrello}, {De Zotti},
  {Gonzalez-Nuevo}, {Bonavera}, {Cosco}, {Guarese}, {Boaretto}, {Salucci},
  {Baccigalupi}, {Clements}, {Danese}, {Lapi}, {Mandolesi}, {Partridge},
  {Perrotta}, {Serjeant}, {Scott}, \& {Toffolatti}}]{Clemens2013}
{Clemens}, M.~S., {Negrello}, M., {De Zotti}, G., {et~al.} 2013, \mnras

\bibitem[{{Corbelli} {et~al.}(2010){Corbelli}, {Lorenzoni}, {Walterbos},
  {Braun}, \& {Thilker}}]{Corbelli2010}
{Corbelli}, E., {Lorenzoni}, S., {Walterbos}, R., {Braun}, R., \& {Thilker}, D.
  2010, \aap, 511, A89

\bibitem[{{Cortese} {et~al.}(2012){Cortese}, {Ciesla}, {Boselli}, {Bianchi},
  {Gomez}, {Smith}, {Bendo}, {Eales}, {Pohlen}, {Baes}, {Corbelli}, {Davies},
  {Hughes}, {Hunt}, {Madden}, {Pierini}, {di Serego Alighieri}, {Zibetti},
  {Boquien}, {Clements}, {Cooray}, {Galametz}, {Magrini}, {Pappalardo},
  {Spinoglio}, \& {Vlahakis}}]{Cortese2012}
{Cortese}, L., {Ciesla}, L., {Boselli}, A., {et~al.} 2012, \aap, 540, A52

\bibitem[{{da Cunha} {et~al.}(2008){da Cunha}, {Charlot}, \&
  {Elbaz}}]{Dacunha2008}
{da Cunha}, E., {Charlot}, S., \& {Elbaz}, D. 2008, \mnras, 388, 1595

\bibitem[{{da Cunha} {et~al.}(2010){da Cunha}, {Eminian}, {Charlot}, \&
  {Blaizot}}]{daCunha2010}
{da Cunha}, E., {Eminian}, C., {Charlot}, S., \& {Blaizot}, J. 2010, \mnras,
  403, 1894

\bibitem[{{Davies} {et~al.}(2010){Davies}, {Wilson}, {Auld}, {Baes}, {Barlow},
  {Bendo}, {Bock}, {Boselli}, {Bradford}, {Buat}, {Castro-Rodriguez},
  {Chanial}, {Charlot}, {Ciesla}, {Clements}, {Cooray}, {Cormier}, {Cortese},
  {Dwek}, {Eales}, {Elbaz}, {Galametz}, {Galliano}, {Gear}, {Glenn}, {Gomez},
  {Griffin}, {Hony}, {Isaak}, {Levenson}, {Lu}, {Madden}, {O'Halloran},
  {Okumura}, {Oliver}, {Page}, {Panuzzo}, {Papageorgiou}, {Parkin},
  {Perez-Fournon}, {Pohlen}, {Rangwala}, {Rigby}, {Roussel}, {Rykala},
  {Sacchi}, {Sauvage}, {Schulz}, {Schirm}, {Smith}, {Spinoglio}, {Stevens},
  {Srinivasan}, {Symeonidis}, {Trichas}, {Vaccari}, {Vigroux}, {Wozniak},
  {Wright}, \& {Zeilinger}}]{Davies2010}
{Davies}, J.~I., {Wilson}, C.~D., {Auld}, R., {et~al.} 2010, \mnras, 409, 102

\bibitem[{{de Vaucouleurs} {et~al.}(1991){de Vaucouleurs}, {de Vaucouleurs},
  {Corwin}, {Buta}, {Paturel}, \& {Fouqu{\'e}}}]{deVaucouleurs1991}
{de Vaucouleurs}, G., {de Vaucouleurs}, A., {Corwin}, Jr., H.~G., {et~al.}
  1991, {Third Reference Catalogue of Bright Galaxies.}

\bibitem[{{Draine}(2003)}]{Draine2003}
{Draine}, B.~T. 2003, \araa, 41, 241

\bibitem[{{Draine} {et~al.}(2014){Draine}, {Aniano}, {Krause}, {Groves},
  {Sandstrom}, {Braun}, {Leroy}, {Klaas}, {Linz}, {Rix}, {Schinnerer},
  {Schmiedeke}, \& {Walter}}]{Draine2014}
{Draine}, B.~T., {Aniano}, G., {Krause}, O., {et~al.} 2014, \apj, 780, 172

\bibitem[{{Draine} {et~al.}(2007){Draine}, {Dale}, {Bendo}, {Gordon}, {Smith},
  {Armus}, {Engelbracht}, {Helou}, {Kennicutt}, {Li}, {Roussel}, {Walter},
  {Calzetti}, {Moustakas}, {Murphy}, {Rieke}, {Bot}, {Hollenbach}, {Sheth}, \&
  {Teplitz}}]{Draine2007}
{Draine}, B.~T., {Dale}, D.~A., {Bendo}, G., {et~al.} 2007, \apj, 663, 866

\bibitem[{{Draine} \& {Li}(2007)}]{DraineLi2007}
{Draine}, B.~T. \& {Li}, A. 2007, \apj, 657, 810

\bibitem[{{Dunne} {et~al.}(2000){Dunne}, {Eales}, {Edmunds}, {Ivison},
  {Alexander}, \& {Clements}}]{Dunne2000}
{Dunne}, L., {Eales}, S., {Edmunds}, M., {et~al.} 2000, \mnras, 315, 115

\bibitem[{{Engelbracht} {et~al.}(2007){Engelbracht}, {Blaylock}, {Su}, {Rho},
  {Rieke}, {Muzerolle}, {Padgett}, {Hines}, {Gordon}, {Fadda},
  {Noriega-Crespo}, {Kelly}, {Latter}, {Hinz}, {Misselt}, {Morrison},
  {Stansberry}, {Shupe}, {Stolovy}, {Wheaton}, {Young}, {Neugebauer},
  {Wachter}, {P{\'e}rez-Gonz{\'a}lez}, {Frayer}, \&
  {Marleau}}]{Engelbracht2007}
{Engelbracht}, C.~W., {Blaylock}, M., {Su}, K.~Y.~L., {et~al.} 2007, \pasp,
  119, 994

\bibitem[{{Fazio} {et~al.}(2004){Fazio}, {Hora}, {Allen}, {Ashby}, {Barmby},
  {Deutsch}, {Huang}, {Kleiner}, {Marengo}, {Megeath}, {Melnick}, {Pahre},
  {Patten}, {Polizotti}, {Smith}, {Taylor}, {Wang}, {Willner}, {Hoffmann},
  {Pipher}, {Forrest}, {McMurty}, {McCreight}, {McKelvey}, {McMurray}, {Koch},
  {Moseley}, {Arendt}, {Mentzell}, {Marx}, {Losch}, {Mayman}, {Eichhorn},
  {Krebs}, {Jhabvala}, {Gezari}, {Fixsen}, {Flores}, {Shakoorzadeh}, {Jungo},
  {Hakun}, {Workman}, {Karpati}, {Kichak}, {Whitley}, {Mann}, {Tollestrup},
  {Eisenhardt}, {Stern}, {Gorjian}, {Bhattacharya}, {Carey}, {Nelson},
  {Glaccum}, {Lacy}, {Lowrance}, {Laine}, {Reach}, {Stauffer}, {Surace},
  {Wilson}, {Wright}, {Hoffman}, {Domingo}, \& {Cohen}}]{IRAC}
{Fazio}, G.~G., {Hora}, J.~L., {Allen}, L.~E., {et~al.} 2004, \apjs, 154, 10

\bibitem[{{Fisher} {et~al.}(2009){Fisher}, {Drory}, \&
  {Fabricius}}]{Fisher2009}
{Fisher}, D.~B., {Drory}, N., \& {Fabricius}, M.~H. 2009, \apj, 697, 630

\bibitem[{{Ford} {et~al.}(2013){Ford}, {Gear}, {Smith}, {Eales}, {Baes},
  {Bendo}, {Boquien}, {Boselli}, {Cooray}, {De Looze}, {Fritz}, {Gentile},
  {Gomez}, {Gordon}, {Kirk}, {Lebouteiller}, {O'Halloran}, {Spinoglio},
  {Verstappen}, \& {Wilson}}]{Ford2013}
{Ford}, G.~P., {Gear}, W.~K., {Smith}, M.~W.~L., {et~al.} 2013, \apj, 769, 55

\bibitem[{{Foyle} {et~al.}(2013){Foyle}, {Natale}, {Wilson}, {Popescu}, {Baes},
  {Bendo}, {Boquien}, {Boselli}, {Cooray}, {Cormier}, {De Looze}, {Fischera},
  {Karczewski}, {Lebouteiller}, {Madden}, {Pereira-Santaella}, {Smith},
  {Spinoglio}, \& {Tuffs}}]{Foyle2013}
{Foyle}, K., {Natale}, G., {Wilson}, C.~D., {et~al.} 2013, \mnras, 432, 2182

\bibitem[{{Fritz} {et~al.}(2012){Fritz}, {Gentile}, {Smith}, {Gear}, {Braun},
  {Duval}, {Bendo}, {Baes}, {Eales}, {Verstappen}, {Blommaert}, {Boquien},
  {Boselli}, {Clements}, {Cooray}, {Cortese}, {De Looze}, {Ford}, {Galliano},
  {Gomez}, {Gordon}, {Lebouteiller}, {O'Halloran}, {Kirk}, {Madden}, {Page},
  {Remy}, {Roussel}, {Spinoglio}, {Thilker}, {Vaccari}, {Wilson}, \&
  {Waelkens}}]{HELGAI}
{Fritz}, J., {Gentile}, G., {Smith}, M.~W.~L., {et~al.} 2012, \aap, 546, A34

\bibitem[{{Galametz} {et~al.}(2012){Galametz}, {Kennicutt}, {Albrecht},
  {Aniano}, {Armus}, {Bertoldi}, {Calzetti}, {Crocker}, {Croxall}, {Dale},
  {Donovan Meyer}, {Draine}, {Engelbracht}, {Hinz}, {Roussel}, {Skibba},
  {Tabatabaei}, {Walter}, {Weiss}, {Wilson}, \& {Wolfire}}]{Galametz2012}
{Galametz}, M., {Kennicutt}, R.~C., {Albrecht}, M., {et~al.} 2012, \mnras, 425,
  763

\bibitem[{{Galametz} {et~al.}(2011){Galametz}, {Madden}, {Galliano}, {Hony},
  {Bendo}, \& {Sauvage}}]{Galametz2011}
{Galametz}, M., {Madden}, S.~C., {Galliano}, F., {et~al.} 2011, \aap, 532, A56

\bibitem[{{Galliano} {et~al.}(2011){Galliano}, {Hony}, {Bernard}, {Bot},
  {Madden}, {Roman-Duval}, {Galametz}, {Li}, {Meixner}, {Engelbracht},
  {Lebouteiller}, {Misselt}, {Montiel}, {Panuzzo}, {Reach}, \&
  {Skibba}}]{Galliano2011}
{Galliano}, F., {Hony}, S., {Bernard}, J.-P., {et~al.} 2011, \aap, 536, A88

\bibitem[{{Gil de Paz} {et~al.}(2007){Gil de Paz}, {Boissier}, {Madore},
  {Seibert}, {Joe}, {Boselli}, {Wyder}, {Thilker}, {Bianchi}, {Rey}, {Rich},
  {Barlow}, {Conrow}, {Forster}, {Friedman}, {Martin}, {Morrissey}, {Neff},
  {Schiminovich}, {Small}, {Donas}, {Heckman}, {Lee}, {Milliard}, {Szalay}, \&
  {Yi}}]{GilDePaz2007}
{Gil de Paz}, A., {Boissier}, S., {Madore}, B.~F., {et~al.} 2007, \apjs, 173,
  185

\bibitem[{{Gonz{\'a}lez Delgado} {et~al.}(2013){Gonz{\'a}lez Delgado},
  {P{\'e}rez}, {Cid Fernandes}, {Garc{\'{\i}}a-Benito}, {de Amorim},
  {S{\'a}nchez}, {Husemann}, {Cortijo-Ferrero}, {L{\'o}pez Fern{\'a}ndez},
  {S{\'a}nchez-Bl{\'a}zquez}, {Bekeraite}, {Walcher}, {Falc{\'o}n-Barroso},
  {Gallazzi}, {van de Ven}, {Alves}, {Bland-Hawthorn}, {Kennicutt}, {Kupko},
  {Lyubenova}, {Mast}, {Moll{\'a}}, {Marino}, {Quirrenbach}, {V{\'{\i}}lchez},
  {Wisotzki}, \& {$\backslash$}}]{Delgado2013}
{Gonz{\'a}lez Delgado}, R.~M., {P{\'e}rez}, E., {Cid Fernandes}, R., {et~al.}
  2013, \href{http://arxiv.org/abs/1310.5517}{arXiv:1310.5517}

\bibitem[{{Gordon} {et~al.}(2006){Gordon}, {Bailin}, {Engelbracht}, {Rieke},
  {Misselt}, {Latter}, {Young}, {Ashby}, {Barmby}, {Gibson}, {Hines}, {Hinz},
  {Krause}, {Levine}, {Marleau}, {Noriega-Crespo}, {Stolovy}, {Thilker}, \&
  {Werner}}]{Gordon2006}
{Gordon}, K.~D., {Bailin}, J., {Engelbracht}, C.~W., {et~al.} 2006, \apjl, 638,
  L87

\bibitem[{{Gordon} {et~al.}(2007){Gordon}, {Engelbracht}, {Fadda},
  {Stansberry}, {Wachter}, {Frayer}, {Rieke}, {Noriega-Crespo}, {Latter},
  {Young}, {Neugebauer}, {Balog}, {Beeman}, {Dole}, {Egami}, {Haller}, {Hines},
  {Kelly}, {Marleau}, {Misselt}, {Morrison}, {P{\'e}rez-Gonz{\'a}lez}, {Rho},
  \& {Wheaton}}]{Gordon2007}
{Gordon}, K.~D., {Engelbracht}, C.~W., {Fadda}, D., {et~al.} 2007, \pasp, 119,
  1019

\bibitem[{{Griffin} {et~al.}(2010){Griffin}, {Abergel}, {Abreu}, {Ade},
  {Andr{\'e}}, {Augueres}, {Babbedge}, {Bae}, {Baillie}, {Baluteau}, {Barlow},
  {Bendo}, {Benielli}, {Bock}, {Bonhomme}, {Brisbin}, {Brockley-Blatt},
  {Caldwell}, {Cara}, {Castro-Rodriguez}, {Cerulli}, {Chanial}, {Chen},
  {Clark}, {Clements}, {Clerc}, {Coker}, {Communal}, {Conversi}, {Cox},
  {Crumb}, {Cunningham}, {Daly}, {Davis}, {de Antoni}, {Delderfield}, {Devin},
  {di Giorgio}, {Didschuns}, {Dohlen}, {Donati}, {Dowell}, {Dowell}, {Duband},
  {Dumaye}, {Emery}, {Ferlet}, {Ferrand}, {Fontignie}, {Fox}, {Franceschini},
  {Frerking}, {Fulton}, {Garcia}, {Gastaud}, {Gear}, {Glenn}, {Goizel},
  {Griffin}, {Grundy}, {Guest}, {Guillemet}, {Hargrave}, {Harwit}, {Hastings},
  {Hatziminaoglou}, {Herman}, {Hinde}, {Hristov}, {Huang}, {Imhof}, {Isaak},
  {Israelsson}, {Ivison}, {Jennings}, {Kiernan}, {King}, {Lange}, {Latter},
  {Laurent}, {Laurent}, {Leeks}, {Lellouch}, {Levenson}, {Li}, {Li},
  {Lilienthal}, {Lim}, {Liu}, {Lu}, {Madden}, {Mainetti}, {Marliani}, {McKay},
  {Mercier}, {Molinari}, {Morris}, {Moseley}, {Mulder}, {Mur}, {Naylor},
  {Nguyen}, {O'Halloran}, {Oliver}, {Olofsson}, {Olofsson}, {Orfei}, {Page},
  {Pain}, {Panuzzo}, {Papageorgiou}, {Parks}, {Parr-Burman}, {Pearce},
  {Pearson}, {P{\'e}rez-Fournon}, {Pinsard}, {Pisano}, {Podosek}, {Pohlen},
  {Polehampton}, {Pouliquen}, {Rigopoulou}, {Rizzo}, {Roseboom}, {Roussel},
  {Rowan-Robinson}, {Rownd}, {Saraceno}, {Sauvage}, {Savage}, {Savini},
  {Sawyer}, {Scharmberg}, {Schmitt}, {Schneider}, {Schulz}, {Schwartz},
  {Shafer}, {Shupe}, {Sibthorpe}, {Sidher}, {Smith}, {Smith}, {Smith},
  {Spencer}, {Stobie}, {Sudiwala}, {Sukhatme}, {Surace}, {Stevens}, {Swinyard},
  {Trichas}, {Tourette}, {Triou}, {Tseng}, {Tucker}, {Turner}, {Vaccari},
  {Valtchanov}, {Vigroux}, {Virique}, {Voellmer}, {Walker}, {Ward}, {Waskett},
  {Weilert}, {Wesson}, {White}, {Whitehouse}, {Wilson}, {Winter}, {Woodcraft},
  {Wright}, {Xu}, {Zavagno}, {Zemcov}, {Zhang}, \& {Zonca}}]{SPIRE}
{Griffin}, M.~J., {Abergel}, A., {Abreu}, A., {et~al.} 2010, \aap, 518, L3

\bibitem[{{Groves} {et~al.}(2012){Groves}, {Krause}, {Sandstrom}, {Schmiedeke},
  {Leroy}, {Linz}, {Kapala}, {Rix}, {Schinnerer}, {Tabatabaei}, {Walter}, \&
  {da Cunha}}]{Groves2012}
{Groves}, B., {Krause}, O., {Sandstrom}, K., {et~al.} 2012, \mnras, 426, 892

\bibitem[{{Herschel Space Observatory}(2011)}]{SPIREmanual}
{Herschel Space Observatory}. 2011, {SPIRE Observer’s Manual},
  \url{http://herschel.esac.esa.int/Docs/SPIRE/html/spire_om.html}

\bibitem[{{Hughes} {et~al.}(2014){Hughes}, {Baes}, {Fritz}, {Smith}, {Parkin},
  {Gentile}, {Bendo}, {Wilson}, {Allaert}, {Bianchi}, {De Looze}, {Verstappen},
  {Viaene}, {Boquien}, {Boselli}, {Clements}, {Davies}, {Galametz}, {Madden},
  {Remy-Ruyer}, \& {Spinoglio}}]{Hughes2014}
{Hughes}, T.~M., {Baes}, M., {Fritz}, J., {et~al.} 2014, ArXiv e-prints

\bibitem[{{Issa} {et~al.}(1990){Issa}, {MacLaren}, \& {Wolfendale}}]{Issa1990}
{Issa}, M.~R., {MacLaren}, I., \& {Wolfendale}, A.~W. 1990, \aap, 236, 237

\bibitem[{{Jarrett} {et~al.}(2011){Jarrett}, {Cohen}, {Masci}, {Wright},
  {Stern}, {Benford}, {Blain}, {Carey}, {Cutri}, {Eisenhardt}, {Lonsdale},
  {Mainzer}, {Marsh}, {Padgett}, {Petty}, {Ressler}, {Skrutskie}, {Stanford},
  {Surace}, {Tsai}, {Wheelock}, \& {Yan}}]{Jarrett2011}
{Jarrett}, T.~H., {Cohen}, M., {Masci}, F., {et~al.} 2011, \apj, 735, 112

\bibitem[{{Jarrett} {et~al.}(2013){Jarrett}, {Masci}, {Tsai}, {Petty},
  {Cluver}, {Assef}, {Benford}, {Blain}, {Bridge}, {Donoso}, {Eisenhardt},
  {Koribalski}, {Lake}, {Neill}, {Seibert}, {Sheth}, {Stanford}, \&
  {Wright}}]{Jarrett2013}
{Jarrett}, T.~H., {Masci}, F., {Tsai}, C.~W., {et~al.} 2013, \aj, 145, 6

\bibitem[{{Karczewski} {et~al.}(2013){Karczewski}, {Barlow}, {Page}, {Kuin},
  {Ferreras}, {Baes}, {Bendo}, {Boselli}, {Cooray}, {Cormier}, {De Looze},
  {Galametz}, {Galliano}, {Lebouteiller}, {Madden}, {Pohlen}, {R{\'e}my-Ruyer},
  {Smith}, \& {Spinoglio}}]{Karczewski2013}
{Karczewski}, O.~{\L}., {Barlow}, M.~J., {Page}, M.~J., {et~al.} 2013, \mnras,
  431, 2493

\bibitem[{{Keel} {et~al.}(2014){Keel}, {Manning}, {Holwerda}, {Lintott}, \&
  {Schawinski}}]{Keel2014}
{Keel}, W.~C., {Manning}, A.~M., {Holwerda}, B.~W., {Lintott}, C.~J., \&
  {Schawinski}, K. 2014, \aj, 147, 44

\bibitem[{{Kennicutt} \& {Evans}(2012)}]{Kennicutt2012}
{Kennicutt}, R.~C. \& {Evans}, N.~J. 2012, \araa, 50, 531

\bibitem[{{Kirk} {et~al.}(2014){Kirk}, {Gear}, {Fritz}, {Smith}, {Ford},
  {Baes}, {Bendo}, {De Looze}, {Eales}, {Gentile}, {Gomez}, {Gordon},
  {O'Halloran}, {Madden}, {Duval}, {Verstappen}, {Viaene}, {Boselli}, {Cooray},
  {Lebouteiller}, \& {Spinoglio}}]{Kirk2014}
{Kirk}, J.~M., {Gear}, W.~K., {Fritz}, J., {et~al.} 2014,
  \href{http://arxiv.org/abs/1306.2913}{arXiv:1306.2913}

\bibitem[{{Krause}(in prep.)}]{Krause}
{Krause}, O. in prep.

\bibitem[{{Leroy} {et~al.}(2008){Leroy}, {Walter}, {Brinks}, {Bigiel}, {de
  Blok}, {Madore}, \& {Thornley}}]{Leroy2008}
{Leroy}, A.~K., {Walter}, F., {Brinks}, E., {et~al.} 2008, \aj, 136, 2782

\bibitem[{{Lisenfeld} \& {Ferrara}(1998)}]{Lisenfeld1998}
{Lisenfeld}, U. \& {Ferrara}, A. 1998, \apj, 496, 145

\bibitem[{{Liu} {et~al.}(2013){Liu}, {Calzetti}, {Hong}, {Whitmore}, {Chandar},
  {O'Connell}, {Blair}, {Cohen}, {Frogel}, \& {Kim}}]{Liu2013}
{Liu}, G., {Calzetti}, D., {Hong}, S., {et~al.} 2013, \apjl, 778, L41

\bibitem[{{Lutz}(2010)}]{PACSpsf}
{Lutz}, D. 2010, {PACS photometer PSF},
  \url{http://herschel.esac.esa.int/twiki/pub/Public/PacsCalibrationWeb/bolopsfv1.01.pdf}

\bibitem[{{Magrini} {et~al.}(2011){Magrini}, {Bianchi}, {Corbelli}, {Cortese},
  {Hunt}, {Smith}, {Vlahakis}, {Davies}, {Bendo}, {Baes}, {Boselli}, {Clemens},
  {Casasola}, {de Looze}, {Fritz}, {Giovanardi}, {Grossi}, {Hughes}, {Madden},
  {Pappalardo}, {Pohlen}, {di Serego Alighieri}, \& {Verstappen}}]{Magrini2011}
{Magrini}, L., {Bianchi}, S., {Corbelli}, E., {et~al.} 2011, \aap, 535, A13

\bibitem[{{Martin} {et~al.}(2005){Martin}, {Fanson}, {Schiminovich},
  {Morrissey}, {Friedman}, {Barlow}, {Conrow}, {Grange}, {Jelinsky},
  {Milliard}, {Siegmund}, {Bianchi}, {Byun}, {Donas}, {Forster}, {Heckman},
  {Lee}, {Madore}, {Malina}, {Neff}, {Rich}, {Small}, {Surber}, {Szalay},
  {Welsh}, \& {Wyder}}]{Martin2005}
{Martin}, D.~C., {Fanson}, J., {Schiminovich}, D., {et~al.} 2005, \apjl, 619,
  L1

\bibitem[{{McConnachie} {et~al.}(2005){McConnachie}, {Irwin}, {Ferguson},
  {Ibata}, {Lewis}, \& {Tanvir}}]{McConnachie2005}
{McConnachie}, A.~W., {Irwin}, M.~J., {Ferguson}, A.~M.~N., {et~al.} 2005,
  \mnras, 356, 979

\bibitem[{{Mentuch Cooper} {et~al.}(2012){Mentuch Cooper}, {Wilson}, {Foyle},
  {Bendo}, {Koda}, {Baes}, {Boquien}, {Boselli}, {Ciesla}, {Cooray}, {Eales},
  {Galametz}, {Lebouteiller}, {Parkin}, {Roussel}, {Sauvage}, {Spinoglio}, \&
  {Smith}}]{MentuchCooper2012}
{Mentuch Cooper}, E., {Wilson}, C.~D., {Foyle}, K., {et~al.} 2012, \apj, 755,
  165

\bibitem[{{Mo} {et~al.}(1998){Mo}, {Mao}, \& {White}}]{Mo1998}
{Mo}, H.~J., {Mao}, S., \& {White}, S.~D.~M. 1998, \mnras, 295, 319

\bibitem[{{Montalto} {et~al.}(2009){Montalto}, {Seitz}, {Riffeser}, {Hopp},
  {Lee}, \& {Sch{\"o}nrich}}]{Montaldo2009}
{Montalto}, M., {Seitz}, S., {Riffeser}, A., {et~al.} 2009, \aap, 507, 283

\bibitem[{{Moorthy} \& {Holtzman}(2006)}]{Moorthy2006}
{Moorthy}, B.~K. \& {Holtzman}, J.~A. 2006, \mnras, 371, 583

\bibitem[{{Morrissey} {et~al.}(2007){Morrissey}, {Conrow}, {Barlow}, {Small},
  {Seibert}, {Wyder}, {Budav{\'a}ri}, {Arnouts}, {Friedman}, {Forster},
  {Martin}, {Neff}, {Schiminovich}, {Bianchi}, {Donas}, {Heckman}, {Lee},
  {Madore}, {Milliard}, {Rich}, {Szalay}, {Welsh}, \& {Yi}}]{Morrissey2007}
{Morrissey}, P., {Conrow}, T., {Barlow}, T.~A., {et~al.} 2007, \apjs, 173, 682

\bibitem[{{Mu{\~n}oz-Mateos} {et~al.}(2009{\natexlab{a}}){Mu{\~n}oz-Mateos},
  {Gil de Paz}, {Boissier}, {Zamorano}, {Dale}, {P{\'e}rez-Gonz{\'a}lez},
  {Gallego}, {Madore}, {Bendo}, {Thornley}, {Draine}, {Boselli}, {Buat},
  {Calzetti}, {Moustakas}, \& {Kennicutt}}]{MunozMateos2009b}
{Mu{\~n}oz-Mateos}, J.~C., {Gil de Paz}, A., {Boissier}, S., {et~al.}
  2009{\natexlab{a}}, \apj, 701, 1965

\bibitem[{{Mu{\~n}oz-Mateos} {et~al.}(2007){Mu{\~n}oz-Mateos}, {Gil de Paz},
  {Boissier}, {Zamorano}, {Jarrett}, {Gallego}, \& {Madore}}]{MunozMateos2007}
{Mu{\~n}oz-Mateos}, J.~C., {Gil de Paz}, A., {Boissier}, S., {et~al.} 2007,
  \apj, 658, 1006

\bibitem[{{Mu{\~n}oz-Mateos} {et~al.}(2009{\natexlab{b}}){Mu{\~n}oz-Mateos},
  {Gil de Paz}, {Zamorano}, {Boissier}, {Dale}, {P{\'e}rez-Gonz{\'a}lez},
  {Gallego}, {Madore}, {Bendo}, {Boselli}, {Buat}, {Calzetti}, {Moustakas}, \&
  {Kennicutt}}]{MunozMateos2009}
{Mu{\~n}oz-Mateos}, J.~C., {Gil de Paz}, A., {Zamorano}, J., {et~al.}
  2009{\natexlab{b}}, \apj, 703, 1569

\bibitem[{{Noll} {et~al.}(2009){Noll}, {Burgarella}, {Giovannoli}, {Buat},
  {Marcillac}, \& {Mu{\~n}oz-Mateos}}]{Noll2009}
{Noll}, S., {Burgarella}, D., {Giovannoli}, E., {et~al.} 2009, \aap, 507, 1793

\bibitem[{{Padmanabhan} {et~al.}(2008){Padmanabhan}, {Schlegel}, {Finkbeiner},
  {Barentine}, {Blanton}, {Brewington}, {Gunn}, {Harvanek}, {Hogg},
  {Ivezi{\'c}}, {Johnston}, {Kent}, {Kleinman}, {Knapp}, {Krzesinski}, {Long},
  {Neilsen}, {Nitta}, {Loomis}, {Lupton}, {Roweis}, {Snedden}, {Strauss}, \&
  {Tucker}}]{Padmanabhan2008}
{Padmanabhan}, N., {Schlegel}, D.~J., {Finkbeiner}, D.~P., {et~al.} 2008, \apj,
  674, 1217

\bibitem[{{Parkin} {et~al.}(2012){Parkin}, {Wilson}, {Foyle}, {Baes}, {Bendo},
  {Boselli}, {Boquien}, {Cooray}, {Cormier}, {Davies}, {Eales}, {Galametz},
  {Gomez}, {Lebouteiller}, {Madden}, {Mentuch}, {Page}, {Pohlen}, {Remy},
  {Roussel}, {Sauvage}, {Smith}, \& {Spinoglio}}]{Parkin2012}
{Parkin}, T.~J., {Wilson}, C.~D., {Foyle}, K., {et~al.} 2012, \mnras, 422, 2291

\bibitem[{{P{\'e}rez} {et~al.}(2013){P{\'e}rez}, {Cid Fernandes}, {Gonz{\'a}lez
  Delgado}, {Garc{\'{\i}}a-Benito}, {S{\'a}nchez}, {Husemann}, {Mast},
  {Rod{\'o}n}, {Kupko}, {Backsmann}, {de Amorim}, {van de Ven}, {Walcher},
  {Wisotzki}, {Cortijo-Ferrero}, \& {Collaboration6}}]{Perez2013}
{P{\'e}rez}, E., {Cid Fernandes}, R., {Gonz{\'a}lez Delgado}, R.~M., {et~al.}
  2013, \apjl, 764, L1

\bibitem[{{Petty} {et~al.}(2013){Petty}, {Neill}, {Jarrett}, {Blain}, {Farrah},
  {Rich}, {Tsai}, {Benford}, {Bridge}, {Lake}, {Masci}, \&
  {Wright}}]{Petty2013}
{Petty}, S.~M., {Neill}, J.~D., {Jarrett}, T.~H., {et~al.} 2013, \aj, 146, 77

\bibitem[{{Pilbratt} {et~al.}(2010){Pilbratt}, {Riedinger}, {Passvogel},
  {Crone}, {Doyle}, {Gageur}, {Heras}, {Jewell}, {Metcalfe}, {Ott}, \&
  {Schmidt}}]{Herschel}
{Pilbratt}, G.~L., {Riedinger}, J.~R., {Passvogel}, T., {et~al.} 2010, \aap,
  518, L1

\bibitem[{{Poglitsch} {et~al.}(2010){Poglitsch}, {Waelkens}, {Geis},
  {Feuchtgruber}, {Vandenbussche}, {Rodriguez}, {Krause}, {Renotte}, {van
  Hoof}, {Saraceno}, {Cepa}, {Kerschbaum}, {Agn{\`e}se}, {Ali}, {Altieri},
  {Andreani}, {Augueres}, {Balog}, {Barl}, {Bauer}, {Belbachir}, {Benedettini},
  {Billot}, {Boulade}, {Bischof}, {Blommaert}, {Callut}, {Cara}, {Cerulli},
  {Cesarsky}, {Contursi}, {Creten}, {De Meester}, {Doublier}, {Doumayrou},
  {Duband}, {Exter}, {Genzel}, {Gillis}, {Gr{\"o}zinger}, {Henning},
  {Herreros}, {Huygen}, {Inguscio}, {Jakob}, {Jamar}, {Jean}, {de Jong},
  {Katterloher}, {Kiss}, {Klaas}, {Lemke}, {Lutz}, {Madden}, {Marquet},
  {Martignac}, {Mazy}, {Merken}, {Montfort}, {Morbidelli}, {M{\"u}ller},
  {Nielbock}, {Okumura}, {Orfei}, {Ottensamer}, {Pezzuto}, {Popesso},
  {Putzeys}, {Regibo}, {Reveret}, {Royer}, {Sauvage}, {Schreiber}, {Stegmaier},
  {Schmitt}, {Schubert}, {Sturm}, {Thiel}, {Tofani}, {Vavrek}, {Wetzstein},
  {Wieprecht}, \& {Wiezorrek}}]{PACS}
{Poglitsch}, A., {Waelkens}, C., {Geis}, N., {et~al.} 2010, \aap, 518, L2

\bibitem[{{Popescu} \& {Tuffs}(2002)}]{PopescuTuffs2002}
{Popescu}, C.~C. \& {Tuffs}, R.~J. 2002, \mnras, 335, L41

\bibitem[{{Popescu} {et~al.}(2011){Popescu}, {Tuffs}, {Dopita}, {Fischera},
  {Kylafis}, \& {Madore}}]{Popescu2011}
{Popescu}, C.~C., {Tuffs}, R.~J., {Dopita}, M.~A., {et~al.} 2011, \aap, 527,
  A109

\bibitem[{{Popescu} {et~al.}(2002){Popescu}, {Tuffs}, {V{\"o}lk}, {Pierini}, \&
  {Madore}}]{Popescu2002}
{Popescu}, C.~C., {Tuffs}, R.~J., {V{\"o}lk}, H.~J., {Pierini}, D., \&
  {Madore}, B.~F. 2002, \apj, 567, 221

\bibitem[{{Richstone} \& {Sargent}(1972)}]{Richstone1972}
{Richstone}, D. \& {Sargent}, W.~L.~W. 1972, \apj, 176, 91

\bibitem[{{Rieke} {et~al.}(2004){Rieke}, {Young}, {Engelbracht}, {Kelly},
  {Low}, {Haller}, {Beeman}, {Gordon}, {Stansberry}, {Misselt}, {Cadien},
  {Morrison}, {Rivlis}, {Latter}, {Noriega-Crespo}, {Padgett}, {Stapelfeldt},
  {Hines}, {Egami}, {Muzerolle}, {Alonso-Herrero}, {Blaylock}, {Dole}, {Hinz},
  {Le Floc'h}, {Papovich}, {P{\'e}rez-Gonz{\'a}lez}, {Smith}, {Su}, {Bennett},
  {Frayer}, {Henderson}, {Lu}, {Masci}, {Pesenson}, {Rebull}, {Rho}, {Keene},
  {Stolovy}, {Wachter}, {Wheaton}, {Werner}, \& {Richards}}]{MIPS}
{Rieke}, G.~H., {Young}, E.~T., {Engelbracht}, C.~W., {et~al.} 2004, \apjs,
  154, 25

\bibitem[{{Rowlands} {et~al.}(2012){Rowlands}, {Dunne}, {Maddox}, {Bourne},
  {Gomez}, {Kaviraj}, {Bamford}, {Brough}, {Charlot}, {da Cunha}, {Driver},
  {Eales}, {Hopkins}, {Kelvin}, {Nichol}, {Sansom}, {Sharp}, {Smith}, {Temi},
  {van der Werf}, {Baes}, {Cava}, {Cooray}, {Croom}, {Dariush}, {de Zotti},
  {Dye}, {Fritz}, {Hopwood}, {Ibar}, {Ivison}, {Liske}, {Loveday}, {Madore},
  {Norberg}, {Popescu}, {Rigby}, {Robotham}, {Rodighiero}, {Seibert}, \&
  {Tuffs}}]{Rowlands2012}
{Rowlands}, K., {Dunne}, L., {Maddox}, S., {et~al.} 2012, \mnras, 419, 2545

\bibitem[{{Salim} {et~al.}(2005){Salim}, {Charlot}, {Rich}, {Kauffmann},
  {Heckman}, {Barlow}, {Bianchi}, {Byun}, {Donas}, {Forster}, {Friedman},
  {Jelinsky}, {Lee}, {Madore}, {Malina}, {Martin}, {Milliard}, {Morrissey},
  {Neff}, {Schiminovich}, {Seibert}, {Siegmund}, {Small}, {Szalay}, {Welsh}, \&
  {Wyder}}]{Salim2005}
{Salim}, S., {Charlot}, S., {Rich}, R.~M., {et~al.} 2005, \apjl, 619, L39

\bibitem[{{Sandstrom} {et~al.}(2012){Sandstrom}, {Leroy}, {Walter}, {Bolatto},
  {Croxall}, {Draine}, {Wilson}, {Wolfire}, {Calzetti}, {Kennicutt}, {Aniano},
  {Donovan Meyer}, {Usero}, {Bigiel}, {Brinks}, {de Blok}, {Crocker}, {Dale},
  {Engelbracht}, {Galametz}, {Groves}, {Hunt}, {Koda}, {Kreckel}, {Linz},
  {Meidt}, {Pellegrini}, {Rix}, {Roussel}, {Schinnerer}, {Schruba}, {Schuster},
  {Skibba}, {van der Laan}, {Appleton}, {Armus}, {Brandl}, {Gordon}, {Hinz},
  {Krause}, {Montiel}, {Sauvage}, {Schmiedeke}, {Smith}, \&
  {Vigroux}}]{Sandstrom2012}
{Sandstrom}, K.~M., {Leroy}, A.~K., {Walter}, F., {et~al.} 2012,
  \href{http://arxiv.org/abs/1212.1208}{arXiv:1212.1208}

\bibitem[{{Schiminovich} {et~al.}(2007){Schiminovich}, {Wyder}, {Martin},
  {Johnson}, {Salim}, {Seibert}, {Treyer}, {Budav{\'a}ri}, {Hoopes},
  {Zamojski}, {Barlow}, {Forster}, {Friedman}, {Morrissey}, {Neff}, {Small},
  {Bianchi}, {Donas}, {Heckman}, {Lee}, {Madore}, {Milliard}, {Rich}, {Szalay},
  {Welsh}, \& {Yi}}]{Schiminovich2007}
{Schiminovich}, D., {Wyder}, T.~K., {Martin}, D.~C., {et~al.} 2007, \apjs, 173,
  315

\bibitem[{{Sick} {et~al.}(2013){Sick}, {Courteau}, {Cuillandre}, {McDonald},
  {de Jong}, \& {Tully}}]{Sick2013}
{Sick}, J., {Courteau}, S., {Cuillandre}, J.-C., {et~al.} 2013,
  \href{http://arxiv.org/abs/1303.6290}{arXiv:1303.6290}

\bibitem[{{Skrutskie} {et~al.}(2006){Skrutskie}, {Cutri}, {Stiening},
  {Weinberg}, {Schneider}, {Carpenter}, {Beichman}, {Capps}, {Chester},
  {Elias}, {Huchra}, {Liebert}, {Lonsdale}, {Monet}, {Price}, {Seitzer},
  {Jarrett}, {Kirkpatrick}, {Gizis}, {Howard}, {Evans}, {Fowler}, {Fullmer},
  {Hurt}, {Light}, {Kopan}, {Marsh}, {McCallon}, {Tam}, {Van Dyk}, \&
  {Wheelock}}]{2MASS}
{Skrutskie}, M.~F., {Cutri}, R.~M., {Stiening}, R., {et~al.} 2006, \aj, 131,
  1163

\bibitem[{{Smith} {et~al.}(2012{\natexlab{a}}){Smith}, {Dunne}, {da Cunha},
  {Rowlands}, {Maddox}, {Gomez}, {Bonfield}, {Charlot}, {Driver}, {Popescu},
  {Tuffs}, {Dunlop}, {Jarvis}, {Seymour}, {Symeonidis}, {Baes}, {Bourne},
  {Clements}, {Cooray}, {De Zotti}, {Dye}, {Eales}, {Scott}, {Verma}, {van der
  Werf}, {Andrae}, {Auld}, {Buttiglione}, {Cava}, {Dariush}, {Fritz},
  {Hopwood}, {Ibar}, {Ivison}, {Kelvin}, {Madore}, {Pohlen}, {Rigby},
  {Robotham}, {Seibert}, \& {Temi}}]{SmithD2012}
{Smith}, D.~J.~B., {Dunne}, L., {da Cunha}, E., {et~al.} 2012{\natexlab{a}},
  \mnras, 427, 703

\bibitem[{{Smith} {et~al.}(2012{\natexlab{b}}){Smith}, {Eales}, {Gomez},
  {Roman-Duval}, {Fritz}, {Braun}, {Baes}, {Bendo}, {Blommaert}, {Boquien},
  {Boselli}, {Clements}, {Cooray}, {Cortese}, {de Looze}, {Ford}, {Gear},
  {Gentile}, {Gordon}, {Kirk}, {Lebouteiller}, {Madden}, {Mentuch},
  {O'Halloran}, {Page}, {Schulz}, {Spinoglio}, {Verstappen}, {Wilson}, \&
  {Thilker}}]{HelgaII}
{Smith}, M.~W.~L., {Eales}, S.~A., {Gomez}, H.~L., {et~al.} 2012{\natexlab{b}},
  \apj, 756, 40

\bibitem[{{Smith} {et~al.}(2010){Smith}, {Vlahakis}, {Baes}, {Bendo},
  {Bianchi}, {Bomans}, {Boselli}, {Clemens}, {Corbelli}, {Cortese}, {Dariush},
  {Davies}, {de Looze}, {di Serego Alighieri}, {Fadda}, {Fritz},
  {Garcia-Appadoo}, {Gavazzi}, {Giovanardi}, {Grossi}, {Hughes}, {Hunt},
  {Jones}, {Madden}, {Pierini}, {Pohlen}, {Sabatini}, {Verstappen}, {Xilouris},
  \& {Zibetti}}]{Smith2010}
{Smith}, M.~W.~L., {Vlahakis}, C., {Baes}, M., {et~al.} 2010, \aap, 518, L51

\bibitem[{{Sodroski} {et~al.}(1997){Sodroski}, {Odegard}, {Arendt}, {Dwek},
  {Weiland}, {Hauser}, \& {Kelsall}}]{sodroski1997}
{Sodroski}, T.~J., {Odegard}, N., {Arendt}, R.~G., {et~al.} 1997, \apj, 480,
  173

\bibitem[{{Tabatabaei} \& {Berkhuijsen}(2010)}]{Tabatabaei2010}
{Tabatabaei}, F.~S. \& {Berkhuijsen}, E.~M. 2010, \aap, 517, A77

\bibitem[{{Tabatabaei} {et~al.}(2014){Tabatabaei}, {Braine}, {Xilouris},
  {Kramer}, {Boquien}, {Combes}, {Henkel}, {Relano}, {Verley}, {Gratier},
  {Israel}, {Wiedner}, {R{\"o}llig}, {Schuster}, \& {van der
  Werf}}]{Tabatabaei2014}
{Tabatabaei}, F.~S., {Braine}, J., {Xilouris}, E.~M., {et~al.} 2014, \aap, 561,
  A95

\bibitem[{{Tamm} {et~al.}(2012){Tamm}, {Tempel}, {Tenjes}, {Tihhonova}, \&
  {Tuvikene}}]{Tamm2012}
{Tamm}, A., {Tempel}, E., {Tenjes}, P., {Tihhonova}, O., \& {Tuvikene}, T.
  2012, \aap, 546, A4

\bibitem[{{Tempel} {et~al.}(2010){Tempel}, {Tamm}, \& {Tenjes}}]{Tempel2010}
{Tempel}, E., {Tamm}, A., \& {Tenjes}, P. 2010, \aap, 509, A91

\bibitem[{{Tempel} {et~al.}(2011){Tempel}, {Tuvikene}, {Tamm}, \&
  {Tenjes}}]{tempel2012}
{Tempel}, E., {Tuvikene}, T., {Tamm}, A., \& {Tenjes}, P. 2011, \aap, 526, A155

\bibitem[{{Thilker} {et~al.}(2005){Thilker}, {Hoopes}, {Bianchi}, {Boissier},
  {Rich}, {Seibert}, {Friedman}, {Rey}, {Buat}, {Barlow}, {Byun}, {Donas},
  {Forster}, {Heckman}, {Jelinsky}, {Lee}, {Madore}, {Malina}, {Martin},
  {Milliard}, {Morrissey}, {Neff}, {Schiminovich}, {Siegmund}, {Small},
  {Szalay}, {Welsh}, \& {Wyder}}]{Thilker2005}
{Thilker}, D.~A., {Hoopes}, C.~G., {Bianchi}, L., {et~al.} 2005, \apjl, 619,
  L67

\bibitem[{{Tuffs} {et~al.}(2004){Tuffs}, {Popescu}, {V{\"o}lk}, {Kylafis}, \&
  {Dopita}}]{Tuffs2004}
{Tuffs}, R.~J., {Popescu}, C.~C., {V{\"o}lk}, H.~J., {Kylafis}, N.~D., \&
  {Dopita}, M.~A. 2004, \aap, 419, 821

\bibitem[{{Walterbos} \& {Schwering}(1987)}]{WalterbosSchwering1987}
{Walterbos}, R.~A.~M. \& {Schwering}, P.~B.~W. 1987, \aap, 180, 27

\bibitem[{{Werner} {et~al.}(2004){Werner}, {Roellig}, {Low}, {Rieke}, {Rieke},
  {Hoffmann}, {Young}, {Houck}, {Brandl}, {Fazio}, {Hora}, {Gehrz}, {Helou},
  {Soifer}, {Stauffer}, {Keene}, {Eisenhardt}, {Gallagher}, {Gautier}, {Irace},
  {Lawrence}, {Simmons}, {Van Cleve}, {Jura}, {Wright}, \&
  {Cruikshank}}]{Spitzer}
{Werner}, M.~W., {Roellig}, T.~L., {Low}, F.~J., {et~al.} 2004, \apjs, 154, 1

\bibitem[{{White} \& {Frenk}(1991)}]{White1991}
{White}, S.~D.~M. \& {Frenk}, C.~S. 1991, \apj, 379, 52

\bibitem[{{Wright} {et~al.}(2010){Wright}, {Eisenhardt}, {Mainzer}, {Ressler},
  {Cutri}, {Jarrett}, {Kirkpatrick}, {Padgett}, {McMillan}, {Skrutskie},
  {Stanford}, {Cohen}, {Walker}, {Mather}, {Leisawitz}, {Gautier}, {McLean},
  {Benford}, {Lonsdale}, {Blain}, {Mendez}, {Irace}, {Duval}, {Liu}, {Royer},
  {Heinrichsen}, {Howard}, {Shannon}, {Kendall}, {Walsh}, {Larsen}, {Cardon},
  {Schick}, {Schwalm}, {Abid}, {Fabinsky}, {Naes}, \& {Tsai}}]{WISE}
{Wright}, E.~L., {Eisenhardt}, P.~R.~M., {Mainzer}, A.~K., {et~al.} 2010, \aj,
  140, 1868

\bibitem[{{York} {et~al.}(2000){York}, {Adelman}, {Anderson}, {Anderson},
  {Annis}, {Bahcall}, {Bakken}, {Barkhouser}, {Bastian}, {Berman}, {Boroski},
  {Bracker}, {Briegel}, {Briggs}, {Brinkmann}, {Brunner}, {Burles}, {Carey},
  {Carr}, {Castander}, {Chen}, {Colestock}, {Connolly}, {Crocker}, {Csabai},
  {Czarapata}, {Davis}, {Doi}, {Dombeck}, {Eisenstein}, {Ellman}, {Elms},
  {Evans}, {Fan}, {Federwitz}, {Fiscelli}, {Friedman}, {Frieman}, {Fukugita},
  {Gillespie}, {Gunn}, {Gurbani}, {de Haas}, {Haldeman}, {Harris}, {Hayes},
  {Heckman}, {Hennessy}, {Hindsley}, {Holm}, {Holmgren}, {Huang}, {Hull},
  {Husby}, {Ichikawa}, {Ichikawa}, {Ivezi{\'c}}, {Kent}, {Kim}, {Kinney},
  {Klaene}, {Kleinman}, {Kleinman}, {Knapp}, {Korienek}, {Kron}, {Kunszt},
  {Lamb}, {Lee}, {Leger}, {Limmongkol}, {Lindenmeyer}, {Long}, {Loomis},
  {Loveday}, {Lucinio}, {Lupton}, {MacKinnon}, {Mannery}, {Mantsch}, {Margon},
  {McGehee}, {McKay}, {Meiksin}, {Merelli}, {Monet}, {Munn}, {Narayanan},
  {Nash}, {Neilsen}, {Neswold}, {Newberg}, {Nichol}, {Nicinski}, {Nonino},
  {Okada}, {Okamura}, {Ostriker}, {Owen}, {Pauls}, {Peoples}, {Peterson},
  {Petravick}, {Pier}, {Pope}, {Pordes}, {Prosapio}, {Rechenmacher}, {Quinn},
  {Richards}, {Richmond}, {Rivetta}, {Rockosi}, {Ruthmansdorfer}, {Sandford},
  {Schlegel}, {Schneider}, {Sekiguchi}, {Sergey}, {Shimasaku}, {Siegmund},
  {Smee}, {Smith}, {Snedden}, {Stone}, {Stoughton}, {Strauss}, {Stubbs},
  {SubbaRao}, {Szalay}, {Szapudi}, {Szokoly}, {Thakar}, {Tremonti}, {Tucker},
  {Uomoto}, {Vanden Berk}, {Vogeley}, {Waddell}, {Wang}, {Watanabe},
  {Weinberg}, {Yanny}, {Yasuda}, \& {SDSS Collaboration}}]{SDSS}
{York}, D.~G., {Adelman}, J., {Anderson}, Jr., J.~E., {et~al.} 2000, \aj, 120,
  1579

\end{thebibliography}

\newpage

\appendix

\section{Multi-wavelength data processing} \label{app:dataprocessing}

In this section we present the road towards spatially resolved, panchromatic SED fitting. The data obtained from various sources and own \textit{Herschel} observations are manipulated in order to make a consistent comparison over such a wide wavelength range (Appendix \ref{app:processing}). These manipulations bring with them a complex uncertainty propagation which is addressed in Appendix \ref{app:errors}. 

	\subsection{Image manipulations} \label{app:processing}

		\subsubsection{Background subtraction}		

The WISE and GALEX subsets had a non-zero average background value due to emission from unresolved sources. The \textit{Herschel} images also come with a flat, non-zero background as a consequence of the data reduction process. The average background for the \textit{Spitzer} and SDSS images was already zero, hence no background subtraction was performed for these frames.

While global background gradients were significant in the WISE frames, no clear gradients were identified in the GALEX or \textit{Herschel} images. We therefore fit and subtract a second--order polynomial to the background in the WISE frames using standard \texttt{ESO-MIDAS} routines. In the other frames, we estimate the background as follows.
 
A set of regions was chosen far enough from visible emission from M31 to avoid contamination by the galaxy, but close enough to make a reliable estimation of the background near the M31. Inside these pre-defined regions, a number of aperture measurements was made. The number of measurements per region was set to be proportional to the number of pixels inside. For PACS and SPIRE, a total of $10000$ measurements were spread over $8$ regions. In the GALEX fields, $20000$ measurements were divided over $23$ regions. From the set of measured fluxes, a sigma clipped median was derived as a reliable estimate for the background flux. We used $3\sigma$ as a threshold and iterated until convergence. This median background value was consequently subtracted from the images.

		\subsubsection{Masking} \label{sec:masking}

Andromeda covers a large part of the sky for a single galaxy and lies close to the Galactic disk (with a Galactic latitude of $-21.6 \degree$). It is consequently contaminated by the light of thousands of foreground stars, especially in the UV and optical part of the spectrum. At longer wavelengths, the infrared emission of background galaxies becomes the main source of contaminating sources. At \textit{Herschel} wavelengths, however, most of the emission from non-M31 point sources is negligible even at scales of the SPIRE $500~\mu\mathrm{m}$ beam. As mentioned before, the extended emission of the Milky Way Galactic Cirrus is prominently visible here. This dust emission can fortunately be associated with HI emission. Using the velocity information of HI maps, the Galactic cirrus can partly be disentangled from the emission of M31. Paper I goes into more detail about this technique. 

We made use of SExtractor v$2.8.6$ \citep{sextractor} to list the location of all point sources above a certain threshold ($5$ times the background noise level). The program simultaneously produces background maps that can be tweaked to represent the diffuse emission from M31. In this way, we could replace the non-M31 point sources with the local M31 background value obtained from these maps. 

For each source an optimal radius was derived by comparing the pixel flux with the local background at increasing distance from the peak location. Once the pixel-to-background flux ratio dropped below $2$, the radius was cut off at that distance. Based on this radius, a total flux was extracted in order to make colour evaluations. We constructed point source masks for the GALEX, SDSS, WISE, and \textit{Spitzer} subsets based on different colour criteria.

The GALEX and SDSS point sources were evaluated based on their UV colour. This technique was applied by \citet{GilDePaz2007} for over $1000$ galaxies and proved successful. In practice, we mask all sources with
\begin{equation}
\mid$FUV-NUV$\mid > 0.75
\end{equation}
if they are detected at the $1\sigma$ level in their particular wavelength band. 
SExtractor identified $58330$ point sources in the UV fields, of which over $51000$ were masked in the FUV and NUV. Many point sources from the UV catalogue were not detected at optical bands, hence only $25000$ sources were masked in the SDSS bands. Around $7000$ sources were identified as extragalactic. They were therefore assumed to belong to M31 and were not masked.

\begin{figure}
	\resizebox{\hsize}{!}{\includegraphics{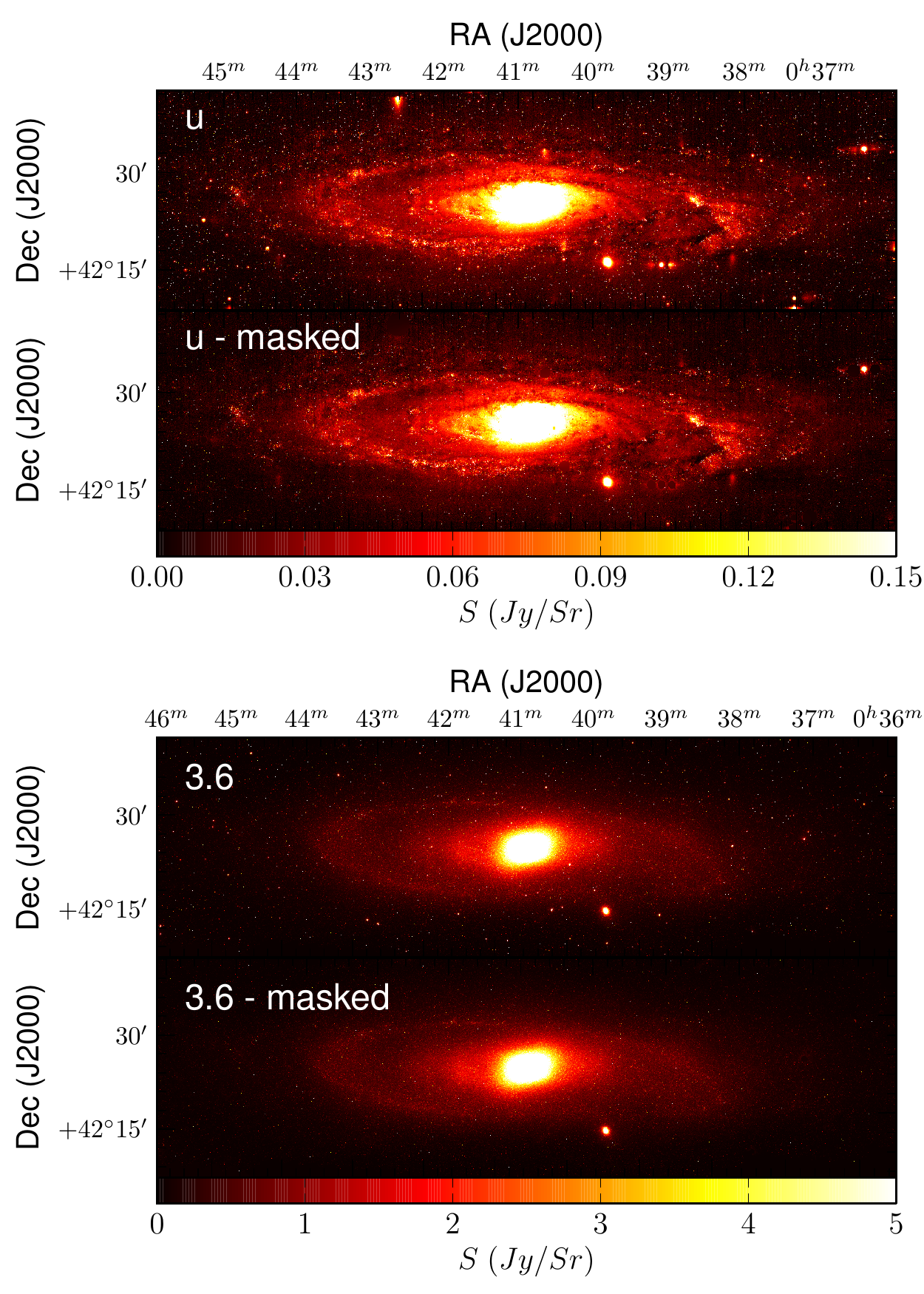}}
	\caption{Masking of point sources that do not belong to M31: before and after view of the galaxy in the $u$ band (top) and IRAC $3.6~\mu\mathrm{m}$ band (bottom).}
	\label{fig:masking}
\end{figure}

As an example, Fig.~\ref{fig:masking} shows the $u$--band image of M31 before and after the mask was applied. The contamination of the image has been significantly reduced using the above technique.

The point sources in the WISE and the \textit{Spitzer} IRAC and MIPS frames were masked analogously, based on their IRAC colours (see below). At these wavelengths, however, the non-M31 point sources are a mix of foreground stars and background galaxies. Furthermore, some bright sources may be associated with H{\sc{ii}} regions in M31 and must not be masked. We designed a scheme based on the technique by \citet{MunozMateos2009}, which was successfully applied to the SINGS galaxies. Foreground stars have almost no PAH emission, while the diffuse ISM in galaxies shows a roughly constant $F_{5.8}/F_{8}$ ratio \citep{DraineLi2007}. Background galaxies are redshifted spirals or ellipticals and can consequently have a wide range in $F_{5.8}/F_{8}$. It is thus possible to construct a rough filter relying on the difference in MIR flux ratios. First, it was checked which point source extracted from the IRAC $3.6~\mu\mathrm{m}$ had a non-detection at $8~\mu\mathrm{m}$. This criterion proved to be sufficient to select the foreground stars in the field. A second, colour-based, criterion disentangled the background galaxies from the H{\sc{ii}} regions:
\begin{equation}
0.29 < F_{5.8}/F_{8} < 0.85
\end{equation}
\begin{equation}
F_{3.6}/F_{5.8} < 1.58.
\end{equation}
\begin{figure}
	\resizebox{\hsize}{!}{\includegraphics{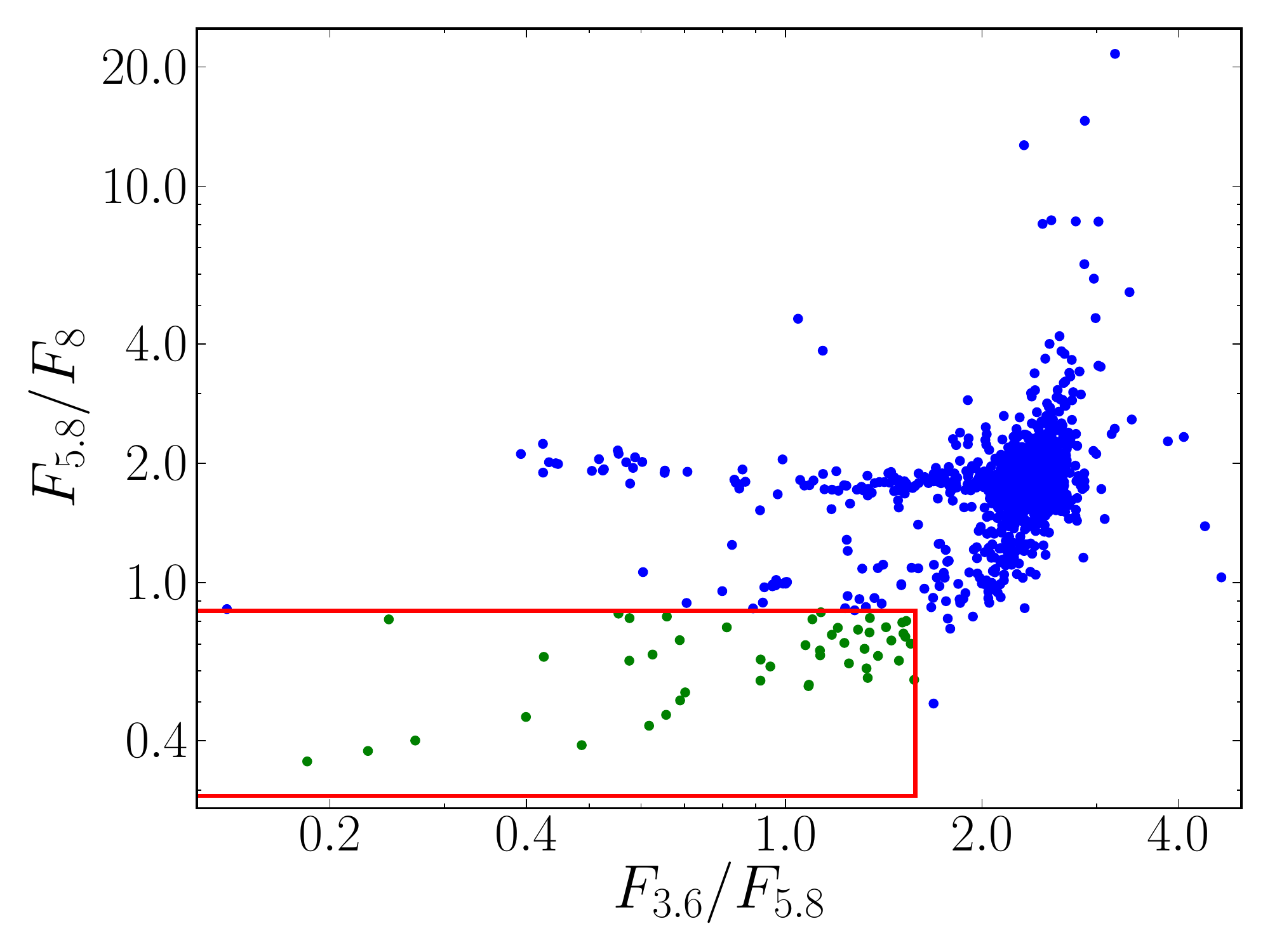}}
	\caption{Colour-colour plot of the IRAC selected bright sources. The sources inside the red rectangle are identified as H{\sc{ii}} regions belonging to M31. They were consequently not masked.}
	\label{fig:IRACcolors}
\end{figure}
Figure~\ref{fig:IRACcolors} shows the colour--colour diagram for these sources. The H{\sc{ii}} regions follow a more or less horizontal track at the lower--left part of the plot. The colour criteria for filtering out these H{\sc{ii}} regions were obtained empirically to ensure effective identification. Once identified, these star forming regions were consequently not masked.
The resulting mask was applied to all IRAC and MIPS bands. Sources that were not detected at longer wavelengths were obviously not masked. Figure~\ref{fig:masking} shows the IRAC $3.6~\mu\mathrm{m}$ image of M31 before and after the mask was applied.
SExtractor identified $1933$ sources in the IRAC bands and $536$ in MIPS. From these catalogues, around $1800$ sources were masked in IRAC and $230$ in MIPS. All masked sources in IRAC were also masked in the WISE frames as both instruments cover roughly the same wavelength range. No additional masking was necessary for the WISE data.

As a way to check the reliability of our masks, we compared the masked regions with the locations of known Andromeda sources, i.e. H{\sc{ii}} regions and planetary Nebula \citep{Azimlu2011} and bright young clusters \citep{Barmby2009}. The overlap between our masked sources and actual M31 sources proved negligible; $0.3 \%$, $0.1\%$, $0.4 \%$, $0.4 \%$, and $1.5 \%$ of the identified sources were incorrectly masked in the GALEX, SDSS, WISE, IRAC, and MIPS bands, respectively. The handful of sources that were incorrectly masked were manually restored.

		\subsubsection{Convolution and rescaling}

The masked images were all brought to the same resolution before extracting the separate pixel values. By doing this, information is lost because of the significantly lower resolutions of the end products. It is, however, a critical step to make a consistent comparison of the fluxes over this wide range of wavelengths. Our working resolution was limited by the SPIRE $500~\mu\mathrm{m}$ point spread function, which is $36.0$ arcseconds. As M31 is the nearest spiral galaxy, this still corresponds to an unprecedented physical scale of $136.8$ pc. \citet{Aniano2011} conducted an in-depth study on convolution kernels for most known telescope PSFs. Additionally they wrote an efficient IDL routine \texttt{convolve\_image.pro} which makes use of their designed kernels and takes NaN values into account when convolving. The kernels for the GALEX, WISE, IRAC, MIPS, PACS, and SPIRE instruments were readily available to make the convolution. For the SDSS images, we used a Gaussian-to-SPIRE 500 kernel which assumes an initial FWHM of $4^{\prime\prime}$ for the SDSS images (see Sect.~\ref{sec:SDSS}). 

\begin{figure*}
	\centering
   	\includegraphics[width=17cm]{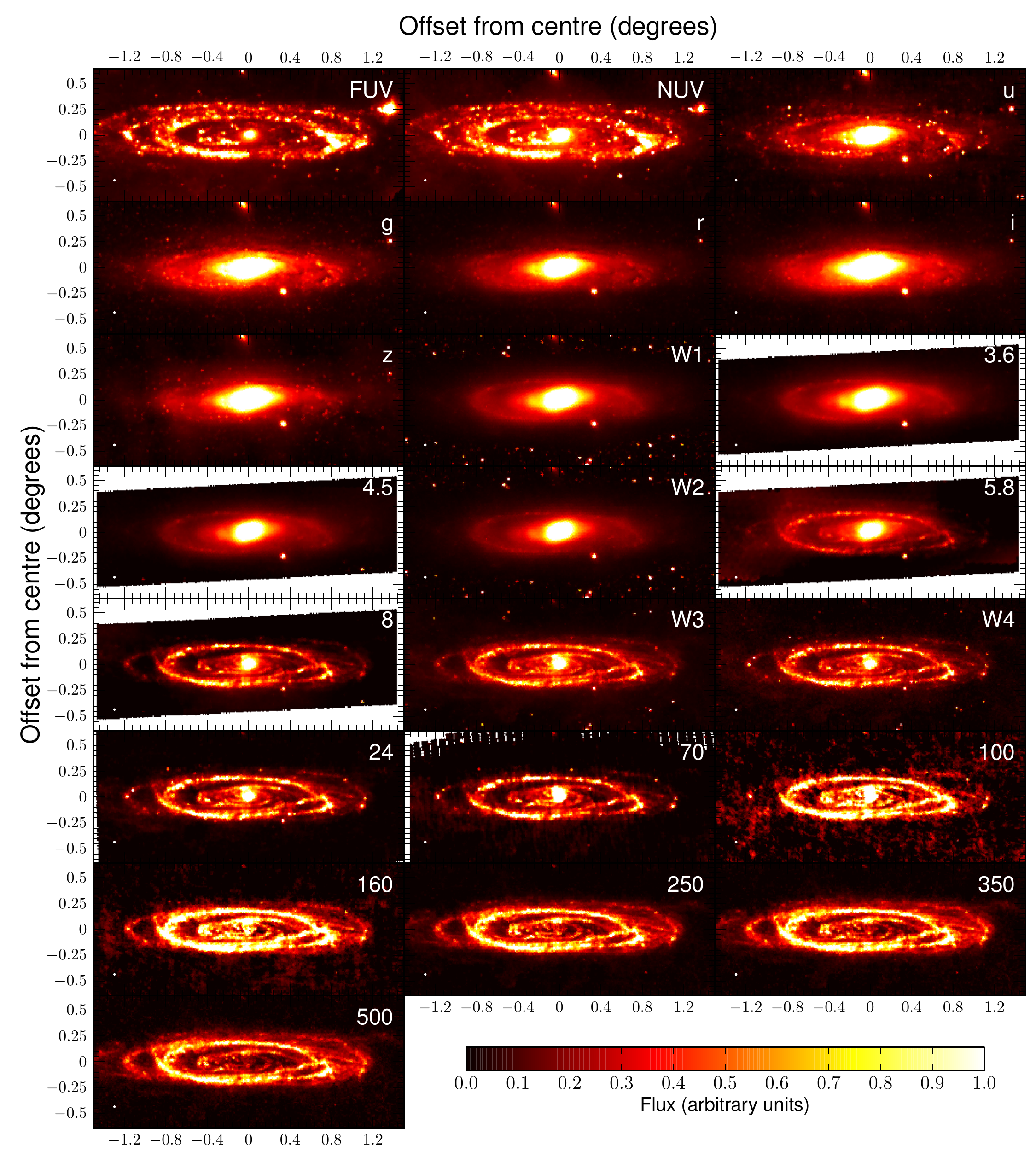}
	\caption{Overview of the FUV to submm dataset at SPIRE 500 resolution, rotated from a position angle of $38\degree$.}
	\label{fig:fitdata}
\end{figure*}

The SPIRE $500~\mu\mathrm{m}$ beam is sampled with a pixel scale of $12^{\prime\prime}$ which means the PSF covers nine pixels which are not independent. The frame thus had to be rebinned to a pixel scale of $36^{\prime\prime}$ to make each pixel correspond to a statistically independent region in M31. The convolved frames were consequently rescaled to match the pixel grid of the SPIRE $500~\mu\mathrm{m}$ rebinned image. Our data cube covers the electromagnetic spectrum from UV to submm wavelengths. Figure~\ref{fig:fitdata} gives an overview of all frames used for the fitting of a panchromatic SED to each pixel.
 
This series of steps results in sets of corresponding pixels which each represent a physical region of $136.8 \times 608.1$ pc along the major and minor axes (using an inclination of $i=77\degree$). Off course, it must be noted that the third dimension, the direction along the line of sight, also contributes to the appearance of each pixel. Spiral galaxies are, however, known to have relatively thin disks compared to their lengths, so even along this axis, the resolution remains subgalactic. The attenuation effects of this larger dimension will, however, be treated during the modelling in terms of optical depth parameters (see Sect.~\ref{subsec:magphys}).
				
\subsection{Single--pixel uncertainties} \label{app:errors}
	
Each pixel comes with several sources of uncertainty, which will have to be estimated and combined to a total error. Uncertainty propagation based on initial errors can become complex and hazardous after masking, convolutions, rebinning, and rescaling. In their Appendix D.1, \citet{Aniano2012} opted to start the uncertainty estimation after all these steps in their resolved analysis for NGC 628 and NGC 6946. They postulate two sources of uncertainties: background variations and calibration errors. Additionally, we add a third source for the UV and optical subsets: the Poisson error. This term is negligible for infrared and submm observations because of the large number of incoming photons.

		\subsubsection{Background variations}
		
Variations of the background can give rise to errors in the flux measurements. To estimate the impact of this term, we select regions around M31 where the background is dominant. A sigma clipping filter is used on the pixels inside each region. Different methods of sigma clipping were evaluated. They proved not to affect the resulting variance by much as the evaluated regions were chosen to be free of point sources. In the end, a low--level clipping was done ($10\sigma$ and only two iterations) to filter out any non-background emission. The variance of the remaining pixels from all of the regions will be a good representation of the background variation error $\sigma_\mathrm{background}$, following Equation D2 from \citet{Aniano2012},

\begin{equation}
\sigma_\mathrm{background} = \sqrt{\frac{1}{N_\mathrm{bg}-3} \sum_\mathrm{(x,y)} [I^\mathrm{obs}(x,y)]^2 },
\end{equation}
where $N_\mathrm{bg}$ is the number of background pixels used and $I^\mathrm{obs}$ the observed background subtracted flux of the pixel with coordinates x and y. For each telescope, a different set of background regions was used. This was needed because the background and Galactic Cirrus features change in morphology and brightness along the electromagnetic spectrum. 

		\subsubsection{Poisson errors}

Photon arrivals are considered a random event following a Poisson distribution; the variability of the counts scales with the square root of the number of photons. Infrared and submm observations deal with huge numbers of photons (all of which have low energy) and consequently have a negligible Poisson error compared to calibration uncertainties and background variations.

Optical and UV observations, however, do not collect as many photons and consequently their Poisson-like nature could start to play a more prominent role.
In the cases of the GALEX and SDSS observations, the images were converted from flux to actual photon counts using the individual exposure time for each pixel and then converted to counts/Sr. This surface density unit was necessary to rebin and rescale the frames to the SPIRE $500~\mu\mathrm{m}$ pixel grid without disrupting the exposure time information. We note that convolution did not take place here. The resulting images were converted back to counts using the new pixel scale and from here the Poisson errors could be computed for each individual pixel by taking the square root of the counts.

		\subsubsection{Calibration errors}

In higher signal-to-noise areas, the calibration of the instrumentation can become a dominant term. We therefore include a fixed percentage as calibration uncertainty for each filter (see Table~\ref{tab:caliberrors}) in addition to the previous error terms.

\begin{table}
\caption{Overview of the adopted relative calibration uncertainties for each pixel. }
\label{tab:caliberrors}
\centering     
\begin{tabular}{>{\centering\arraybackslash}m{2.5cm}>{\centering\arraybackslash}m{1.5cm}>{\centering\arraybackslash}m{1.5cm}}
\hline
\hline
Filter & Error & Reference \\  [1ex]
\hline
GALEX-FUV & $5 \%$ & a \\ [1ex]
GALEX-NUV & $3 \%$ & a \\ [1ex]
SDSS & $2 \%$ & b \\ [1ex]
WISE-W1 & $2.4 \%$ & c \\[1ex]
WISE-W2 & $2.8 \%$ & c \\[1ex]
WISE-W3 & $4.5 \%$ & c \\[1ex]
WISE-W4 & $5.7 \%$ & c \\[1ex]
IRAC-$3.6$ & $8.3 \%$ & d \\ [1ex]
IRAC-$4.5$ & $7.1 \%$ & d \\ [1ex]
IRAC-$5.8$ & $22.1 \%$ & d \\ [1ex]
IRAC-$8$ & $16.7 \%$ & d \\ [1ex]
MIPS 24& $4 \%$ & e \\ [1ex]
MIPS 70& $10 \%$ & f \\ [1ex]
PACS & $10 \%$ & g \\ [1ex]
SPIRE & $7 \%$ & h \\ [1ex]
\hline 
\end{tabular}
\tablebib{
(a) \citet{Morrissey2007}; (b) \citet{Padmanabhan2008}; (c) \citet{Jarrett2011}; (d) \citet{Aniano2012}; (e) \citet{Engelbracht2007}; (f) \citet{Gordon2007}; (g) paper I; (h) \citet{SPIREmanual}.
} 
\end{table}

Finally, all error sources are added in quadrature for each pixel $i$ and each wavelength band $\lambda$ to obtain its total photometric uncertainty

\begin{equation}
\sigma_{\lambda,i}^\mathrm{Tot} = \sqrt{(\sigma_{\lambda,i}^\mathrm{bg})^2 + (\sigma_{\lambda,i}^\mathrm{cal})^2 + (\sigma_{\lambda,i}^\mathrm{Pois})^2} \; .
\end{equation}

\section{Integrated photometry of the separate regions} \label{app:fluxes}

\begin{table*}
\caption{Overview of the obtained fluxes for the different regions of Andromeda. All measurements are in units of Jansky.}
\label{tab:fluxes}
\centering     
\begin{tabular}{>{\centering\arraybackslash}m{2.0cm}>{\centering\arraybackslash}m{2.0cm}>{\centering\arraybackslash}m{2.5cm}>{\centering\arraybackslash}m{2.5cm}>{\centering\arraybackslash}m{2.0cm}>{\centering\arraybackslash}m{2.0cm}>{\centering\arraybackslash}m{3.0cm}}
\hline
\hline
Band	&	Global			&	Bulge			&	Inner disk			&	Ring			&	Outer disk			&	M32		\\
\hline	
FUV	&	1.483	$\pm$	0.074	&	0.042	$\pm$	0.0021	&	0.1742	$\pm$	0.0087	&	0.451	$\pm$	0.023	&	0.811	$\pm$	0.041	&	0.00587	$\pm$	0.00029	\\
NUV	&	2.67	$\pm$	0.08	&	0.1403	$\pm$	0.0042	&	0.447	$\pm$	0.013	&	0.763	$\pm$	0.023	&	1.304	$\pm$	0.039	&	0.01807	$\pm$	0.00054	\\
$u$	&	18.71	$\pm$	0.37	&	3.418	$\pm$	0.068	&	5.59	$\pm$	0.11	&	3.691	$\pm$	0.074	&	5.77	$\pm$	0.12	&	0.259	$\pm$	0.0052	\\
$g$	&	84.2	$\pm$	1.7	&	17.55	$\pm$	0.35	&	27.88	$\pm$	0.56	&	15.34	$\pm$	0.31	&	22.37	$\pm$	0.45	&	1.17	$\pm$	0.023	\\
$r$	&	183.4	$\pm$	3.7	&	38.88	$\pm$	0.78	&	63.5	$\pm$	1.3	&	33.73	$\pm$	0.67	&	45.1	$\pm$	0.9	&	2.394	$\pm$	0.048	\\
$i$	&	279.6	$\pm$	5.6	&	60.7	$\pm$	1.2	&	99	$\pm$	2	&	51	$\pm$	1	&	65	$\pm$	1.3	&	3.596	$\pm$	0.072	\\
$z$	&	342	$\pm$	6.8	&	80	$\pm$	1.6	&	120.6	$\pm$	2.4	&	56.3	$\pm$	1.1	&	80.9	$\pm$	1.6	&	4.52	$\pm$	0.09	\\
W1	&	287.3	$\pm$	6.9	&	63.4	$\pm$	1.5	&	103.2	$\pm$	2.5	&	56.2	$\pm$	1.3	&	61.4	$\pm$	1.5	&	3.372	$\pm$	0.081	\\
IRAC 3.6	&	286	$\pm$	24	&	63.1	$\pm$	5.2	&	103.2	$\pm$	8.6	&	56.1	$\pm$	4.7	&	60	$\pm$	5	&	3.37	$\pm$	0.28	\\
IRAC 4.5	&	161	$\pm$	11	&	36.7	$\pm$	2.6	&	58.9	$\pm$	4.2	&	32.3	$\pm$	2.3	&	31.5	$\pm$	2.2	&	2.08	$\pm$	0.15	\\
W2	&	158.5	$\pm$	4.4	&	33.7	$\pm$	0.94	&	56.2	$\pm$	1.6	&	32.1	$\pm$	0.9	&	34.77	$\pm$	0.97	&	1.855	$\pm$	0.052	\\
IRAC 5.8	&	222	$\pm$	49	&	34.4	$\pm$	7.6	&	68	$\pm$	15	&	59	$\pm$	13	&	59	$\pm$	13	&	1.75	$\pm$	0.39	\\
IRAC 8	&	211	$\pm$	35	&	21.2	$\pm$	3.5	&	50	$\pm$	8.3	&	94	$\pm$	16	&	44.4	$\pm$	7.4	&	1.53	$\pm$	0.25	\\
W3	&	209.3	$\pm$	9.4	&	14.48	$\pm$	0.65	&	48.4	$\pm$	2.2	&	79.6	$\pm$	3.6	&	66	$\pm$	3	&	1.015	$\pm$	0.046	\\
W4	&	164.2	$\pm$	9.4	&	9.51	$\pm$	0.54	&	36	$\pm$	2	&	69	$\pm$	3.9	&	49.3	$\pm$	2.8	&	0.671	$\pm$	0.038	\\
MIPS 24	&	119	$\pm$	48	&	8	$\pm$	3	&	26	$\pm$	10	&	56	$\pm$	22	&	29	$\pm$	12	&	0.5	$\pm$	0.2	\\
MIPS 70	&	1051	$\pm$	110	&	97.9	$\pm$	9.8	&	228	$\pm$	23	&	492	$\pm$	49	&	233	$\pm$	23	&	1.49	$\pm$	0.15	\\
PACS 100	&	3500	$\pm$	350	&	195	$\pm$	19	&	805	$\pm$	81	&	1601	$\pm$	160	&	893	$\pm$	89	&	8.13	$\pm$	0.82	\\
PACS 160	&	7526	$\pm$	750	&	199	$\pm$	20	&	1630	$\pm$	160	&	3588	$\pm$	360	&	2094	$\pm$	210	&	18.9	$\pm$	1.9	\\
SPIRE 250	&	5952	$\pm$	420	&	87.4	$\pm$	6.1	&	1126	$\pm$	79	&	2736	$\pm$	190	&	1992	$\pm$	140	&	13.25	$\pm$	0.93	\\
SPIRE 350	&	3122	$\pm$	220	&	33.8	$\pm$	2.4	&	520	$\pm$	36	&	1395	$\pm$	98	&	1168	$\pm$	82	&	7.29	$\pm$	0.51	\\
SPIRE 500	&	1350	$\pm$	94	&	11.76	$\pm$	0.82	&	201	$\pm$	14	&	583	$\pm$	41	&	551	$\pm$	39	&	3.37	$\pm$	0.24	\\
\hline 
\end{tabular}

\end{table*}

\end{document}